\RequirePackage{fix-cm}

\documentclass[twocolumn]{svjour3}

\smartqed

\usepackage{booktabs} 

\usepackage{graphicx}
\usepackage{balance}  
\usepackage{times}
\usepackage{multirow}
\usepackage{algorithmicx}

\usepackage{appendix}

\usepackage{hyperref}

\usepackage{caption}
\usepackage{subfigure}

\usepackage{algpseudocode}

\usepackage[linesnumbered, ruled, vlined]{algorithm2e}

\SetKwRepeat{Do}{do}{while}

\usepackage{tabu}
\usepackage[export]{adjustbox}
\usepackage{amsmath}

\usepackage{amssymb}

\usepackage{color}
\definecolor{Xiang}{rgb}{1,0,0}
\definecolor{Weilong}{rgb}{0,0,1}

\newcommand{\comment}[1]{}
\newcommand{\nop}[1]{}

\begin{document}
\title{Skyline Queries Over Incomplete Data Streams (Technical Report)}

\author{Weilong Ren$^1$ \and
        Xiang Lian$^1$ \and
        Kambiz Ghazinour$^{1,2}$}

\authorrunning{W. Ren et al.}

\institute{Weilong Ren, Xiang Lian, and Kambiz Ghazinour \at
\email{\{wren3, xlian, kghazino\}@kent.edu} \\
$^1$ Department of Computer Science, Kent State University, Kent, OH 44242, USA \\
$^2$ Center for Criminal Justice, Intelligence and Cybersecurity, State University of New York, Canton, NY 13617, USA
}

\date{}

\maketitle

\begin{abstract}
Nowadays, efficient and effective processing over massive stream data has attracted much attention from the database community, which are useful in many real applications such as sensor data monitoring, network intrusion detection, and so on. In practice, due to the malfunction of sensing devices or imperfect data collection techniques, real-world stream data may often contain missing or incomplete data attributes. In this paper, we will formalize and tackle a novel and important problem, named \textit{skyline query over incomplete data stream} (Sky-iDS), which retrieves skyline objects (in the presence of missing attributes) with high confidences from incomplete data stream. In order to tackle the Sky-iDS problem, we will design efficient approaches to impute missing attributes of objects from incomplete data stream via \textit{differential dependency} (DD) rules. We will propose effective pruning strategies to reduce the search space of the Sky-iDS problem, devise cost-model-based index structures to facilitate the data imputation and skyline computation at the same time, and integrate our proposed techniques into an efficient Sky-iDS query answering algorithm. Extensive experiments have been conducted to confirm the efficiency and effectiveness of our Sky-iDS processing approach over both real and synthetic data sets.
\end{abstract}

\nop{
\begin{abstract}

Nowadays, efficient and effective processing over massive stream data has attracted much attention from the database community, which are useful in many real applications such as sensor data monitoring, network intrusion detection, and so on. In practice, due to the malfunction of sensing devices or imperfect data collection techniques, real-world stream data may often contain missing or incomplete data attributes. In this paper, we will formalize and tackle a novel and important problem, named \textit{skyline query over incomplete data stream} (Sky-iDS), which retrieves skyline objects (in the presence of missing attributes) with high confidences from incomplete data stream. In order to tackle the Sky-iDS problem, we will design efficient approaches to impute missing attributes of objects from incomplete data stream via \textit{differential dependency} (DD) rules. We will propose effective pruning strategies to reduce the search space of the Sky-iDS problem, devise cost-model-based index structures to facilitate the data imputation and skyline computation at the same time, and integrate our proposed techniques into an efficient Sky-iDS query answering algorithm. Extensive experiments have been conducted to confirm the efficiency and effectiveness of our Sky-iDS processing approach over both real and synthetic data sets.
\end{abstract}
}



\section{Introduction}

\begin{figure}[t!]
\centering\vspace{-2ex}
\hspace{-4ex}\includegraphics[scale=0.48]{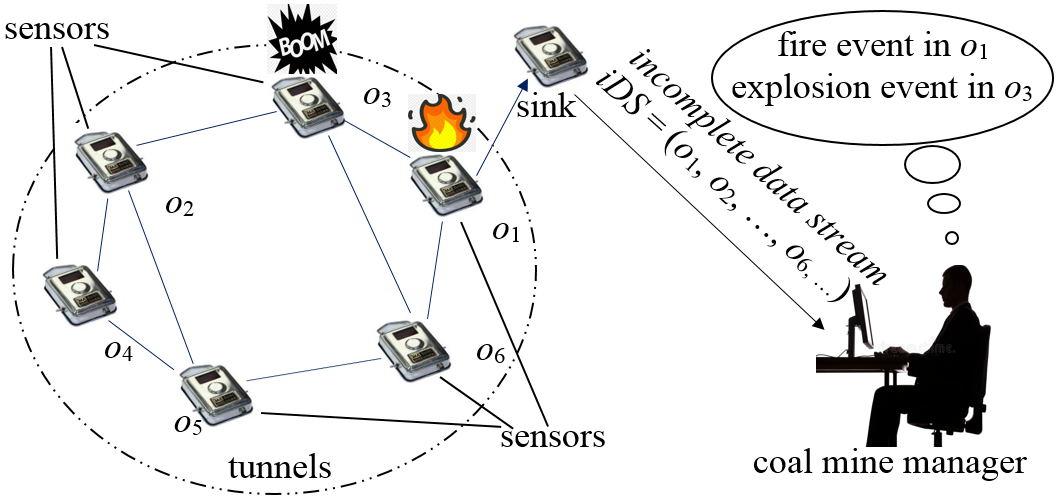}\vspace{-2ex}
\caption{\small An example of the coal mine surveillance}
\label{fig:surveillanceEx}
\end{figure}

For decades, efficient management over massive data streams has received much attention in many real applications such as IP network traffic analysis \cite{cranor2003gigascope}, network intrusion detection \cite{igbe2016distributed}, sensor networks \cite{aberer2007infrastructure}, telephone call record management \cite{golab2003issues}, Web log and clickstream mining \cite{srivastava2000web}, and so on. As an example, Figure \ref{fig:surveillanceEx} shows an application of the coal mine surveillance \cite{Xue06}, where sensors are deployed at different sites in tunnels of the coal mine, and collect data attributes such as the densities of gas/oxygen/dust and temperature. These sensory samples are periodically obtained from each sensor, and transmitted back to a \textit{sink} in a streaming manner for real-time analysis, for example, detecting potentially abnormal events such as fire or gas explosion. 

Table \ref{example_table1} depicts the sensory data stream, $iDS = (o_1, o_2,$ $o_3, o_4,$ $o_5, o_6, o_1, o_2, ...)$, collected from sensors and received by the sink (as shown in Figure \ref{fig:surveillanceEx}) in the order of their arrival times. Each record with sensor ID $o_i$ (for $1\leq i\leq 6$) has four sampled attributes such as temperature and densities of gas/oxygen/dust, which is associated with record arrival time and expiration time. For example, sensor (object) $o_1$ sends a sample record with attributes temperature 100 $^\circ$F, and the densities of gas, oxygen, and dust all equal to $3$, which arrives at the sink at timestamp $1$ and will expire at timestamp $6$, with a valid duration $5$ ($=6-1$). Similarly, objects $o_2 \sim o_6$ arrive at different times in a streaming fashion, and may have distinct valid durations (due to different sensor sampling rates). 

\setlength{\textfloatsep}{0pt}
\hspace{-3ex}


In order to timely detect dangerous events such as fire or explosion in the coal mine, one important query type in such a streaming scenario is the \textit{skyline query} \cite{Borzsonyi01}, which returns those sensors (and their locations in the coal mine) with high risks of incurring abnormal events (e.g., explosion event with both high temperature and density of gas). Specifically, given a database $D$, a skyline query retrieves those objects $o\in D$ that are not \textit{dominated} by other objects in $D$, where we say an object $o$ \textit{dominates} another object $o'$ (denoted as $o \prec o'$), iff two conditions hold: (1) $o[A_i]\geq o'[A_i]$, for \textit{all} attributes $A_i$, and; (2) $o[A_j] > o'[A_j]$, for \textit{at least one} attribute $A_j$. 

\begin{table}[t!]
\caption{\small An incomplete data stream, $iDS$, collected from sensor networks in Figure \ref{fig:surveillanceEx}.}
\label{example_table1}\vspace{-1ex}
\hspace{-2ex}\scriptsize
\begin{tabular}{|c|c|c||c|c|c|c|}
\hline
\textbf{sensor ID} & \textbf{arr.} & \textbf{exp.} & \textbf{temperature} & \textbf{density} & \textbf{density} & \textbf{density}\\
\textbf{(object)} & \textbf{time} & \textbf{time} & \textbf{($^\circ$F)} & \textbf{of gas} & \textbf{of oxygen} & \textbf{of dust}\\
\hline
\hline
$o_1$ & 1 & 6 & 100 & 3 & 3 & 3 \\\hline
$o_2$ & 2 & 6 & 50 & 1 & 1 & 1 \\\hline
$o_3$ & 3 & 9 & 90 & 2 & $-$ & 3 \\\hline
$o_4$ & 3 & 9 & 60 & $-$ & 1 & $-$ \\\hline
$o_5$ & 6 & 11 & 70 & 2 & 2 & $-$ \\\hline
$o_6$ & 6 & 10 & $-$ & 2 & 3 & 2 \\\hline
$o_1$ & 7 & 12 & 80& 2 & 2 & $2$ \\\hline
$o_2$ & 8 & 12 & $90$ & 1 & 3 & 3 \\\hline
... & ... & ... & ... & ... & ... & ... \\\hline
\end{tabular}
\end{table}

\begin{table}[t!]\vspace{-2ex}
\caption{\small An incomplete data stream, $iDS$, collected from computer networks.}
\label{intrusion_example_table2}\vspace{-1ex}
\hspace{-2ex}\scriptsize
\begin{tabular}{|c|c|c||c|c|c|}
\hline
\textbf{router} & \textbf{arr.} & \textbf{exp.} & \textbf{[A] No. of} & \textbf{[B] connection} & \textbf{[C] transferred}\\
\textbf{ID} & \textbf{time} & \textbf{time} & \textbf{connections} & \textbf{duration} & \textbf{data size}\\
\textbf{(object)} &  &  & \textbf{($\times 10^3$)} & \textbf{(min)} & \textbf{(GB)}\\
\hline
\hline
$T_1$ & 1 & 6 & 0.5 & 0.5 & 0.2 \\\hline
$T_2$ & 2 & 6 & 0.5 & 0.2 & 0.5 \\\hline
$T_3$ & 3 & 9 & 0.5 & 0.5 & 0.5 ($-$) \\\hline
... & ... & ... & ... & ... & ... \\\hline
\end{tabular}
\end{table}

Note that, in this example of the coal mine surveillance, to detect sensors with high risks, one straightforward solution is to look at sensory values from each sensor using existing methods \cite{song2015screen,zhang2016sequential,zhang2017time}. However, such a solution may encounter the problem of setting the alarming thresholds for different attributes, which are difficult to tune by the coal mine manager. In contrast, our skyline query does not require the specification of such thresholds, and can directly return users with the most probable objects (i.e., sensor locations) in danger (e.g., sensors with fire/explosion events). The skyline considers multiple attributes (rather than just the value of one single attribute), which can be used for multi-criteria decision making. For skylines, we can obtain the locations of sensors that may have the most dangerous events (not dominated by other sensors). Under the dominance semantics between sensory objects, if a sensor $S_1$ dominates another sensor $S_2$, then we consider that the location of sensor $S_1$ is more dangerous than that of sensor $S_2$. 

\nop{
{\color{Weilong} Reviewer 1 mentioned $n$ typically denotes the number of objects, do we need to use another notation (e.g., $v$) to replace $n$?} {\color{Xiang} use $o'$} {\color{Weilong} I replaced $n$ by $o'$ in Sections 2 and 3. But I didn't change it in Section 4, since notations like $o'^p.min$ and $o_i^p.min$ may be not easy to be distinguished in a printed version.}  {\color{Xiang} ok, please remove blue part}
}


In the previous example of Table \ref{example_table1}, object $o_1$ dominates object $o_2$, since each of the four attributes (i.e., temperature and densities of gas/oxygen/dust) in object $o_1$ are greater than that of object $o_2$. Thus, up to timestamp $2$, the sink has only received two objects $o_1$ and $o_2$, and $o_1$ is the skyline answer (since it is not dominated by other object like $o_2$). Intuitively, the skyline answers, for example, sensor (object) $o_1$, indicate high risks of abnormal events (i.e., high temperature and/or density measures compared with other sensors), which require immediate attentions from the coal mine manager (for potential evacuation to save the lives of workers). Therefore, it is very critical, yet challenging, to study efficient and effective processing of skyline queries over such data streams. 


Due to transmission errors, packet losses, low battery power, or environmental factors, some sensory data attributes may be missing and thus incomplete. For example, in Table \ref{example_table1}, object $o_3$ has an incomplete attribute, the density of oxygen, whose missing value is denoted by ``$-$''. Similarly, objects $o_4 \sim o_6$ contain 1 or 2 missing attributes each. Due to the missing information, inaccurate skyline answers over incomplete streams may lead to wrong decision making about the coal mine evacuation, or even false alarms that incur losses of millions of dollars resulting from unnecessary evacuation. In such a scenario with incomplete data, it is even more challenging and important to process skyline queries efficiently and accurately over incomplete data streams. 

Inspired by the example above, in this paper, we will formally propose the problem of the \textit{skyline query over incomplete data streams} (Sky-iDS), which retrieves those skyline objects from incomplete data streams with high confidences. The Sky-iDS problem has many other real applications such as the network intrusion detection \cite{igbe2016distributed}. 

Specifically, in computer networks, spatially distributed routers often suffer from malicious network intrusion, where each router is connected with a number of servers. Since the network intrusion may lead to serious consequences such as virus installation, network congestion, and leakage of users' information, it is very crucial to online monitor and prevent the network intrusion, based on network statistics such as \textit{No. of connections} (denoted as $A$), \textit{connection duration} (denoted as $B$), and \textit{transferred data size} (denoted as $C$) \cite{dhanabal2015study} (as depicted in Table \ref{intrusion_example_table2}). In reality, there are many routers in IP networks, and a large volume of the collected streaming network statistics arrive at fast speed, which is rather challenging for network security people to efficiently and accurately monitor. What is more, some network statistics may be missing/lost, for reasons such as the network failure, cyber attacks, or network congestion. Therefore, in this case, network security users can issue a skyline query over such incomplete network statistics from the data stream. 


As an example in Table \ref{intrusion_example_table2}, for each router, $T$, we use $T=(A, B, C)$ to represent its collected network statistics, where $A$, $B$ and $C$ are normalized to $[0, 1]$. At each timestamp, given the collected network statistics from three routers, $T_1=(0.5, 0.5, 0.2)$, $T_2=(0.5, 0.2, 0.5)$, and $T_3 = (0.5, 0.5, 0.5)$, network security people can obtain router $T_3$ as the only skyline router, based on dominance relationships among $T_1\sim T_3$. Intuitively, $T_3$ is the router that may be under attack with the highest probability among the three routers, and should be reported to network security people. If $T_3$ is safe (i.e., not under attack), then network security people may not need to monitor the rest two routers (i.e., $T_1$ and $T_2$), since $T_1$ and $T_2$ are dominated by $T_3$. However, in practice, these network statistics may be potentially unavailable (e.g., missing due to the network failure or network congestion). For instance, when \textit{transferred data size} (i.e., attribute $C$) of $T_3$ is not available (i.e., $T_3=(0.5, 0.5, -)$), it is not trivial how to retrieve skylines over such incomplete data from the stream. In this scenario, we can exactly issue a Sky-iDS query to monitor skylines over such a (incomplete) network data stream, which correspond to the routers with high risks of being under cyber attacks.



Note that, while prior works \cite{gao2014processing,khalefa2008skyline} studied the skyline query over \textit{static} incomplete databases, their proposed approaches compute skylines by simply ignoring those missing attributes (when considering dominance relationships), which may incur biased or wrong skyline results (Please refer to Section \ref{sec:related_work} for a detailed example).
Instead, in this paper, we will consider the imputation of missing attributes in data streams via \textit{differential dependency} (DD) rules \cite{song2011differential}, which allows the skyline computation with all (complete or imputed) attributes and results in unbiased skylines with high confidences. Moreover, to the best of our knowledge, this is the first work to study the skyline operator over incomplete data in the streaming environment.

Specifically, in the streaming scenario, Sky-iDS query processing requires high efficiency, which is critical and important in many real applications. For example, as shown in Fig. \ref{fig:surveillanceEx}, the coal mine manager needs to quickly and timely detect dangerous fire events (i.e., Sky-iDS answers), and immediately take actions. If Sky-iDS query answering is slow, then it may lead to enormous economic loss or even threaten people's lives. Similarly, in the scenario of network intrusion detection, high Sky-iDS processing cost may cause more servers and computers under attack. Therefore, it is important that we can efficiently retrieve Sky-iDS answers from incomplete data streams in these scenarios (otherwise, serious consequences like economic/life losses or network intrusion may occur). While a straightforward method can conduct the skyline query \textit{after} the data imputation, it may still take a long time to obtain the Sky-iDS answers, which is not suitable for fast stream processing. Thus, in our work, we design an efficient Sky-iDS approach that integrates data imputation and skyline query at the same time, which can perform much better than the straightforward method.



Therefore, due to stream processing requirements such as efficient stream processing and limited memory consumption, in this paper, we will design cost-model-based and space-efficient index structures for both data imputation and query processing, devise effective pruning methods to greatly reduce the Sky-iDS search space, and propose efficient Sky-iDS answering algorithms to perform the attribute imputation and incremental skyline computation at the same time (i.e., ``imputation and query processing at the same time'' style).

In this paper, we make the following major contributions. 
\vspace{-1ex}
\begin{enumerate}
\item We formalize a novel and important problem of the \textit{skyline query over incomplete data stream} (Sky-iDS) in Section \ref{sec:problemDefinition}.

\item We design effective and efficient data imputation techniques via DD rules in Section \ref{sec:impuation_of_io}.


\item We propose effective pruning strategies to reduce the search space of the Sky-iDS problem in Section \ref{sec:pruning_strategies}.

\item We devise effective indexes and efficient algorithms to tackle the Sky-iDS problem on incomplete data stream in Section \ref{sec:dsp_over_ids}.

\item We demonstrate through extensive experiments the effectiveness and efficiency of our Sky-iDS approach in Section
\ref{sec:experimental_eval}.

\end{enumerate}

In addition, Section \ref{sec:related_work} reviews related works on stream processing, differential dependency, skyline queries, stream outlier detection and repair, and incomplete data management. Section \ref{sec:conclusions} concludes this paper. 

\section{Problem Definition}
\label{sec:problemDefinition}
In this section, we formally define the problem of a \textit{skyline query over incomplete data streams} (Sky-iDS), which takes into account the missing attribute values during the skyline query processing. 

\vspace{-2ex}
\subsection{Incomplete Data Streams}

We first define the data model for incomplete data streams.

\begin{definition} {\textbf{(Incomplete Data Streams)}} An \textit{incomplete da-ta stream}, $iDS$, is an ordered sequence of objects, $\{o_1, o_2, o_3$, $...$, $o_r$, $...\}$, where objects $o_i$ arrive at timestamp $o_i.arr$, and expire at timestamp $o_i.exp$. Each object $o_i$ contains $d$ attributes $A_j$ (for $1\leq j\leq d$), some of which have missing attribute values $o_i[A_j]$, represented by ``$-$''.
\label{def:iDS}
\end{definition}

In Definition \ref{def:iDS}, an incomplete data stream $iDS$ dynamically keeps in memory all objects that are currently valid (i.e., not expired). When a new object $o_t$ arrives, $o_t$ will be inserted into $iDS$; whenever an old object $o_i \in iDS$ expires at timestamp $o_i.exp$, it will be evicted from $iDS$. Each object $o_i \in iDS$ has a valid period from timestamp $o_i.arr$ to timestamp $o_i.exp$, with a duration $o_i.dur$ ($= o_i.exp - o_i.arr$). 

In the example of Figure \ref{fig:surveillanceEx} and Table \ref{example_table1}, the incomplete data stream is given by $iDS = (o_1, o_2, ...)$, in which objects like $o_3$ contain incomplete attributes (e.g., the missing attribute, the density of oxygen, for object $o_3$). At timestamp 6, new objects $o_5$ and $o_6$ are added to $iDS$, whereas old expired objects $o_1$ and $o_2$ are removed from $iDS$, which results in valid objects $\{o_3,o_4,o_5,o_6\}$.

Without loss of generality, in this paper, we use $W_t$ to denote a set of objects in $iDS$ that are valid (i.e., not expired) at timestamp $t$. As shown in the example of Table \ref{example_table1}, at timestamp $t=2$, we have $W_2 = \{o_1, o_2\}$. At timestamp $t=6$, we have $W_6 = \{o_3, o_4, o_5, o_6\}$. Similarly, at timestamp $t=8$, we have $W_8 = \{o_3, o_4, o_5, o_6, o_1, o_2\}$. Note that, here objects $o_1$ and $o_2$ in $W_8$ are new updates at timestamp $8$ from sensors $o_1$ and $o_2$, respectively, which are different from that in $W_2$ at timestamp 2.

\subsection{Imputation Over Incomplete Data Stream}
\label{Imputation_over_iDS}

\noindent {\bf Imputed Data Stream.} To leverage the processing on incomplete data streams, in this paper, we will impute and model incomplete data stream $iDS$ by \textit{probabilistic data stream} \cite{Ding12}, by estimating possible values of missing attributes in objects from $iDS$.

\begin{definition} {\textbf{(Imputed Data Stream)}} Given an incomplete data stream $iDS = (o_1, o_2, ..., o_r, ...)$, its imputed (complete) data stream, $pDS$, is given by an ordered sequence of objects, $(o_1^p, o_2^p,$ $...,$ $o_r^p, ...)$. 

Each object $o_i^p \in pDS$, obtained from object $o_i\in iDS$ with missing attribute(s) $o_i[A_j]$ (=``$-$''), is a probabilistic object, which consists of instances $o_{il}$ (with the imputed attribute values). Each instance $o_{il}$ is associated with an existence probability $o_{il}.p$, where $\sum_{\forall o_{il}\in o_i^p}$ $o_{il}.p$ $= 1$. \label{def:imputed_data_stream}
\end{definition}

Definition \ref{def:imputed_data_stream} defines a probabilistic data stream $pDS$, imputed from incomplete data stream $iDS$. Specifically, we can estimate and impute possible values of each missing attribute $o_i[A_j]$ in objects $o_i \in iDS$, and represent the resulting probabilistic object $o_i^p$ by several instances $o_{il}$. Each instance $o_{il}$ contains complete/imputed attribute values, associated with an existence probability $o_{il}.p \in (0, 1]$, which indicates the confidence that instance $o_{il}$ actually exists in reality (i.e., truly representing object $o_i$).

\nop{

for any missing attribute $o_i[A_j]$ (=``$-$''), the corresponding attribute $o_i^p[A_j]$ of object $o_i^p$ has $m$ (mutually exclusive) possible values, $o_{ij}[1]$, $o_{ij}[2]$, ..., and $o_{ij}[m]$, each associated with an existence probability $o_{ij}[k].p$ (for $1\leq k\leq m$), where it holds that $\sum_{k=1}^m o_{ij}[k].p = 1$.
}

\nop{

\begin{definition} {\textbf{(Data Imputations, DI)}} Given the object $o_i$ in the \textit{sliding window} $W$ and complete historical databases $HD$, \textit{data imputation}, $DI$, will first find statistical inference rules from the $HD$, and then compare and find some similar data objects $o_s$ from the $HD$ by using the detected rules. 
Leveraging the attribute values $o_s.A_j$ of objects $o_s$, we can fill the missing attribute value $o_i.A_j=`-'$ of object i with candidates with relative probabilities to be correct.
\end{definition}

}

\nop{

Specifically, each probabilistic object $o_i^p\in pDS$ contains uncertain attribute(s) (i.e., imputed missing attributes) $o_i^p[A_j]$. Each uncertain attribute $o_i^p[A_j]$ is represented by its possible attribute values $o_{ij}[k]$, associated with existence probabilities $o_{ij}[k].p$, indicating the confidences that attributes $o_i^p[A_j]$ takes values $o_{ij}[k]$ in reality. 

Equivalently, by considering all possible combinations of uncertain attribute values, each probabilistic object $o_i^p$ in imputed data stream $pDS$ can be captured by several instances, denoted by $o_{il}$, each associated with an existence probability $o_{il}.p$.

}

\begin{table}[t!]
\caption{\small The imputed data stream, $pDS$, at timestamp 6 (i.e., $W_6$) in the example of Table \ref{example_table1}.}\vspace{-1ex}
\label{example_table2}
\scriptsize
\hspace{-3ex}
\begin{tabular}{|c|c||c|c|c|c||c|}
\hline
\textbf{object} &\textbf{instance} & \textbf{temper-} & \textbf{density} & \textbf{density} & \textbf{density} & \textbf{prob.}\\
\textbf{} & \textbf{} & \textbf{ature($^\circ$F)}  & \textbf{of gas} & \textbf{of oxygen} & \textbf{of dust} &\textbf{}\\
\hline
\hline
$o_3^p$ & $o_{31}$ & 90 & 2 & {\bf 2} & 3  & 0.4 \\
    & $o_{32}$ & 90 & 2 & {\bf 3} & 3 & 0.6 \\ \hline
        & $o_{41}$ & 60 & {\bf 1} & 1 & {\bf 1} & 0.56 \\
$o_4^p$ & $o_{42}$ & 60 & {\bf 1} & 1 & {\bf 2}  & 0.24 \\
    & $o_{43}$ & 60 & {\bf 2} & 1 & {\bf 1} & 0.14 \\ 
    & $o_{44}$ & 60 & {\bf 2} & 1 & {\bf 2} & 0.06 \\ \hline
$o_5^p$ & $o_{51}$ & 70 & 2 & 2 & {\bf 2} & 1.0 \\ \hline
$o_6^p$ & $o_{61}$ & {\bf 90} & 2 & 3 & 2 & 0.6 \\
& $o_{62}$ & {\bf 80} & 2 & 3 & 2 & 0.4 \\
\hline
\end{tabular}\vspace{2ex}
\end{table}

Table \ref{example_table2} shows an example of the imputed data stream $pDS$ at timestamp $t=6$ (i.e., $W_6 = (o_3^p, o_4^p, o_5^p, o_6^p$)), obtained from incomplete data stream $iDS$ in Table \ref{example_table1}. As an example, probabilistic object $o_3^p$ has two instances $o_{31}$ and $o_{32}$, with the imputed possible values $2$ and $3$ for attribute ``density of oxygen'', which are associated with existence probabilities 0.4 and 0.6,  respectively. Similarly, probabilistic object $o_4^p$ contains 4 instances $o_{41} \sim o_{44}$, where each missing attribute, ``density of gas'' or ``density of dust'', has two possible (imputed) values (i.e., 1 or 2). In particular, instance $o_{41}$ has ``density of gas'' equal to 1 with probability 0.8, and ``density of dust'' equal to 1 with probability 0.7. Thus, the instance $o_{41}$ has the existence probability 0.56 ($=0.8\times 0.7$).

The cases of probabilistic objects $o_5^p$ and $o_6^p$ are similar, and thus omitted here.

\vspace{1ex} \noindent {\bf Possible Worlds Over Imputed Data Stream.} Following the literature of probabilistic databases \cite{Dalvi07}, we consider the \textit{possible worlds} semantics over (imputed) probabilistic data stream $pDS$ at timestamp $t$, that is, a set, $W_t$, of valid (not expired) objects, where each possible world is a materialized instance of $W_t \in pDS$ that can appear in the real world.

\begin{definition} {\textbf{(Possible Worlds of the Imputed Data Stream, $pw(W_t)$)}} Given an imputed data stream $pDS$ at timestamp $t$ (i.e., $W_t$), a possible world, $pw(W_t)$, of $W_t$ is a set of object instances $o_{il}$, where $o_{il}$ is an instance of probabilistic object $o_i^p \in W_t$ (i.e., satisfying $o_i.exp > t$). 

Each possible world, $pw(W_t)$, has an appearance probability,  $Pr\{pw(W_t)\}$, given as follows:
\begin{eqnarray}
Pr\{pw(W_t)\} = \prod_{\forall o_{il} \in pw(W_t)} o_{il}.p.
\label{eq:eq1}
\end{eqnarray}
\end{definition}

\begin{table}[t!]
\caption{\small Possible worlds, $pw(W_6)$, of $W_6$ from the imputed data stream, $pDS$, at timestamp $6$ in Table \ref{example_table2}.}\vspace{0ex}
\label{example_table3}
\centering\scriptsize
\begin{tabular}{|c|c|c|c|}
\hline
\textbf{possible world of $W_6$} & \textbf{content of $pw(W_6)$} & \textbf{appearance probability}\\
\hline
\hline
$pw_1(W_6)$ & ($o_{31}$, $o_{41}$, $o_{51}$, $o_{61}$) & 0.1344 \\
\hline
$pw_2(W_6)$ & ($o_{31}$, $o_{41}$, $o_{51}$, $o_{62}$) & 0.0896 \\
\hline
$pw_3(W_6)$ & ($o_{32}$, $o_{41}$, $o_{51}$, $o_{61}$) & 0.2016 \\
\hline
$pw_4(W_6)$ & ($o_{32}$, $o_{41}$, $o_{51}$, $o_{62}$) & 0.1344 \\
\hline
... & ... & ...\\
\hline
$pw_{16}(W_6)$ & ($o_{32}$, $o_{44}$, $o_{51}$, $o_{62}$) & 0.0144 \\
\hline
\end{tabular}\vspace{2ex}
\end{table}

In the example of Table \ref{example_table2}, probabilistic objects $o_3^p$, $o_4^p$, $o_5^p$, and $o_6^p$ in $W_6$ have 2, 4, 1, and 2 possible instances, respectively. Therefore, there are totally 16 ($= 2 \times 4 \times 1 \times 2$) possible worlds of $W_6$ over imputed data stream $pDS$ at timestamp 6, as depicted in Table \ref{example_table3}. The appearance probability of each possible world can be computed by Eq.~(\ref{eq:eq1}), for example, 
$Pr\{pw_1(W_6)\} = o_{31}.p \times o_{41}.p \times o_{51}.p \times o_{61} = 0.4 \times 0.56 \times 1 \times 0.6 = 0.1344$.

\vspace{-3ex}
\subsection{Skyline Queries on Incomplete Data Stream}
\label{subsec:skyline_query_on_iDS}
\vspace{-2ex}In this subsection, we will define the skyline query over incomplete data streams (Sky-iDS). Before we introduce the Sky-iDS query, we first provide the definition of the dominance between two certain (or imputed probabilistic) objects. 

\nop{
\begin{definition} {\textbf{(Dominance Between Certain Objects $t$ and $o$ \cite{Borzsonyi01})}} 
Given a fixed query point $q$ and two objects $t$ and $o$, we say that object $t$ dominates object $o$ with respect to $q$, denoted by $t \prec_q o$, if two conditions are satisfied: 
\begin{itemize}
    \item for any dimension $1\leq i\leq d$, it holds that $|t[A_i] - q[A_i]| \leq |o[A_i] - q[A_i]|$, and;
    \item for some dimension $1 \leq j \leq d$, $|t[A_j] - q[A_j]| < |o[A_j] - q[A_j]|$.
\end{itemize} \label{def:dominance_certain_object}
\end{definition}
}

\begin{definition} {\textbf{(Dominance Between Certain Objects $o$ and $o'$ \cite{Borzsonyi01})}} 
Given two objects $o$ and $o'$, we say that object $o$ dominates object $o'$, denoted by $o \prec o'$, if two conditions are satisfied: 
\begin{itemize}
    \item for any dimension $1\leq i\leq d$, $o[A_i] \ge o'[A_i]$ holds, and;
    \item for some dimension $1 \leq j \leq d$, $o[A_j] > o'[A_j]$ holds.
\end{itemize} \label{def:dominance_certain_object}
\end{definition} 

Without loss of generality, in this paper, we use ``the larger, the better'' semantics (i.e., larger attribute values are better) for the dominance definition (and skyline as discussed later). Intuitively, as given in Definition \ref{def:dominance_certain_object}, object $o$ dominates object $o'$, if and only if two conditions hold: (1) $o$ is not worse than $o'$ for all attributes $A_i$, and (2) $o$ is strictly better than $o'$ on at least one attribute $A_j$. If only the first condition is satisfied, we denote it as $o \preccurlyeq o'$. 
\nop{{\color{Xiang} (Weilong, please swap $o$ with $o'$ in definition and descriptions)}
 \color{Weilong} Hi professor, it is addressed in Definitions 4 and 5.}

In the example of Table \ref{example_table1}, object $o_1$ dominates $o_2$, since all the four attribute values of $o_1$ are larger than that of $o_2$, respectively.

Next, we define the dominance probability between two imputed probabilistic objects $o^p$ and $o'^p$.

\nop{
\begin{definition} {\textbf{(Dominance Probability Between the Imputed Probabilistic Objects $t^p$ and $o_i^p$)}} 
Given a fixed query point $q$ and two imputed probabilistic objects $t^p$ and $o_i^p$, the \textit{dominance probability}, $Pr\{t^p\prec_q o_i^p\}$, between $t^p$ and $o_i^p$ is given by:
\begin{eqnarray}
Pr\{t^p\prec_q o_i^p\} = \sum_{\forall t\in t^p}\sum_{\forall o\in o_i^p} t.p \cdot o.p \cdot \chi(t \prec_q o),
\label{eq:eq2}
\end{eqnarray}
where $t$ and $o$ are instances of probabilistic objects $t^p$ and $o_i^p$, respectively, and $\chi(z)$ is either 1 (if $z$ is $true$) or 0 (if $z$ is $false$).
\label{def:dominance_prob}
\end{definition}
}
\begin{definition} {\textbf{(The Dominance Probability Between the Imputed Probabilistic Objects $o^p$ and $o'^p$)}} 
Given two imputed probabilistic objects $o^p$ and $o'^p$, the \textit{dominance probability}, $Pr\{o^p\prec o'^p\}$, between $o^p$ and $o'^p$ is given by:
\begin{eqnarray}
Pr\{o^p\prec o'^p\} = \sum_{\forall o\in o^p}\sum_{\forall o'\in o'^p} o.p \cdot o'.p \cdot \chi(o \prec o'),
\label{eq:eq2}
\end{eqnarray}
where $o$ and $o'$ are instances of probabilistic objects $o^p$ and $o'^p$, respectively, and $\chi(z)$ is either 1 (if $z$ is $true$) or 0 (if $z$ is $false$).
\label{def:dominance_prob}
\end{definition}

As an example in Table \ref{example_table2}, we compute the dominance probability,  $Pr\{o_3^p \prec o_6^p\}$, between two probabilistic objects $o_3^p$ and  $o_6^p$. In particular, we first consider the dominance relationships between instances from $o_3^p$ and $o_6^p$ (based on Definition \ref{def:dominance_certain_object}), and thus have: $\chi(o_{31} \prec o_{61})=0$, $\chi(o_{31} \prec o_{62})=0$, $\chi(o_{32} \prec o_{61})=1$, and $\chi(o_{32} \prec o_{62})=1$ . Then, by Eq.~(\ref{eq:eq2}), we can obtain the dominance probability: $Pr\{o_3^p \prec o_6^p\} = 
o_{31}.p \times o_{61}.p \times 0 + o_{31}.p \times o_{62}.p \times 0 + o_{32}.p \times o_{61}.p \times 1 + o_{32}.p \times o_{62}.p \times 1  = 0.6$.

\begin{definition} {\textbf{(Skyline Queries Over Incomplete Data Str-\\eam, Sky-iDS)}} Given an incomplete data stream $iDS$ and a probabilistic threshold $\alpha$, a \textit{skyline query over incomplete data stream} (Sky-iDS) continuously monitors those objects $o_i\in W_t$ from $iDS$ at any timestamp $t$, such that their imputed probabilistic objects $o_i^p$ are not dominated by other imputed objects $o_j^p \in W_t$ with skyline probabilities, $P_{Sky\text{-}iDS}(o_i^p)$, greater than threshold $\alpha$, that is, 
\begin{eqnarray}
&&\hspace{-3ex}P_{Sky\text{-}iDS}(o_i^p)\label{eq:eq3}\\
&\hspace{-3ex}=&\hspace{-3ex}\sum_{\forall pw(W_t)} Pr\{pw(W_t)\} \cdot \chi\left(\bigwedge_{\forall o_j^p\ne o_i^p \text{ and } o_{il}, o_{js}\in pw(W_t)} o_{js} \nprec o_{il} \right)\hspace{-3ex}\notag\\
&\hspace{-3ex}>&\hspace{-1ex}\alpha,\notag
\end{eqnarray}
\noindent where $pw(W_t)$ is a possible world of $W_t$ containing instances $o_{il}$ or $o_{js}$ of objects $o_i^p, o_j^p \in W_t$, respectively, $o_{js} \nprec o_{il}$ indicates that $o_{js}$ is not dominated by $o_{il}$, and $\chi(z)$ is given in Definition \ref{def:dominance_prob}.
\label{def:sky-iDS}
\end{definition}

Intuitively, users can register a Sky-iDS query in Definition \ref{def:sky-iDS} by specifying a parameter $\alpha$, which will continuously monitor those skyline objects over incomplete data stream $iDS$ with high confidences (i.e., satisfying Inequality~(\ref{eq:eq3})).

As an example in Table \ref{example_table2}, at timestamp $t=6$, the Sky-iDS query will compute skyline answers over $W_6 = \{o_3^p, o_4^p,$ $o_5^p, o_6^p\}$. Specifically,  as given in Definition \ref{def:sky-iDS}, we need to enumerate all possible worlds $pw_1(W_6) \sim pw_{16}(W_6)$ (as shown in Table \ref{example_table3}), and compute the skyline probability, for example, $P_{Sky\text{-}iDS}(o_3^p)$, of each object over all possible wor-lds in Inequality~(\ref{eq:eq3}). In $W_6$, we obtain $P_{Sky\text{-}iDS}(o_3^p)$ $= 1$. If the user-specified probabilistic threshold $\alpha$ is 0.45, then we have $P_{Sky\text{-}iDS}(o_3^p) > \alpha$, which indicates that object $o_3^p$ is one of our Sky-iDS query answers at timestamp $t=6$.

\noindent {\bf Challenges.} To tackle the Sky-iDS problem, there are three major challenges. First, many existing works \cite{lin2005stabbing,das2009randomized} on str-\\eam processing usually assume that the underlying data are complete. However, this assumption does not always hold in practice (e.g., sensory data attributes may be missing or not available). Directly discarding incomplete data objects may lead to the bias of skyline query results over the purged data stream. Thus, we cannot directly apply skyline query processing techniques over complete data to solve our Sky-iDS problem over incomplete data stream, and we should design an effective and efficient approach to impute possible missing attribute values of incomplete data objects.

Second, in the stream environment, it is rather challenging to efficiently process the imputed probabilistic data stream under \textit{possible worlds} semantics \cite{Dalvi07}. In particular, as shown in Inequality~(\ref{eq:eq3}), there are an exponential number of possible worlds, which are inefficient, or even infeasible, to enumerate. Thus, we need to design an effective approach to reduce the problem to the one over imputed objects in probabilistic data stream.

Third, it is not trivial either how to efficiently process the Sky-iDS query in incomplete data stream. In other words, we need to dynamically and incrementally maintain the Sky-iDS query answer set, upon insertions and deletions in incomplete data stream. Therefore, in this paper, we should design effective pruning or indexing mechanisms to reduce the problem search space and enable efficient Sky-iDS query answering.

\subsection{Sky-iDS Processing Framework}
Algorithm \ref{alg:Sky_iDS_framework} illustrates a framework for our Sky-iDS query processing, which consists of three phases. In the first \textit{offline pre-computation phase}, we offline build indexes $\mathcal{I}_j$ over a static (complete) data repository $R$ for imputing attributes $A_j$, respectively (line 1). In the second \textit{imputation and incremental Sky-iDS computation phase}, upon deletions (lines 2-3) and insertions (lines 4-7), we dynamically maintain a data synopsis, called \textit{skyline tree} $ST$, over incomplete data stream $iDS$, which stores potential Sky-iDS candidates. For insertions in particular, we use indexes $\mathcal{I}_j$ over $R$ to facilitate data imputation via DDs, and apply our pruning strategies to rule out false alarms of Sky-iDS candidates (lines 5-6). Note that, in this paper, we focus on DDs, and leave other imputation methods as our future work. Finally, in the \textit{refinement phase}, we refine Sky-iDS candidates in the skyline tree $ST$, and return actual Sky-iDS answers (line 8).


\begin{algorithm}[t!]\scriptsize
\KwIn{an incomplete data stream $iDS$, a static (complete) data repository $R$, a timestamp $t$, and a probabilistic threshold $\alpha$}
\KwOut{a Sky-iDS query answer set over $W_t$}

\tcp{Offline Pre-Computation Phase}

construct indexes, $\mathcal{I}_j$, over data repository $R$

\tcp{Imputation and Incremental Sky-iDS Computation Phase}

\For{each expired object $o_i'$ at timestamp $t$}{
    update a \textit{skyline tree}, $ST$, over $W_t$ with $o_i'$ and evict $o_i'$ from $W_t$
}

\For{each new object $o_i$ with missing attributes $A_j$ arriving at $W_t$}{
    traverse index, $I_j$, over $R$ and the \textit{skyline tree}, $ST$, over $W_t$ at the same time to enable DD attribute imputation and skyline computation, resp.
    
    \If{object $o_i^p$ cannot be pruned by spatial, max-corner, and min-corner pruning strategies}{
        incrementally update the \textit{skyline tree}, $ST$, with new object $o_i^p$ 
    }
}

\tcp{Refinement Phase}

refine Sky-iDS candidates in the $ST$ index and return actual Sky-iDS answers

\caption{Sky-iDS Processing Framework}
\label{alg:Sky_iDS_framework}
\end{algorithm}

Table \ref{table1} depicts the commonly used symbols and their descriptions in this paper.

\begin{table}\vspace{-4ex}\hspace{-2ex}
\caption{\small Symbols and descriptions.} \label{table1}\vspace{1ex}
{\small\scriptsize
    \begin{tabular}{l|l} \hline
    {\bf Symbol} & \qquad\qquad\qquad\qquad{\bf Description} \\ \hline \hline
    $iDS$   & an incomplete data stream \\ \hline
    $pDS$   & an imputed (probabilistic) data stream \\ \hline
    $o_i$   & an object arriving at timestamp $i$ from stream $iDS$ \\ \hline    
    $o_i^p$ & an imputed probabilistic object in the imputed stream $pDS$ \\ \hline
    $W_t$   & a set of valid objects from stream $iDS$ or $pDS$ at timestamp $t$ \\ \hline
    $pw(W_t)$ & a possible world of imputed probabilistic objects in $W_t$ \\ \hline
    $t \prec o_i$ & object $t$ dominates object $o_i$\\ \hline
    $t \preccurlyeq o_i$ & $t \prec o_i$ or $t \equiv o_i$\\ \hline
    \end{tabular}
}
\end{table}

\section{Incomplete Object Imputation}
\label{sec:impuation_of_io}

In this section, we will discuss how to impute missing attributes in incomplete data stream $iDS$ by using rules such as \textit{differential dependencies} (DDs)  \cite{song2011differential}. In the sequel, we will first briefly introduce DD rules, and then present an effective approach to impute missing attributes by a historical complete data repository with the help of conceptual lattices.

\subsection{Preliminary: Differential Dependency}
\label{imputing_of_DD}

\begin{table}[t!]
\caption{\small An example of a complete data repository $R$ with 2 DD rules, $DD_1: (A \to D, \{[0,10],[0,2]\})$ and $DD_2: (BC \to D, \{[0,1],[0,1],[0,1]\})$.}\vspace{0ex}
\label{example_table5}
\centering\scriptsize
\begin{tabular}{|c||c|c|c|c|}
\hline
\textbf{object} & \textbf{$A$} & \textbf{$B$} & \textbf{$C$} & \textbf{$D$}\\
\hline
\hline
$s_1$ & 90 & 2 & 2 & 3  \\ \hline
$s_2$ & 60 & 1 & 1 & 1  \\ \hline
$s_3$ & 70 & 2 & 2 & 2  \\ \hline
$s_4$ & 90 & 2 & 3 & 2  \\ \hline
\end{tabular}\vspace{2ex}
\end{table}

Attributes of real-world objects often have inherent value correlations. The \textit{differential dependency} (DD) technique \cite{song2011differential} is a useful and important tool to explore such attribute correlations among objects. Specifically, given a data repository, $R$, with complete data objects, we can obtain a set, $\Omega$, of DD rules \cite{song2011differential} over $R$. Each DD rule, denoted as $DD_s\in \Omega$, is represented in the form of $(X \to A_j, \phi [X A_j])$,  where $X$ are \textit{determinant attribute(s)}, $A_j$ is a \textit{dependent attribute} ($A_j \notin X$), and $\phi[X A_j]$ is a \textit{differential function} on attributes $X$ and $A_j$. Here, the differential function $\phi[Y]$ specifies distance range restrictions on attributes $Y$, which contain a number of distance intervals, $A_y.I$, for attributes $A_y \in Y$, where $A_y.I=[0,\epsilon_{A_y}]$. In this paper, we have the assumption that a data repository $R$ containing complete data is available for data imputation via DD rules. This data repository can be obtained from historical data (e.g., from data streams or other external sources). The data repository is used as a source to impute missing attributes from other non-missing attributes, and we do not assume that we can obtain all stream data coming in the future. We will leave this interesting topic of detecting DD rules from data streams as our future work.

Table \ref{example_table5} shows an example of a data repository $R$, which contains 4 attributes $A$, $B$, $C$, and $D$, and follows a set, $\Omega$, of two DD rules, $DD_1$ and $DD_2$, below:

$DD_1: (A \to D, \{[0,10],[0,2]\}), \text{and}$

$DD_2: (B C \to D, \{[0,1],[0,1],[0,1]\})$.

In Table \ref{example_table5}, for $DD_1$ ($A\to D, \{[0, 10], [0, 2]\}$), if two objects, such as $s_2$ and $s_3$, have attribute $A$ satisfying the distance constraint $A.I = [0, 10]$ (i.e., $|s_2[A] - s_3[A]| = 10 \in [0, 10]$), then they must have similar values of attribute $D$ (i.e., $|s_2[D]-s_3[D]| = 1 \in [0, 2]$). The case of $DD_2$ is similar. 

DD \cite{song2011differential} is quite useful for many real applications, such as fraud detection over transaction records (e.g., two transactions of a credit card within an hour must occur within 100 miles). DD can be also used for imputing missing attributes, as will be discussed in the next subsection. 

\noindent{\bf The advantages of using DDs as the imputation approach.} In this paper, we use DDs as our imputation approach, which has following advantages. Compared with imputation methods requiring exact matching (e.g., editing rule \cite{fan2010towards}), DD-based imputation approach can tolerate differential differences (e.g., $\phi [A]$=[0, 10] for $DD_1$ in Table \ref{example_table5}) between attribute values, which can lead to a good imputation result even in sparse data sets \cite{song2011differential}. Compared with the state-of-the-art constraint-based imputation approach \cite{zhang2017time} (requiring labelled data in data streams), DD-based imputation approach does not require any labelled data and imputes missing values via complete historical data records (i.e., data repository $R$). Specifically, many existing imputation approaches (e.g., the constraint-based approaches \cite{song2015screen,zhang2017time}) usually impute data based on incomplete data themselves only, which may lead to the imputation failure. For example, \cite{song2015screen} requires that any two consecutive tuples cannot be missing at the same time. Nevertheless, imputation via DDs does not have such limitations. Moreover, imputation via DDs can lead to good query accuracy for skyline operator over incomplete data streams, which can be confirmed in Section \ref{subsec:accuracy}.

\subsection{Data Imputation via DDs}
\label{subsec:data_imputation_via_DDs}

\noindent{\bf Data imputation with one single DD: $X \to A_j$.} Given an incomplete object $o_i\in iDS$ with missing attributes $A_j$ (for $1 \le j \le d$), a (complete) data repository $R$, and a single DD rule $DD_s \in \Omega$ in the form $X \to A_j$, our goal is to impute the missing attribute $A_j$ in object $o_i$ by utilizing $R$ and $DD_s$. 

Intuitively, if some object $s_r$ from complete data repository $R$ has attribute values $s_r[X]$ the same as or similar to that of incomplete object $o_i$, then, according to the $DD_s$ rule, their values of attribute $A_j$ should also be similar. In other words, we use attribute value $s_r[A_j]$ of complete object $s_r\in R$ as one possible imputed value of missing attribute $o_i[A_j]$ for incomplete object $o_i$. 


In particular, given an incomplete object $o_i\in iDS$, if $o_i$ has complete attributes $X$, then we can obtain all objects $s_r$ from data repository $R$ such that their attribute values of $s_r[X]$ satisfy distance constraints with $o_i[X]$ based on $DD_s$, that is, for each attribute $A_x \in X$, it holds that $|s_r[A_x] - o_i[A_x]| \in A_x.I$. Next, all the retrieved objects $s_r\in R$ will contribute their attribute values $s_r[A_j]$ (i.e., samples) to imputing the missing attributes $o_i[A_j]$ for object $o_i$.

Without loss of generality, we assume that all the imputed values $s_r[A_j]$ via objects $s_r \in R$ have equal chances to represent actual attribute value $o_i[A_j]$ of incomplete object $o_i$. Therefore, we will count the frequency, $v.freq$, of each distinct imputed value, $v$, for attribute $o_i[A_j]$, and then we can calculate the probability that the missing attribute $o_i[A_j]$ of incomplete object $o_i$ equals to $z$ (for some $s_r[A_j]$) from complete object $s_r$ as: $Pr\{o_i[A_j] = z\}=\frac{z.freq}{\sum_{\forall v} v.freq}$.

\nop{
{\bf (Weilong, please describe how to calculate the probability via frequency here, with some notations...)}
}

Let us consider an incomplete object $o_5 = (70, 2, 2, -)$ in the example of Table \ref{example_table1}. Based on a DD rule $DD_1: (A \to D, \{[0,10],$ $[0,2]\})$, we will find all objects from the data repository $R$ in Table \ref{example_table5} whose attribute $A$ values are within $10$-distance from $o_5[A] = 70$, that is, falling into interval $[60, 80]$ ($=[70-10, 70+10]$). In Table \ref{example_table5}, objects $s_2$ and $s_3$ from $R$ will be selected (since both $s_2[A]$ and $s_3[A]$ are within interval $[60, 80]$). Then, we will use their attributes $D$ values, $s_2[D]$ ($=1$) and $s_3[D]$ ($=2$), to impute the missing attribute $o_5[D]$ of incomplete object $o_5$. Thus, $o_5[D]$ will be imputed with two possible values $1$ and $2$, each with a probability $0.5$ ($=\frac{1}{1+1}$).

Note that, in this paper, we do not use the imputed attributes to further estimate other missing attributes. We will leave this interesting topic as our future work.

\vspace{1ex}\noindent{\bf Data imputation with multiple DDs.}  In practice, we may have multiple DD rules with the same dependent attribute $A_j$ over data repository $R$, for example, $X_1 \to A_j$, $X_2 \to A_j$, ..., and $X_l \to A_j$. Given an incomplete object $o_i$ with missing attribute $A_j$, assume that attribute sets $X_1 \sim X_l$ from DD rules are all complete in object $o_i$. Then, we will utilize attributes $o_i[X_1]$, $o_i[X_2]$, ..., and $o_i[X_l]$ to impute the missing attribute $o_i[A_j]$ (via $R$ and DDs). In other words, we can apply a combined DD rule, $X_1X_2...X_l \to A_j$, to efficiently impute $o_i[A_j]$. Here, if two attribute sets $X_a$ and $X_b$ share the same attributes $A_y$, then we will use the intersection of their intervals $A_y.I$ as the distance constraint in $X_1X_2...X_l \to A_j$. 

Note that, one straightforward method is to use $l$ individual DD rules to separately impute $o_i[A_j]$. However, this method may lead to low efficiency and, most importantly, biased estimates of attribute value $o_i[A_j]$ (due to the correlations among determinant attributes in $X_1 \sim X_l$). On the other hand, if we apply all attributes $X_1X_2...X_l$ to impute $o_i[A_j]$ (though it is efficient), due to the limited number of samples in data repository $R$, it is possible that none of objects (samples) in $R$ satisfy the distance constraints for all attributes $X_1X_2...X_l$, which cannot perform the imputation at all. Alternatively, in this paper, we will consider appropriate selection of attributes (e.g., a subset of $X_1X_2 ... X_l$) to impute attribute $o_i[A_j]$, making a balance between efficiency and accuracy.

\begin{figure}[t!]
\centering
\subfigure[][{\small conceptual lattice $Lat_j$}]{                    
\hspace{-4ex}\scalebox{0.34}[0.36]{\includegraphics{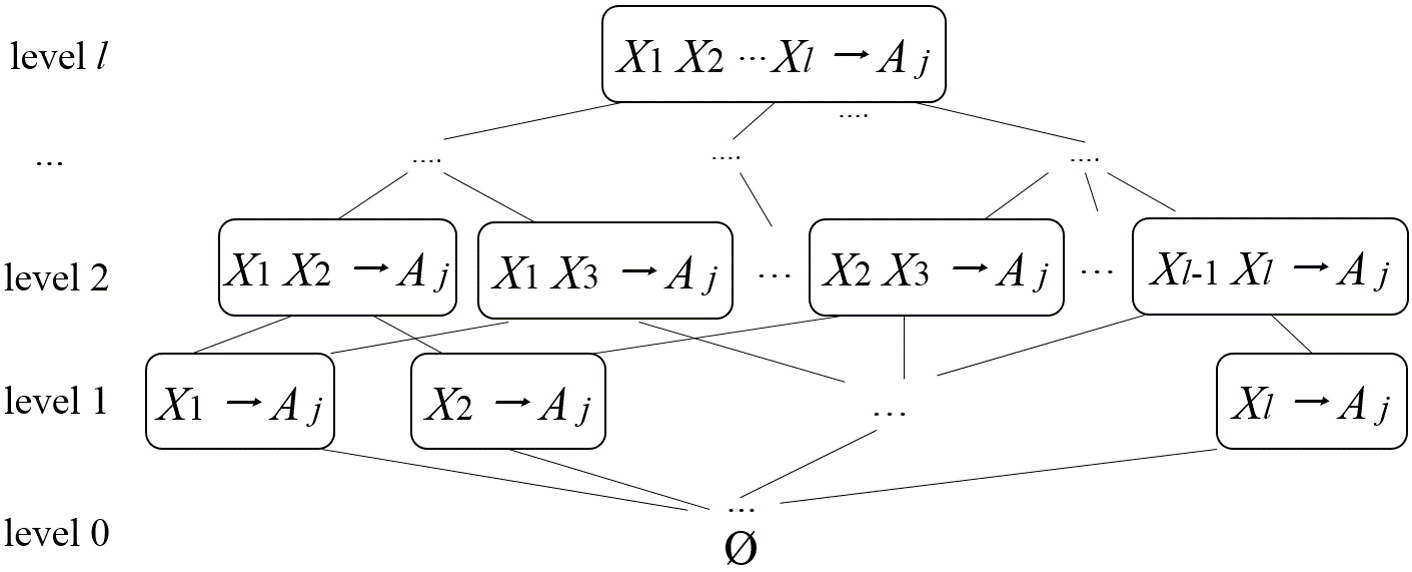}}\label{subfig:lattice}          
}\\\vspace{-2ex}
\subfigure[][{\small lattice $Lat_D$ from 2 DDs in Table \ref{example_table5}}]{
\scalebox{0.47}[0.4]{\includegraphics{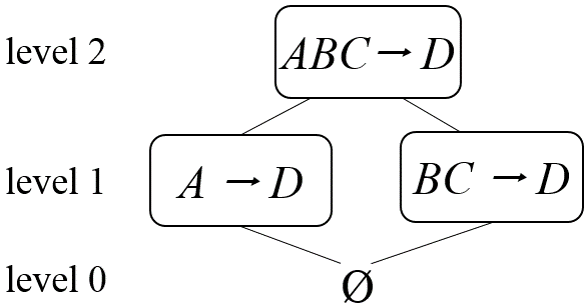}}\label{subfig:lattice_example}       
}\vspace{-2ex}
\caption{\small Illustration of a conceptual lattice and its example.} \label{fig:conceptualLattice}
\end{figure}

\underline{\it Conceptual lattice:} Inspired by the reason above, in the sequel, we will propose a \textit{conceptual lattice}, denoted by $Lat_j$ (for $1\leq j \leq d$), which can facilitate the decision of selecting DD rules for imputing the missing attribute $A_j$. Figure \ref{subfig:lattice} shows the logical structure of the conceptual lattice $Lat_j$, which consists of $(l+1)$ levels. Specifically, on level 0, we  have an empty set, $\emptyset$, indicating that we cannot use any DD rules to infer attribute $A_j$; on level 1, we have $l$ nodes, each corresponding to a DD rule $DD_s$: $X_s \to A_j$; on level 2, lattice nodes contain rules in the form of $X_aX_b \to A_j$; and so on. Finally, on level $l$, we have one node with a combined DD rule $X_1X_2...X_l \to A_j$. Figure \ref{subfig:lattice_example} depicts the conceptual lattice $Lat_D$ for the example in Table \ref{example_table5} (with 2 DDs).

\underline{\it DD selection via lattice:} Given a conceptual lattice $Lat_j$ and an incomplete object $o_i$ with missing attribute $A_j$ (i.e., $o_i[A_j] = $``$-$''), we need to decide which (combined) DD rule from lattice $Lat_j$ should be selected for imputing $o_i[A_j]$. Algorithm \ref{alg:DD_selection_using_lattice} illustrates the pseudo code of the DD selection algorithm, which traverses the lattice $Lat_j$ in a breadth-first manner. Specifically, we start the traversal of the lattice $Lat_j$ from level $l$ to level $0$ (line 1). Intuitively, higher level of lattice $Lat_j$ involves more determinant attributes (e.g., level $l$ has the largest number of attributes in $X_1X_2...X_l$), which will lead to more accurate imputation results and higher imputation efficiency (i.e., handling fewer candidates in $R$). Thus, here, we will start from higher level first.

When we access level $lv$, for each node with DD rule, $Y \to A_j$, on this level, we offline rank these DDs in increasing order of the imputation cost (defined as the expected number of possible samples from $R$). Intuitively, DDs with high ranks will have both low imputation cost and smaller imputation errors. Thus, for DDs on the same level, we will consider DDs with high ranks first. Then, we will check if this combined DD rule can be used for imputing $o_i[A_j]$ (lines 2-4). In particular, if some complete objects $s_r$ in $R$ satisfy the distance constraints with incomplete object $o_i$ on attributes $Y$ (i.e., the number of samples for imputation is nonzero, estimated from histograms), then we will terminate the loop and return the DD rule $Y \to A_j$ as the best DD rule for imputing the missing attribute $o_i[A_j]$ (lines 3-4). Note that, if multiple combined DD rules on the same level $lv$ satisfy distance constraints, then we will only return the one with higher rank. If the lattice traversal descends to level $0$, this indicates that none of DDs can be used for imputation. In this case, we can only apply a statistics-based method \cite{mayfield2010eracer} to impute $o_i[A_j]$ with possible values of attribute $A_j$ over $R$, following some probabilistic distribution, where the probability of each possible value can be calculated by the count of this value in attribute $A_j$ over $R$ divided by the size of $R$ (lines 5-6). For instance, given a value set $\{0.1, 0.1, 0.1, 0.2\}$ on attribute $A_j$ over data repository $R$ (assuming $R$ only having 4 complete data records), and an incomplete data object $o_i$ with missing value on the attribute $A_j$, if no DD can be used for imputation $o_i[A_j]$, we will fill $o_i[A_j]$ with 0.1 and 0.2 with probabilities 0.75 and 0.25, respectively. Note that, if no DD can be used for imputing $o_i[A_j]$, we may not be able to use other imputation approaches (e.g., editing rule \cite{fan2010towards}) to impute $o_i[A_j]$.


\begin{algorithm}[t!]\scriptsize
\KwIn{Lattice $Lat_j$, incomplete object $o_i$ with missing attribute $A_j$, and data repository $R$}
\KwOut{the best DD rule from $Lat_j$ to do the imputation}
\For{level $lv$ $=$ $l$ to $0$}{
    \For{each node, $Y \to A_j$, on level $lv$ in increasing order of the imputation cost}{
        \If{the number of samples in R satisfying the distance constraints on attributes $Y$ is not zero}{
            return the DD rule, $Y \to A_j$, in this node
        }
    }
    \If{$lv=0$}{
        apply a statistics-based method \cite{mayfield2010eracer} to impute $o_i[A_j]$ with possible values of attribute $A_j$ over $R$
    }
    
}
\caption{DD Selection Using Conceptual Lattice}
\label{alg:DD_selection_using_lattice}
\end{algorithm}

Once we select an appropriate (combined) DD rule, $Y\to A_j$, we can use this rule to impute the missing attribute, $o_i[A_j]$, of incomplete object $o_i$, similar to the aforementioned case of the data imputation with a single DD.

\begin{figure*}
\centering \vspace{-2ex} 
\subfigure[][{\small spatial pruning}]{                    
\scalebox{0.32}[0.32]{\includegraphics{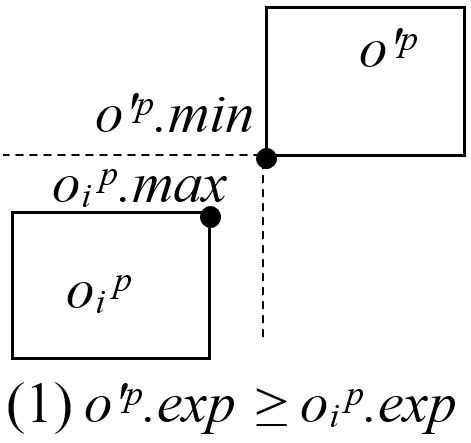}}\label{subfig:pruningRule_1}          
}\qquad\qquad
\subfigure[][{\small max-corner pruning}]{
\scalebox{0.28}[0.28]{\includegraphics{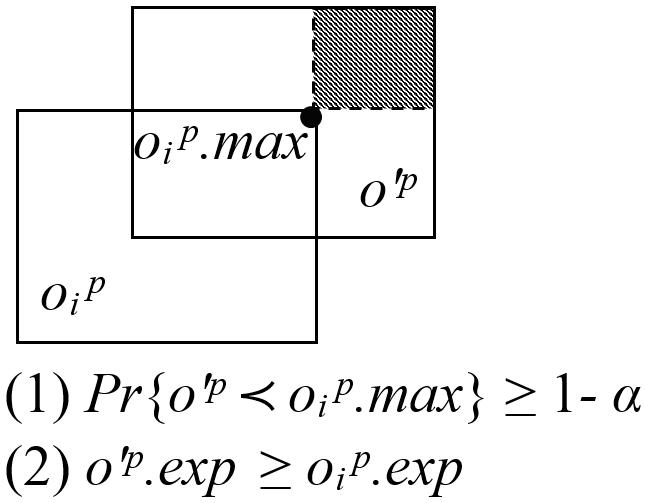}}\label{subfig:pruningRule_2}       
}\qquad\qquad
\subfigure[][{\small min-corner pruning}]{
\scalebox{0.28}[0.28]{\includegraphics{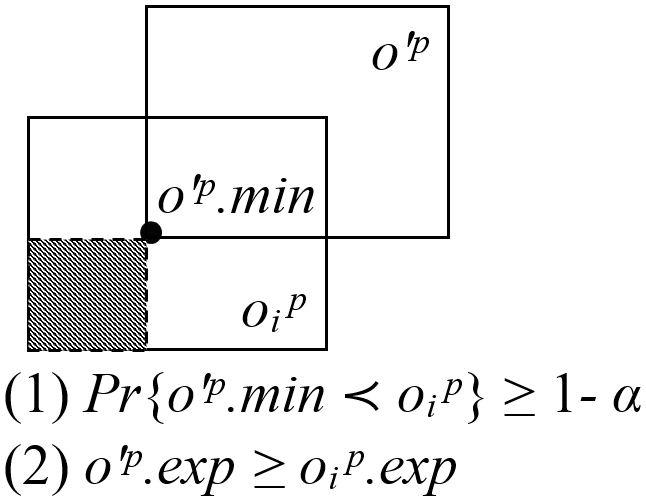}}\label{subfig:pruningRule_3} 
}\vspace{-2ex}  
\caption{\small Illustration of pruning strategies.} \label{fig:pruning} \vspace{-4ex}                                
\end{figure*}

\section{Pruning Strategies}
\label{sec:pruning_strategies}

\noindent {\bf Problem reduction.} As mentioned in Section \ref{sec:problemDefinition}, it is not efficient, or even not feasible, to compute the skyline probability, $P_{Sky\text{-}iDS} (o_i^p)$, in Inequality~(\ref{eq:eq3}) by enumerating an exponential number of possible worlds $pw(W_t)$. In order to speed up the efficiency, we will reduce our Sky-iDS problem over possible worlds to the one on uncertain objects. In particular, we will rewrite the skyline probability $P_{Sky\text{-}iDS} (o_i^p)$ as the probability that instances, $o_{il}$, of $o_i^p$ are not dominated by other (imputed) objects $o_j^p$, which is given in an equivalent form below:\vspace{-2ex}

\begin{eqnarray}
P_{Sky\text{-}iDS} (o_i^p)&=&\hspace{-2ex}\sum_{\forall o_{il} \in o_i^p} o_{il}.p \cdot \hspace{-3ex}\prod_{\forall o_j^p \in W_t \wedge o_j\ne o_i} \hspace{-4ex}(1- Pr\{o_j^p \prec o_{il}\}).\label{eq:eq14}\vspace{-3ex}
\end{eqnarray}\vspace{-2ex}

Since it is still not efficient for stream processing to calculate the probability in Eq.~(\ref{eq:eq14}) for every object $o_i^p \in W_t$, in this paper, we will provide pruning lemmas below to filter out false alarms (i.e., objects with low skyline probabilities) and reduce the search space of the Sky-iDS problem.

\noindent {\bf Spatial pruning.} We first present an effective \textit{spatial pruning} method, which utilizes the interval of each imputed attribute to rule out objects that can never be Sky-iDS answers (i.e., with zero skyline probabilities) over data stream.

\nop{
Next, instead of checking each object instance, we will introduce several lemma (pruning rules) to prune the objects that cannot be skylines till they expire from the data stream.
}

\nop{
Given a new coming incomplete object $o_i$, we will impute the missing attribute values of $o_i$ by first going through the R*-tree index layer by layer, and then through the grid cell, and finally through the exact complete objects in each cell.
However, before reaching the exact complete objects in each cell, the incomplete object $o_i$ may be classified as a probabilistic skyline or not by some spatial and probabilistic pruning rules. 
}

\nop{
\begin{figure}[ht!]
\centering
\includegraphics[scale=0.38]{SPPruningRules.png}
\caption{The scenarios of \textit{Pruning Rule 1} (a), \textit{Pruning Rule 2} (b) and \textit{Pruning Rule 3}(c)}
\label{fig:spatialPruning}
\end{figure}
}

\nop{
\begin{figure}[t!]
\centering
\includegraphics[scale=0.25]{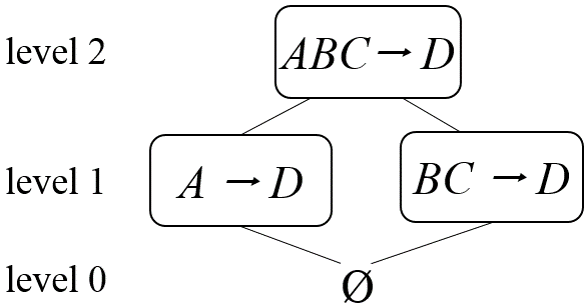}
\caption{\small The imputation lattice for attribute $D$ based on the DD set $\Omega$ in Table \ref{example_table5}}
\label{fig:rankingLattice}
\end{figure}
}

Specifically, for each incomplete object $o_i$ from data stream $iDS$, we use a \textit{minimum bounding rectangle} (MBR), $o_i^p.MBR$, to represent its imputed object $o_i^p$. We denote $o_i^p.min$ and $o_i^p.max$ as minimum and maximum corners of MBR $o_i^p.MBR$, respectively, which have minimum and maximum possible coordinates on all attributes $A_j$ in $o_i^p.MBR$. \vspace{-1ex} 

\nop{
bound all its instances, and let $o_i^p.min$ and $o_i^p.max$ be the possible minimum and maximum instances of the imputed object $o_i^p$, where attributes $A_j$ in $o_i^p.min$ and $o_i^p.max$ are the minimum and maximum values of all instances in MBRs of $o_i^p$ on attribute $A_j$, respectively.
}

\begin{lemma} {\bf (Spatial Pruning)}
Given two incomplete objects $o_i$ and $o'$ from incomplete data stream $iDS$, if $o'^p.min$ $\prec o_i^p.max$ and $o'.exp \ge o_i.exp$ hold, then object $o_i$ can be safely pruned.\label{lemma:lem1}
\end{lemma}
\textit{Proof: }
Please refer to Appendix \ref{subsec:proof_lemmas_pruning_strategies1}. $\hfill\square$\vspace{0ex}


As illustrated in Figure \ref{subfig:pruningRule_1}, (imputed) object $o'^p$ dominates $o_i^p$, since the corner point, $o'^p.min$, of $o'^p$ dominates that, $o_i^p.max$, of $o_i^p$. Moreover, since $o'.exp \ge o_i.exp$ holds, object $o_i^p$ can never be the skyline in its lifetime, and can be safely pruned (as given by Lemma \ref{lemma:lem1}).

\nop{
Since $o_i^p.min$ and $t^p.max$ are the possible minimum and maximum instances of imputed objects $o_i^p$ and $t^p$, respectively, we can get $P_{Sky\text{-}iDS}(o_i^p) \ge P_{Sky\text{-}iDS}(o_i^p.min)$ and $P_{Sky\text{-}iDS}(t^p) \le P_{Sky\text{-}iDS}(t^p.max)$. If $o_i^p.min \prec t^p.max$, we can get $P_{Sky\text{-}iDS}(t^p.max) = 0$, so $P_{Sky\text{-}iDS}(t^p) = 0$ due to existence of imputed object $o_i^p$. And if object $o_i$ expires from the stream no earlier than object $t$, then object $t$ cannot become a skyline till it expires and can be safely pruned.}

\nop{
In figure (a) of Figure \ref{fig:spatialPruning}, the two squares represent two MBRs enclosed by all possible instances of imputed objects of $o_i^p$ and $t^p$, respectively.
If $o_i^p.min \prec t^p.max$, imputed object $t^p$ is definitely dominated by imputed object $o_i^p$.
Besides, object $o_i$ will stay in data stream longer than or equal to object $t$ ($o_i.exp \ge t.exp$), so the imputed object $t^p$ will never become a skyline due to the existence of the imputed object $o_i^p$ and the object $t$ can be safely pruned.
}


\noindent {\bf Max-corner pruning.} Next, we present a \textit{max-corner pruning} method, which uses the max-corner, $o_i^p.max$, of MBR $o_i^p.MBR$ to prune the false alarm. 

\begin{lemma} {\bf (Max-Corner Pruning)}
Given two incomplete objects $o_i$ and $o'$ from incomplete data stream $iDS$, and a max-corner $o_i^p.max$ of the imputed object $o_i^p$, if $Pr\{o'^p \prec o_i^p.max\} \ge 1-\alpha$ and $o'.exp \ge o_i.exp$ hold, then object $o_i$ can be safely pruned.
\label{lemma:lem2}
\end{lemma}
\textit{Proof : }
Please refer to Appendix \ref{subsec:proof_lemmas_pruning_strategies2}.
$\hfill\square$

\nop{If $Pr\{t^p \prec o_i^p.max\} > 1-\alpha$, according to Eq.~(\ref{eq:eq14}), we can get $P_{Sky\text{-}iDS}(o_i^p.max) \le \alpha$ by replacing all instances of object $o_i^p$ by $o_i^p.max$ in Eq.~(\ref{eq:eq14}).
And due to $P_{Sky\text{-}iDS}(o_i^p) \le P_{Sky\text{-}iDS}(o_i^p.max)$, we get $P_{Sky\text{-}iDS}(o_i^p) \le \alpha$, thus object $o_i$ cannot be a skyline when object $t$ is alive.
Meanwhile, , so object $o_i$ can be safely pruned.
}

In Lemma \ref{lemma:lem2}, the probability $Pr\{o'^p \prec o_i^p.max\}$ is given by the probability that object $o'^p$ falls into the shaded region w.r.t. max-corner $o_i^p.max$ (as shown in Figure \ref{subfig:pruningRule_2}). Intuitively, if $Pr\{o'^p \prec o_i^p.max\}  \ge 1-\alpha$ holds, then object $o_i^p$ is not dominated by $o'^p$ with probability less than $\alpha$, and in turn, the skyline probability, $P_{Sky\text{-}iDS} (o_i^p)$, of $o_i^p$  is less than $\alpha$. Moreover, object $o'^p$ expires from $pDS$ later than $o_i^p$. Thus, $o_i^p$ cannot be a skyline in its lifetime and can be safely pruned.


\nop{
if the probability sum of instances of imputed object $o_i^p$ falling into the shaded area is bigger than or equal to $1-\alpha$, that is, $\sum_{o_{il} \in o_i^p} o_{il}.p \ge 1-\alpha$, where $o_{il}$ is the instance of imputed object $o_i^p$.
And if $t.exp \ge o_i.exp$, then imputed object $o_i^p$ cannot be a skyline till it expires from data stream and can be safely pruned.
}

\nop{
As shown in figure (b) of Figure \ref{fig:spatialPruning}, instances of imputed object $o_i^p$ falling into the shaded area of MBR dominate all instances of the imputed object $t^p$.
Especially, if the frequency ratio of the instances of $o_i^p$ between shaded area and whole MBR is bigger than or equal to $1-\alpha$, imputed object $o_i^p$ will dominate imputed object $t^p$ with the probability $Pr(o_i^p \prec t^p)\ge 1-\alpha$.
At the same time, according to \textit{pruning rule 2}, if $o_i.exp \ge t.exp$, imputed object $t^p$ cannot be a skyline in its whole period and can be safely pruned.
}

\noindent {\bf Min-corner pruning.} Finally, we provide a \textit{min-corner pruning} method, which uses min-corner, $o'^p.min$, of the MBR $o'^p.MBR$ to filter out object $o_i^p$ with low skyline probability.

\begin{lemma} {\bf (Min-Corner Pruning)}
Given two incomplete objects $o_i$ and $o'$ from incomplete data stream $iDS$, and the min-corner, $o'^p.min$, of the imputed object $o'^p$, if $Pr\{o'^p.min$ $\prec o_i^p\} \ge 1-\alpha$ and $o'.exp \ge o_i.exp$ hold, then object $o_i$ can be safely pruned.
\label{lemma:lem3}
\end{lemma}
\textit{Proof : }
Please refer to Appendix \ref{subsec:proof_lemmas_pruning_strategies3}. $\hfill\square$

\nop
{If $Pr\{t^p.min \prec o_i^p\} \ge 1-\alpha$, then we can get $Pr\{t^p \prec o_i^p\} \ge 1-\alpha$ based on \textit{lemma} \ref{lemma:lem1}, similarly with the proof of \textitP{lemma} \ref{lemma:lem2}, according to Eq.~(\ref{eq:eq14}), we can get $P_{Sky\text{-}iDS}(o_i^p) \le \alpha$, and object $o_i$ cannot be a skyline when object $t$ is alive.
Meanwhile, $t.exp \ge o_i.exp$, object $o_i$ then can be safely pruned. 
}

As an example in Figure \ref{subfig:pruningRule_3}, the probability $Pr\{o'^p.min$ $\prec o_i^p\}$ in Lemma \ref{lemma:lem3} is given by the probability that object $o_i$ falls into the shaded region w.r.t. min-corner $o'^p.min$. Similar to Lemma~\ref{lemma:lem2}, in Figure \ref{subfig:pruningRule_3}, object $o_i^p$ is not dominated by $o'^p$ with probability less than $\alpha$ (i.e., with low skyline probability), and $o_i^p$ expires before $o'^p$. Thus, object $o_i^p$ never has a chance to be the skyline during its lifespan, and can be safely pruned.

Note that, for these three pruning rules, we will first apply the \textit{spatial pruning}, and then consider the \textit{max-corner} and \textit{min-corner} pruning rules if the \textit{spatial pruning} fails.

\section{Skyline Processing on Incomplete Data Stream}
\label{sec:dsp_over_ids}

\vspace{-2ex}In this section, we will first propose a novel data synopsis, namely \textit{skyline tree} ($ST$), which dynamically maintains Sky-iDS candidates over incomplete data stream $iDS$. Then, we will present index structures, $\mathcal{I}_j$, constructed over complete data repository $R$ to facilitate missing data imputation. Next, we will discuss how to use data synopsis $ST$ and indexes $\mathcal{I}_j$ to continuously monitor Sky-iDS query answers from incomplete data stream $iDS$, following the style of ``imputation and query processing at the same time''. Finally, we will provide cost models for index construction and parameter tuning.


\nop{
Skyline processing over incomplete data streams has a high requirement of time efficiency, since an object can only stay in data streams for a limited time.
To accelerate skyline processing, in this section, we will introduce a new data structure named \textit{skyline tree}, denoted by $ST$, which stores all the valid (not expired) data objects in imputed data stream $pDS$ that have the chance to be skylines before they expire.  
We will formally define the \textit{skyline tree} and demonstrate how it is dynamically updated when new objects come into the $pDS$ or old objects expire from the $pDS$.
Leveraging the $ST$, we will propose corresponding algorithms to do the skyline proceeding over the incomplete data streams.
And {\bf index, skyline query algorithm, cost model.}

}

\subsection{Skyline Tree}
\label{subsec:skyline_tree}

\vspace{-2ex}In this subsection, we will present the data structure of the skyline tree $ST$, and then discuss properties of $ST$.

\vspace{1ex}\noindent {\bf Data structure of the skyline tree.} In the sequel, we propose a multi-layer tree structure, namely \textit{skyline tree} ($ST$), which is incrementally maintained over valid (imputed) objects (potential skyline candidates) $o_i^p\in W_t$ from incomplete data stream $iDS$. Intuitively, the skyline tree $ST$ stores all possible Sky-iDS candidates over $iDS$ that have chances to be skylines over time. If a Sky-iDS candidate (node) $o_i^p$ on a layer of skyline tree $ST$ expires, then its children (child nodes) $o_c^p$ will become new skyline candidates.

Specifically, each node of the skyline tree $ST$ corresponds to an (imputed) object, $o_i^p \in W_t$, which has one or multiple pointers pointing to its children $o_c^p$, such that: (1) each child $o_c^p$ is dominated by its parent node $o_i^p$ with probability greater than or equal to $(1-\alpha)$ (i.e., $Pr\{o_i^p \prec o_c^p\}\geq 1-\alpha$), and (2) $o_c^p$ expires after $o_i^p$ (i.e., $o_i^p.exp < o_c^p.exp$). 

Moreover, for any two sibling nodes $o_i^p$ and $o_j^p$ on the same layer of the tree $ST$, they should dominate each other with probabilities less than $(1-\alpha)$, that is, (1) $Pr\{o_i^p \prec o_j^p\} < 1-\alpha$, and (2) $Pr\{o_j^p \prec o_i^p\} < 1-\alpha$. 

Further, to obtain a tree structure, we use a virtual node (root) $\emptyset$ to point to all objects (skyline candidates) on the first layer of $ST$. In order to facilitate dynamic updates (e.g., deletions) in the streaming environment, for each layer of the $ST$ tree, we will maintain the list of objects (nodes) $o_i^p$ in non-descending order of their expiration times (i.e., $o_i^p.exp$).

\nop{
\begin{definition} {\textbf{(Skyline Tree, $ST$)}} Given an incomplete data stream $iDS$, timestamp $t$, and a probabilistic threshold $\alpha$, \textit{skyline tree} (ST) is a tree with empty root and contains all valid imputed objects $o_i^p \in pDS$ ($i.e. \ o_i^p.exp > t$) that have chance to be skylines before they expire from the imputed data stream $pDS$; except the empty root node, each node (imputed object) $o_i^p \in ST$ dominates all its descendant nodes $c^p \in ST$ with probabilities $Pr\{o_i^p \prec c^p\} \ge 1-\alpha$, and expires from the $pDS$ before its descendant nodes (i.e. $o_i^p.exp < c^p.exp$); any two nodes (objects), $o_i^p \in ST$ and $s^p \in ST$, in the same layer of the \textit{ST} cannot dominate each other with probabilities greater than or equal to $1-\alpha$, that is, $Pr\{o_i^p \prec s^p\} < 1-\alpha$ and $Pr\{s^p \prec o_i^p\} < 1-\alpha$.

Especially, for a node (object) $o_i^p \ST$, its father node is the last one, within the nodes in $W_t$ dominating $o_i^p$, to expire; and the nodes in the same layer are sorted based on the their expiration times in an incremental order.
\label{def:skyline_tree}
\end{definition}

}

\begin{figure}[t!]
\centering
\includegraphics[scale=0.5]{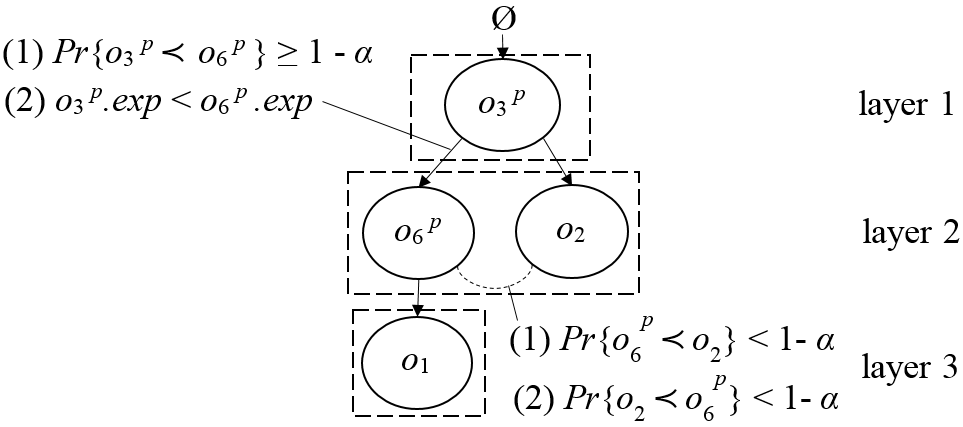}
\caption{\small An example of a skyline tree over incomplete data stream $iDS$ at timestamp 8 (i.e., $W_8$) in Table \ref{example_table1} ($\alpha = 0.45$).}
\label{fig:skyline_tree}\vspace{0ex}
\end{figure}

Figure \ref{fig:skyline_tree} illustrates a skyline tree $ST$ over $W_8 = \{o_3, o_4,$ $o_5, o_6,$ $o_1, o_2\}$ in the example of Table \ref{example_table1}, where $\alpha = 0.45$. This $ST$ tree has 3 layers, $\{o_3^p\}$, $\{o_6^p,\ o_2\}$, and $\{o_1\}$. Consider objects (nodes) $o_3^p$ and $o_6^p$ in the tree structure. Node $o_3^p$ is a parent of node $o_6^p$, since two conditions hold: (1) $Pr\{o_3^p \prec o_6^p\} =0.6  \ge 1-\alpha = 0.55$, and (2) $o_3^p.exp < o_6^p.exp$.

Similarly, objects $o_6^p$ and $o_2$ are sibling nodes on layer 2 of the $ST$ tree. This is because (1) $Pr\{o_6^p \prec o_2\} = 0< 1-\alpha = 0.55$, and (2) $Pr\{o_2 \prec o_6^p\} = 0 <1-\alpha = 0.55$.

Moreover, objects $o_4$ and $o_5$ are not in the $ST$ tree. This is because object $o_4$ (or $o_5$) is dominated by $o_1$ with probability 1 (in layer 3 of $ST$) and expires before $o_1$, which implies that $o_4$ (or $o_5$) can never be the skyline during its lifetime (i.e., always dominated by $o_1$ during the lifetime). 

\nop{
Based on \textit{lemma} \ref{lemma:lem2}, objects $o_4$ and $o_5$ do not have chance to be skylines in their life time, since $Pr\{o_1 \prec o_4\} = 1$, $Pr\{o_1 \prec o_5\} = 1$, $o_1.exp \ge o_4.exp$ and $o_1.exp \ge o_5.exp$, so objects $o_4$ and $o_5$ will not be shown in the skyline tree.
The $ST$ in Figure \ref{fig:skyline_tree} has three layers: $sLL_1,\ sLL_2$ and $sLL_3$, where $sLL_1=\{o_3^p\}$, $sLL_2=\{o_6^p,\ o_2\}$ and $sLL_3=\{o_1\}$.
An object (e.g. $o_6^p$) will expire after and before its father node (e.g. $o_3^p$) and its child node (e.g. $o_1$), respectively, with dominating probabilities not smaller than $1-\alpha$ (0.45), respectively, that is, $Pr\{o_3^p \prec o_6^p\} =0.6  \ge 0.45$ and $Pr\{o_6^p \prec o_1\} = 1 \ge 0.45$.
What is more, the nodes in the same layer (e.g. $o_6^p$ and $o_2$) is sorted based on their expiration time in an incremental order (e.g. $o_6^p.exp \le o_2^p.exp$), and each of them do not dominate each other (i.e. dominance probability smaller than $1-\alpha$), that is, $Pr\{o_6^p \prec o_2\} = 0 < 0.55$ and $Pr\{o_2 \prec o_6^p\} = 0 < 0.55$.

}

\vspace{1ex}\noindent {\bf Properties of the skyline tree.} Next, we will provide the properties of the skyline tree $ST$.

\nop{
{\bf (Weilong, please move proofs of properties to appendix, and move Properties 3-4 to dynamic maintenance section)}
}

\textit{Property 1.} {\bf (Completeness)}
The skyline tree $ST$ contains all the objects $o_i^p$ from $iDS$ that have the chance to be skylines before they expire.

\nop{
\begin{proof}
We can prove this property by showing no such imputed object $t^p \in W_t$ and $t^p \notin ST$, which is actually a current skyline or may become a skyline before it expires, exists.
First, assuming the object $t^p$ having probability $P_{Sky\text{-}iDS}(t^p)>\alpha$, according to Eq.~(\ref{eq:eq14}), $t^p$ should be in the first layer of $ST$ since all objects $o_i^p \ST$ have the dominance probability $Pr\{o^t \prec t^p\} < 1-\alpha$.
Second, assuming the object $t^p$ is dominated by some objects $o_i^p \ST$ ($t^p.exp > o_i^p.exp$), and may become the skyline after these objects $o_i^p$ expire, in this case, object $t^p$ should be the child node of one object within objects $o_i^p$, since $Pr\{o_i^p \prec t^p\} \ge 1-\alpha$ and $o_i^p.exp < t^p.exp$. So the $ST$ contains all the objects $o_i^p \in pDS$ that have the chance to be skylines before they expire.
\end{proof}
}

\textit{Property 2.} {\bf (No False Dismissals)}
If an imputed object $o_i^p$ is not on the first layer of the skyline tree $ST$ over $W_t$, then $o_i^p$ cannot be a skyline at current timestamp $t$.

\nop{
\begin{proof}
Given an imputed object $o_i^p \in ST$, and its non-empty father node (object) $t^p \ST$, since $Pr\{t^p \prec o_i^p\} \ge 1-\alpha$, according to Eq.~(\ref{eq:eq14}), we get $P_{Sky\text{-}iDS}(t^p)<\alpha$, which is not satisfied with the skyline definition (Definition \ref{def:sky-iDS}).
So the object $o_i^p$ cannot become a skyline before its father node expires from the $iDS$.
\end{proof}

}

\textit{Property 3.} {\bf (Superset of Sky-iDS Answers)}
The set of objects $o_i^p$ on the first layer of the skyline tree $ST$ is a superset of Sky-iDS answers at current timestamp $t$.

\nop{

\begin{proof}
According to \textit{Property 2}, we can deduce that current skyline answers must be within the first layer of the \textit{skyline tree}, if we use $skylines$ to represent the set of current skyline answers, we can get $skylines$ is a subset of the set of objects in first layer of $ST$. 
\end{proof}
}

From the three properties above, we can see that the skyline tree $ST$ contains a superset of Sky-iDS answers on the first layer of $ST$ without any false dismissals. We will discuss later how to incrementally maintain this $ST$ tree over incomplete data stream $iDS$. Please refer the proofs of these three properties to Appendix \ref{subsec:proof_propeties_1_3}.

\vspace{-3ex}
\subsection{Cost-Model-Based Indexes on Data Repository \textsl{R} for Imputation}
\label{sec:index_over_R_for_imp}


In this subsection, we will present indexes, $\mathcal{I}_j$, constructed from complete data repository $R$, which can facilitate quick imputation of missing attributes in data stream $iDS$. 

\noindent {\bf Index structure.} In order to facilitate efficient data imputation, in this paper, we will devise $d$ (i.e., the dimensionality of data sets, or the number of attributes in objects) effective indexes, $\mathcal{I}_j$ (for $1\leq j\leq d$), each of which can help quickly access candidates $s_r$ from data repository $R$, and impute missing attributes $o_i[A_j]$. Specifically, given $l$ DD rules $X_1 \to A_j$, $X_2 \to A_j$, ..., and $X_l \to A_j$ from $\Omega$, we build an index $\mathcal{I}_j$ over those objects in $R$ projected on attributes $U_j = X_1\cup X_2\cup ... \cup X_l$ as follows. 

\begin{figure}[t!]
\centering\hspace{-1ex}
\includegraphics[scale=0.28]{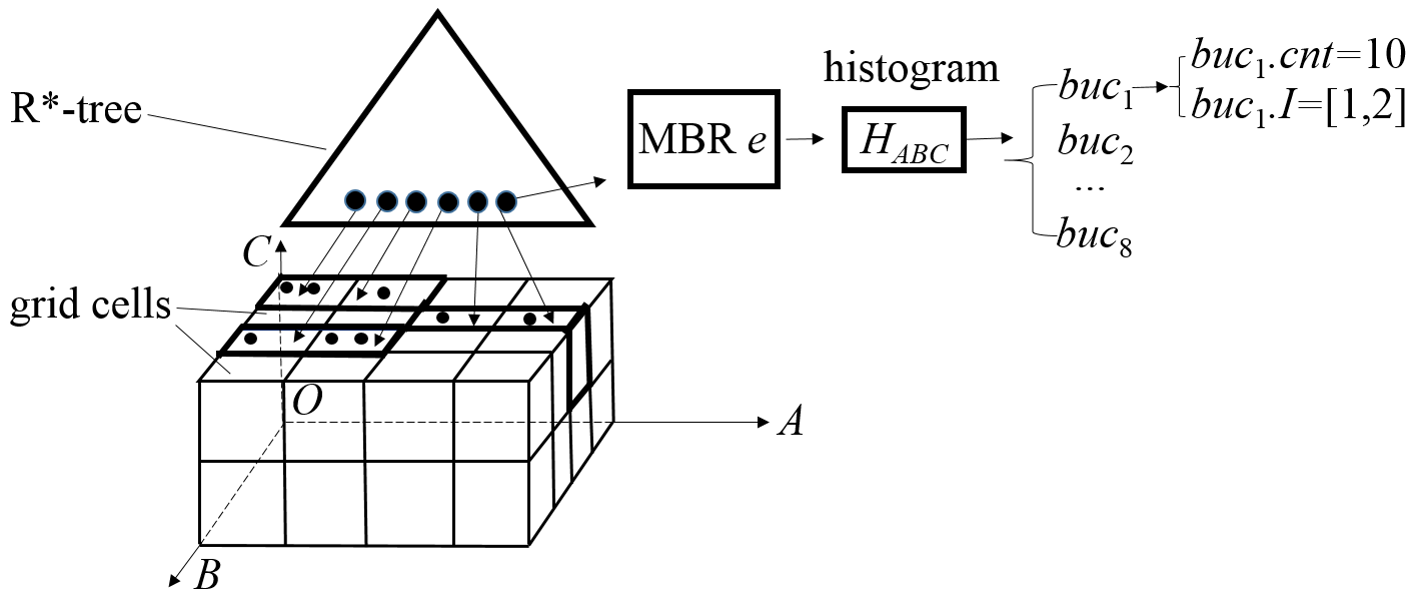}\vspace{-2ex}
\caption{\small The histogram associated with each MBR node $e \in \mathcal{I}_j$ w.r.t. $DD_1: A\to D$ and $DD_2: BC\to D$ in Table \ref{example_table5} ($\lambda = 2$).}
\label{fig:MBRStoredInfo}
\end{figure}

As illustrated in Figure \ref{fig:MBRStoredInfo}, we first divide the data space over attributes $U_j$ into grid cells of equal size \cite{li2014parallelizing}, where the side length of each cell is given by $u$. We will discuss later in Section \ref{sec:cost_model_for_parameter_tuning} how to tune this parameter $u$, in light of our proposed cost model, for minimizing the imputation cost. Then, we insert each object $s_r \in R$ into a cell containing $s_r[U_j]$. Finally, we build an R$^*$-tree \cite{beckmann1990r}  over those cells with objects, by invoking normal ``insert'' method. This way, the R$^*$-tree over non-empty grid cells can be constructed, denoted as index $\mathcal{I}_j$, which can be used for imputing attribute $A_j$.

Note that, compared with directly using R$^*$-tree \cite{beckmann1990r} for imputation, our proposed index $\mathcal{I}_j$ can achieve better imputation cost. This is because, all objects in a non-empty grid cell are stored in a single leaf node in $\mathcal{I}_j$ (rather than multiple leaf nodes in the R$^*$-tree), which incurs lower index traversal (DD imputation) cost than R$^*$-tree.


Furthermore, each entry (MBR) $e$ in nodes of index $\mathcal{I}_j$ is associated with a histogram, $H_{U_j}$, over attributes $U_j = X_1\cup X_2\cup ... \cup X_l$, which stores a summary of objects in $e$.



\textit{\underline{Histogram construction:}} To build a histogram $H_{U_j}$ for node $e$, we first divide each dimension $A_x \in U_j$ of the data space into $\lambda$ intervals of equal size, and obtain $\lambda^{|U_j|}$ buckets, denoted as $buc_q$ (for $1 \le q \le \lambda^{|U_j|}$), where $|U_j|$ is the number of attributes in $U_j$ (e.g., if $U_j=ABC$, then $|U_j|=3$). Then, each bucket, $buc_q$, stores two items: (1) a COUNT aggregate, $buc_q.cnt$, of objects from $R$ that fall into bucket $buc_q$, and; (2) an interval, $buc_q.I = [buc_q.A_j^-, buc_q.A_j^+]$, of attribute values $s_r[A_j]$ for any objects $s_r \in R$ that fall into $buc_q$. Intuitively, the information stored in each bucket of the histogram can be used for spatial, max-corner, and min-corner pruning (as mentioned in Section \ref{sec:pruning_strategies}).

As an example in Figure \ref{fig:MBRStoredInfo}, for data repository with attributes $(A, B, C, D)$ and $DD_1$ and $DD_2$ in Table \ref{example_table5}, index $\mathcal{I}_j$ over attributes $U_j = \{A, B, C\}$ contains a number of MBRs $e$, each of which is associated with a histogram, $H_{ABC}$. Given $\lambda = 2$, the histogram $H_{ABC}$ has 8 ($=2^{|ABC|}$ $= 2^3$) buckets. Each bucket, $buc_q$, contains the number of objects in it (e.g., $buc_1.cnt = 10$ for bucket $buc_1$), and a value bound, $buc_q.I$ of attribute $D$ for those objects in bucket $buc_q$ (e.g., $buc_1.I = [1, 2]$).

\textit{\underline{Updates of index $\mathcal{I}_j$:}} Next, we consider how to maintain index $\mathcal{I}_j$ upon the appending of new objects for data repository $R$ (though we consider $R$ as static data set in our Sky-iDS problem). When a new complete object $s_r$ comes in,
we will insert this object $s_r$, by traversing indexes $\mathcal{I}_j$ from the root node to leaf nodes. During the index traversal, if we insert object $s_r$ into an index node $e$, then we will: (1) increase the COUNT aggregate, $buc_q.cnt$, of bucket $buc_q$ (containing $s_r$) in histogram $H_{U_j}$ by 1; (2) update the minimum and maximum values of attribute $A_j$ for the interval, $buc_q.I$, of bucket $buc_q$, and; (3) recursively insert $s_r$ into one of children under node $e$. When we access a leaf node, we will insert object $s_r$ into this leaf node (maintaining the index structure, if necessary), and update the information of a cell that contains object $s_r$.

\noindent {\bf Data imputation via indexes.} Next, we consider how to efficiently use indexes, $\mathcal{I}_j$, and DD rules, $X\to A_j$, to impute missing attribute $A_j$ of an incomplete object $o_i\in iDS$. As discussed in Section \ref{subsec:data_imputation_via_DDs}, we will utilize the conceptual lattice to decide an appropriate DD rule $Y \to A_j$ (Algorithm \ref{alg:DD_selection_using_lattice}), and then perform a range (aggregate) query over index $\mathcal{I}_j$ for attributes $Y$ (note: range predicates on other attributes are wildcard $*$), where a query range $Q$ is given by an MBR with $[o_i[A_x]-\epsilon_{A_x}, o_i[A_x]+\epsilon_{A_x}]$ on each dimension $A_x \in Y$. 

Specifically, given a range query $Q$, we traverse index $\mathcal{I}_j$ over attributes $Y$ (in the selected DD for imputation), starting from the root, $root(\mathcal{I}_j)$. When we encounter a non-leaf node $e$, we will check whether or not its children are intersecting with the query range $Q$ (ignoring attributes other than $Y$). If the answer is yes, then we will access those intersecting children. When we encounter a leaf node $e$, we will obtain those cells intersecting with query range $Q$, and retrieve objects $s_r\in R$ from cells that fall into $Q$.


After we retrieve all objects in the query range $Q$ from $\mathcal{I}_j$, we can use their corresponding attribute $A_j$ values (and confidences as well) to impute the missing attribute $o_i[A_j]$ of incomplete object $o_i\in iDS$. 

\begin{figure}[t!]
\centering
\includegraphics[scale=0.28]{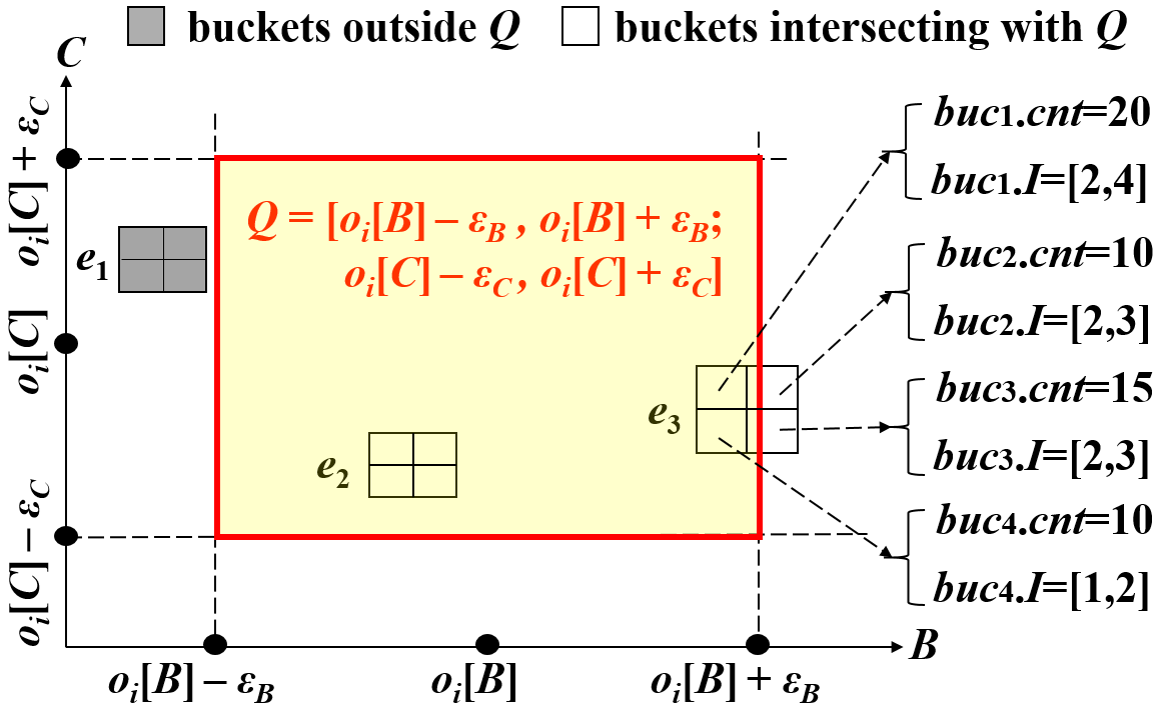}\vspace{-1ex}
\caption{\small The usage of index $\mathcal{I}_j$ for imputing $o_i[D]$ based on $DD_2: BC \to D$}
\label{fig:dominance_by_index}
\end{figure}

As an example in Figure \ref{fig:dominance_by_index}, 
we can use index $\mathcal{I}_j$ (in Figure \ref{fig:MBRStoredInfo}) over attributes $ABC$ to impute missing attribute $D$ for an incomplete object $o_i$. In particular, with the help of a selected DD rule $DD_2: BC \to D$, we can specify a query range: $$Q=\big[o_i[B]-\epsilon_B, o_i[B]+\epsilon_B; o_i[C]-\epsilon_C, o_i[C]+\epsilon_C\big],$$ over attributes $BC$ (wildcard ``$*$'' for other attribute $A$). In Figure \ref{fig:dominance_by_index}, we can obtain two nodes, $e_2$ and $e_3$, from index $\mathcal{I}_j$ intersecting with $Q$, each of which has four (projected) buckets, $buc_q$, intersecting with the query region $Q$. Correspondingly, we can retrieve the value bounds, $buc_q.I$, in these buckets $buc_q$ to impute attribute $D$ for incomplete object $o_i$. For example, in $e_3$, since all the 4 buckets are intersecting with $Q$, we can obtain lower/upper bounds of possible imputed attribute $D$ w.r.t. $e_3$, that is, $[1, 4]$ ($=buc_1.I \cup buc_2.I \cup buc_3.I \cup buc_4.I$).

\noindent {\bf Object pruning via indexes.} As discussed in Section~\ref{sec:pruning_strategies}, we can apply spatial, max-corner, and min-corner pruning to filter out an (imputed) object $o_i^p$ by using another object $n^p$, where the missing attributes in $o_i^p$ and $n^p$ are imputed by their possible values (inferred from data repository $R$). In the sequel, we will briefly discuss how to enable the pruning by traversing indexes $\mathcal{I}_j$ over $R$. 

Specifically, when we access a level of index $\mathcal{I}_j$ for imputing attribute $A_j$ of object $o_i^p$ (or $n^p$), we can retrieve several possible value intervals of attribute $A_j$. Then, we can compute value boundaries, $buc_q.I$, of attributes $A_j$ for object $o_i^p$ (or $n^p$), and thus obtain corners $o_i^p.max$ and $n^p.min$, which can be used in the spatial pruning (as mentioned in Lemma~\ref{lemma:lem1}). Similarly, we can also obtain COUNT aggregates, $buc_q.cnt$, for attribute $A_j$ intervals from buckets $buc_q$, and compute probabilities $Pr\{n^p \prec o_i^p.max\}$ and $Pr\{n^p.min$ $\prec o_i^p\}$, which are used for max-corner and min-corner pruning (Lemmas~\ref{lemma:lem2} and \ref{lemma:lem3}), respectively. Similar to the pruning on the object level, we omit the pruning details via indexes.

\begin{algorithm}[t!]\scriptsize
\KwIn{the skyline tree $ST$ and a new object $o_i$} 
\KwOut{the updated $ST$}
    $parentNode \leftarrow null$ \tcp{parent node of $o_i^p$ in $ST$}  
    
    $isPruned \leftarrow false$ \text{ }\text{ }\tcp{whether $o_i^p$ can be pruned}  
    
    $isAdded \leftarrow false$ \text{ }\text{ }\text{ }\text{ }\tcp{whether $o_i^p$ has been inserted} 
    
    \For{each object $n^p$ on layer $1$}{
        \If{$Pr\{o_i^p \prec n^p\} \ge 1-\alpha$}{
            $isAdded \leftarrow true$ \tcp{insert $o_i^p$ into layer 1} 
            
            \If{$o_i^p.exp \ge n^p.exp$}{
                replace $n^p$ with $o_i^p$ in $ST$ \tcp{$n^p$ is pruned}
            }\Else{
                add $o_i^p$ to the first layer of $ST$ 
                
                move $n^p$ and all its descendant nodes from their current layer $L$ to layer $(L+1)$ 
                
                let $o_i^p$ be the parent node of $n^p$ 
            }
        }
    }
    
\If{$isAdded = false$}{
    queue $\mathcal{Q} \leftarrow$ all objects $n^p$ on the first layer of $ST$ \\ 
    \While{$\mathcal{Q}$ is not empty}{
        remove $n^p$ from $\mathcal{Q}$
        
        \If{$Pr\{n^p \prec o_i^p\} \ge 1-\alpha$}{
            \If{$n^p.exp \ge o_i^p.exp$}{
                $isPruned \leftarrow true$ \tcp{$o_i^p$ is pruned} 
                
                break; \tcp{terminate the while loop}
            }\Else { \tcp{find the parent of $o_i^p$}
                \If{$parentNode$ $=$ $null$ or $n^p.exp$ $>$ $parentNode.exp$}{
                    $parentNode \leftarrow n^p$ 
                }
                add all child nodes of $n^p$ to $\mathcal{Q}$ 
            }
        }
    }
    \If{$isPruned = false$}{
        \If{$parentNode = null$}{\tcp{$o_i^p$ is a skyline candidate}
            add $o_i^p$ to the first layer of $ST$ 
            
        }\Else{
            let $o_i^p$ be the child node of $parentNode$
        }
    }
}

\tcp{If $o_i^p$ is inserted into $ST$, find children of $o_i^p$ and use $o_i^p$ to prune other objects in $ST$}
\If{$isPruned = false$}{
    \For{each object $n^p$ from layer $o_i^p.layer$ to $height(ST)$}{
        \If{$Pr\{o_i^p \prec n^p\} \ge 1-\alpha$}{
            \If{$o_i^p.exp \ge n^p.exp$}{
                remove $n^p$ from layer $n^p.layer$ 
                
                move up all descendant nodes $o_c^p$ of $n^p$ by $(o_c^p.layer-o_i^p.layer-1)$ layer(s) 
                
                let $o_i^p$ be the new parent for child nodes of $n^p$
            }\ElseIf{$o_i^p.exp > par(n^p).exp$}{
                let $o_i^p$ be the new parent of $n^p$  
                
                move up $n^p$ and all its descendant nodes $o_c^p$ by $(o_c^p.layer-o_i^p.layer-1)$ layer(s) 
                
                delete the edge between $n^p$ and its old parent $par(n^p)$ \\
            }
        }
    }
}
\vspace{-0.1cm}
\caption{Insertion}
\label{alg:alg_insertion}
\end{algorithm}

\subsection{Sky-iDS Query Processing Algorithm}
\label{sec:subsec:baseline_algorithm}

\vspace{-2ex}
As discussed in Section \ref{subsec:skyline_tree} (Properties 1-3), the skyline tree $ST$ always contains a superset of Sky-iDS query answers on its first layer. Therefore, in order to efficiently process Sky-iDS queries over incomplete data stream, one important issue is how to dynamically maintain this skyline tree $ST$ in the streaming environment, upon object insertions and deletions. Then, we will discuss how to refine skyline candidates from (the first layer of) $ST$.

\vspace{-2ex}
\subsubsection{Dynamic Maintenance of the Skyline Tree}
\label{subsubsec:dynamic_maintenance_of_skyline_tree}

\noindent{\bf Insertion.} When a new object $o_i$ arrives from incomplete data stream $iDS$, we will consider how to update the skyline tree $ST$ with this (incomplete) object $o_i$. Specifically, Algorithm \ref{alg:alg_insertion} illustrates the pseudo code to decide appropriate location to insert the imputed object $o_i^p$ (if $o_i$ is incomplete), and incrementally maintain the data structure of the $ST$ index. 

\underline{\it Basic idea.} In Algorithm \ref{alg:alg_insertion}, we initialize 3 variables, that is, $parentNode$, $isPruned$, and $isAdded$, which store the parent node of object $o_i^p$ after the insertion, whether $o_i^p$ can be pruned by some object in $ST$, and whether object $o_i^p$ has been added to $ST$, respectively (lines 1-3). Then, we will find appropriate location in $ST$ to insert object $o_i^p$, either on the first layer or on another layer pointed by a parent node, $parentNode$ (lines 4-29). Finally, we will update the $ST$ index by removing those objects dominated by $o_i^p$ and finding children of object $o_i^p$ in $ST$ (lines 30-40).

\underline{\it Finding the location to insert new object $o_i^p$.} First, we will check whether or not new (imputed) object $o_i^p$ dominates any object $n^p$ (i.e., $Pr\{o_i^p \prec n^p\} \ge 1-\alpha$ holds) on layer 1 of $ST$ (lines 4-5). If the answer is yes, then $o_i^p$ can be inserted into layer 1 and the variable $isAdded$ is set to $true$ (line 6). Moreover, if object $o_i^p$ expires after $n^p$ (i.e., $o_i^p.exp \ge n^p.exp$), it indicates that $n^p$ cannot be skyline any more (i.e., always dominated by $o_i^p$ during its lifetime), and thus we replace $n^p$ with $o_i^p$ in $ST$ (note: if there are duplicate objects $o_i^p$ on the first layer, we will keep only one copy and merge their children; lines 7-8). Otherwise (i.e., $o_i^p.exp < n^p.exp$ holds; line 9), $o_i^p$ should be a parent node of $n^p$. Therefore, we will add $o_i^p$ to the first layer (line 10), move layers of $n^p$ and all its descendant nodes from current layer $L$ to $(L+1)$ (line 11), and let $o_i^p$ point to $n^p$ (line 12). 

In the case that new object $o_i^p$ has not been added to layer 1 (i.e., $isAdded = false$; line 13), we will utilize a queue, $\mathcal{Q}$, to search an appropriate parent node, $parentNode$, for this new object $o_i^p$ (lines 14-24). Initially, we insert all objects on layer 1 of $ST$ into the query $\mathcal{Q}$ (line 14). Each time we pop out one object, $n^p$, from queue $\mathcal{Q}$ (line 16). If $n^p$ dominates $o_i^p$ with probabilities greater than $(1-\alpha)$ and $n^p$ expires after $o_i^p$, in this case, $o_i^p$ can never be a skyline during its lifetime, that is, new object $o_i^p$ should not be inserted into $ST$. Thus, we set variable $isPruned$ to $true$, and terminate the search loop (lines 17-20). When $o_i^p$ cannot be pruned by $n^p$ (as $n^p.exp < o_i^p.exp$ holds; line 21), we will set $n^p$ as a temporary (best-so-far) parent, $parentNode$, of $o_i^p$, under one of the two conditions: (1) $n^p$ is the first potential parent node we encounter (i.e., $parentNode = null$), or (2) $n^p$ expires later than a best-so-far parent node, $parentNode$, of $o_i^p$ (intuitively, $o_i^p$ should be inserted under a parent node with the largest expiration time) (lines 22-23). Moreover, we will add children of node $n^p$ to query $\mathcal{Q}$ for further searching (since these children may also be potential parent node of $o_i^p$ in $ST$; line 24). 

The loop of finding parent node $parentNode$ repeats, until queue $\mathcal{Q}$ becomes empty (line 15) or new object $o_i^p$ can be pruned (line 20). If $o_i^p$ cannot be pruned and $parentNode \\= null$ holds, it implies that no object can dominate $o_i^p$ with high probability, and we can insert $o_i^p$ into layer 1 as a skyline candidate (lines 25-27). On the other hand, if any parent node is found in variable $parentNode$, then we will let new object $o_i^p$ be the child of $parentNode$ (lines 28-29).

\underline{\it Finding children of new object $o_i^p$ and pruning objects in} \underline{$ST$}.  After we find the parent node of newly inserted object $o_i^p$ in the $ST$ index, we will next update the children of this new object $o_i^p$, as well as using $o_i^p$ to prune/purge some (dominated) objects in $ST$ (lines 30-40). That is, we will consider all objects, $n^p$, from the layer of $o_i^p$ (i.e., $o_i^p.layer$) to the height of the $ST$ index, and check whether $o_i^p$ dominates $n^p$ during $n^p$'s lifetime (lines 31-33). If the answer is yes, then we will remove object $n^p$ from $ST$, and let descendant nodes, $o_c^p$, of $n^p$ be that of $o_i^p$ (lines 34-36). Otherwise (i.e., $n^p$ is not pruned), if $o_i^p$ expires after the parent node, $par(n^p)$, of $n^p$, then we should let $o_i^p$ be the new parent of $n^p$, move up $n^p$ and all its descendants in $ST$, and remove the link from old parent, $par(n^p)$, to $n^p$ (lines 37-40).

\underline{\it Correctness of the insertion algorithm.}
Please refer to the discussions about the correctness of the insertion algorithm in Appendix~\ref{sec:proof_of_lemma_for_dm_of_st_and_ra}.

\begin{algorithm}[t!]\scriptsize
\KwIn{the skyline tree $ST$ and current timestamp $t$} 
\KwOut{the updated $ST$}
\For{each expired object $n^p$ on layer $1$}{
        remove object $n^p$ from $ST$ \\
        move all descendant nodes of $n^p$ from their current layer $L$ to layer $(L-1)$ \\
    
}
\caption{Deletion}
\label{alg:alg_deletion}
\end{algorithm}

\noindent{\bf Deletion.} At timestamp $t$, some objects $o_i^p$ from incomplete data stream $iDS$ are expired (i.e., $o_i^p.exp \leq t$). Algorithm \ref{alg:alg_deletion} will remove all the expired objects from the skyline tree $ST$ over $iDS$. In fact, we can prove that all the expired objects reside on layer 1 of $ST$ (since objects on layers other than layer 1 will always expire after their parents). Since objects in each layer are sorted in ascending order of their expiration times, we will only check those expired objects on layer 1 (line 1). In particular, for each expired object $n^p$, we first remove it from $ST$ (line 2), and move up all its descendant nodes $o_c^p$ from their current layer $L$ to layer ($L-1$) (line 3).


\noindent{\bf Complexity analysis.} The object insertion in Algorithm \ref{alg:alg_insertion} requires $O\big(|W_t| \cdot \frac{1-fanout(ST)^{height(ST)}}{1-fanout(ST)}\big)$ time complexity, where $|W_t|$ is the size of sliding window $W_t$ (i.e., the number of descendants of $o_i^p$ in index $ST$), $height(ST)$ is the height of the tree $ST$, and $fanout(ST)$ is the average number of children per node in $ST$. Similarly, the object deletion in Algorithm \ref{alg:alg_deletion} needs $O\big(\theta\cdot \frac{1-fanout(ST)^{height(ST)}}{1-fanout(ST)}\big)$ time cost, where $\theta$ is the maximum number of expired objects on layer 1 of index $ST$.

\begin{algorithm}[t!]\scriptsize 
\KwIn{the skyline tree $ST$, timestamp $t$, and a data stream $W_t$} 
\KwOut{the updated skyline answer set, $A_t$, at timestamp $t$}

\If{there is no update with $W_t$ at timestamp $t$}{
    return $A_{t-1}$;
}

$A_t=\emptyset$;

\If{there is no new object added to $W_t$ at timestamp $t$}{
    let $A_t$ be $A_{t-1}$ excluding all expired objects at timestamp $t$\\
    \tcp{objects in $A_t$ are definitely skylines} 
    let $V$ be all objects on layer 1 of $ST$, but not in $A_t$
}
\Else{
    \tcp{objects on layer 1 are potential skylines} 
    let $V$ be all objects on layer 1 of $ST$\\
}

\For{each object $o_i^p \in V $}{
    obtain a lower bound, $lb\_P(o_i^p)$, of probability $P_{Sky\text{-}iDS}(o_i^p)$ \\
    \If{$lb\_P(o_i^p) > \alpha$}{
        add $o_i^p$ to $A_t$
    }\Else{
        compute exact Sky-iDS probability, $P_{Sky\text{-}iDS}(o_i^p)$, of $o_i^p$\\
        \If {$P_{Sky\text{-}iDS}(o_i^p) > \alpha$}{
            add $o_i^p$ to $A_t$
        }
    }
}
return $A_{t}$;

\caption{Sky-iDS Refinement}
\label{alg:refinement_algorithm}
\end{algorithm}

\subsubsection{Sky-iDS Refinement}
\label{subsubsec:refinement}

After dynamic maintenance of the skyline tree $ST$ over $iDS$, the first layer of $ST$ always contains a superset of Sky-iDS answers at timestamp $t$, as guaranteed by Property 3 of $ST$ (in Section \ref{subsec:skyline_tree}). Thus, we will incrementally refine Sky-iDS candidates and return actual Sky-iDS query answers in a skyline answer set $A_t$.

Algorithm \ref{alg:refinement_algorithm} provides the pseudo code of refining Sky-iDS candidates upon stream updates. In particular, if there is no update (insertion or deletion) at timestamp $t$, then skyline answers remain the same and we simply return skylines at previous timestamp $(t-1)$ in $A_{t-1}$ (lines 1-2). In the case that there are deletions but no insertions, those objects in $A_{t-1}$ (excluding expired objects) are still skylines at timestamp $t$. Thus, we add these non-expired objects in $A_{t-1}$ to $A_t$, and objects on layer 1 of $ST$, but not in $A_t$, will form a candidate set $V$ that should be refined (lines 3-6). On the other hand, if both insertions and deletions occur, then we will assign all objects on layer 1 to candidate set $V$ (lines 7-8). 

Next, we will refine objects $o_i^p$ in the candidate set $V$ by checking their Sky-iDS probabilities $P_{Sky\text{-}iDS}(o_i^p)$ (as given by Eq.~(\ref{eq:eq3}); lines 9-16). Specifically, we will first calculate a lower bound, $lb\_P(o_i^p)$, of the skyline probability $P_{Sky\text{-}iDS}(o_i^p)$ (line 10). Here, the lower bound probability can be obtained by calculating the skyline probability of min-corner, $o_i^p.min$, of object $o_i^p$. If $lb\_P(o_i^p) > \alpha$ holds, object $o_i^p$ will definitely be a skyline, and we add $o_i^p$ to the skyline answer set $A_t$ (lines 11-12). Otherwise, we need to compute exact skyline probability, $P_{Sky\text{-}iDS}(o_i^p)$, of $o_i^p$, and add $o_i^p$ to $A_t$ if the Sky-iDS probability is greater than $\alpha$ (lines 13-16). Finally, we return actual Sky-iDS query answers in set $A_t$ (line 17).

\underline{\it Correctness of the refinement algorithm.}
Please refer to discussions on the correctness of the refinement algorithm in Appendix~\ref{sec:proof_of_lemma_for_dm_of_st_and_ra} (Lemmas~\ref{lemma:lem7} and \ref{lemma:lem8}).

\noindent{\bf Complexity analysis.} Algorithm \ref{alg:refinement_algorithm} has $O(|W_t| \cdot \theta)$ time complexity in the worst case, where
$|W_t|$ is the number of valid objects in sliding window $W_t$ at timestamp $t$, and $\theta$ is the number of new objects per timestamp in data stream. At timestamp $t$, we need to update skyline probabilities of (at most $|W_t|$) objects on the first layer of the skyline tree $ST$, due to the insertion of at most $\theta$ new objects and the deletion of at most $\theta$ expired objects. Therefore, the worst-case refinement cost is given by $O(|W_t| \cdot \theta)$. Note that, in practice, the expected number of objects on the first layer of $ST$ is much smaller than $|W_t|$. From our experiments over real/synthetic data sets (as discussed later in Section \ref{subsec:effectiveness_pruning_methods}), the average number of objects on layer 1 of $ST$ is about 2.8\%-11.76\% of $|W_t|$. Thus, the refinement algorithm (Algorithm \ref{alg:refinement_algorithm}) is empirically quite efficient in the average case.

\begin{figure}[t!]
\centering
\includegraphics[scale=0.25]{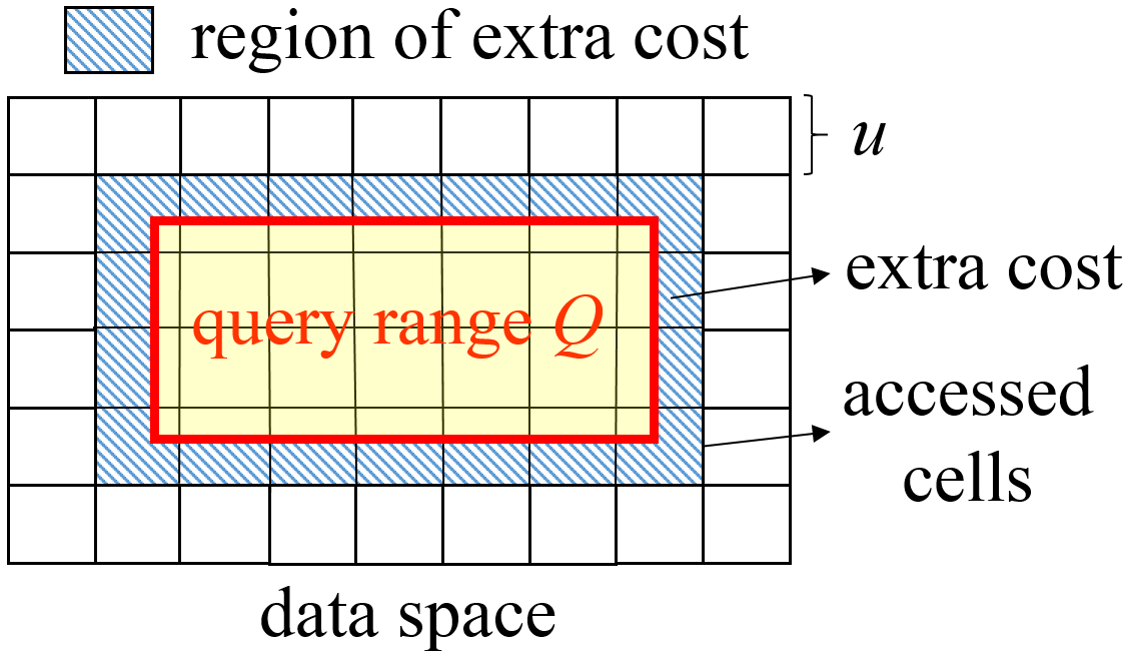}\vspace{-2ex}
\caption{\small Derivation of the cost model.}\vspace{1ex}
\label{fig:extracost}
\end{figure}

\subsection{Cost Model for Parameter Tuning}
\label{sec:cost_model_for_parameter_tuning}

We provide a cost model to tune the parameter $u$ (i.e., the side length of each cell in the grid) for index $\mathcal{I}_j$ over $R$ (discussed in Section \ref{sec:index_over_R_for_imp}). The basic idea is to derive a cost model for the total cost, $Cost$, to access the grid (w.r.t., parameter $u$). Then, we take the derivative of $Cost$ to $u$, and let it be 0, that is, $\frac{\partial Cost}{\partial u}=0$, in order to find the optimal $u$ that minimizes $Cost$. For the details, please refer to Appendix~\ref{sec:derivation_of_cost_model}.

\section{Experimental Evaluation}
\label{sec:experimental_eval}

\subsection{Experimental Settings}

\noindent {\bf Real/synthetic data sets.} We evaluate the performance of our Sky-iDS approach on both real and synthetic stream data. Specifically, for real data, we use Intel lab data\footnote{\scriptsize\url{http://db.csail.mit.edu/labdata/labdata.html}}, UCI gas sensor data for home activity monitoring\footnote{\scriptsize\url{http://archive.ics.uci.edu/ml/datasets/gas+sensors+for+home+activity+monitoring}}, Antallagma time series data for trading goods\footnote{\scriptsize\url{https://www.kaggle.com/abkedar/times-series-kernel}}, and Pump sensor data for predictive maintenance\footnote{\scriptsize\url{https://www.kaggle.com/nphantawee/pump-sensor-data/version/1}}, denoted as $Intel$, $Gas$ $Bid$, and $Pump$, respectively. $Intel$ data are collected every 31 $sec$ from 54 sensors deployed in Intel Berkeley Research lab on Feb. 28-Apr. 5, 2004, including 2.3 million readings. $Gas$ data contain 919,438 sensory instances from 8 MOX gas sensors, a temperature and humidity sensor. $Bid$ data contains 882K operation transactions between buyers and sellers from Jan. 2014 to Jun. 2016. $Pump$ has 220K data, collected from 52 sensors on Apr. 1-Aug. 31, 2018. We extract 4 attributes from $Intel$ data: temperature, humidity, light, and voltage; 10 attributes from $Gas$ data: resistance of sensors 1-8, temperature, and humidity; 8 attributes from $Bid$ data: price\_sd, price\_mean, price\_max, price\_min, mean, max, min, sd; and 10 attributes from $Pump$: sensor\_01-sensor\_10. We normalize all the attributes of each real data set to an interval $[0, 10]$. We obtain DD rules (as depicted in Table \ref{table:real_DDs}), by scanning all complete objects $s_r$ in data repository $R$ and all possible combinations of any two determinant/dependent attributes in the data schema \cite{song2011differential}, and selecting the ones with minimum interval for each dependent attribute $A_j$.





For synthetic data, we generate data repository $R$ and incomplete data stream $iDS$ as follows. Following the convention \cite{Borzsonyi01}, we generate three types of $d$-dimensional data sets: $Uniform$, $Correlated$, and $Anti$-$correlated$, which correspond to different data distributions. Specifically, we first generate 5,000 seeds following uniform, correlated, or anti-correlated distribution \cite{Borzsonyi01}. Then, based on these seeds, we produce the remaining data objects, following DD rules as depicted in Table \ref{table:real_DDs}. 


For real/synthetic data above, given a missing rate $\xi$ (i.e., the probability that objects in the sliding window have missing attributes), for each incomplete object, we randomly set $m$ out of $d$ attributes to ``$-$'' (i.e., missing attributes), and obtain incomplete data stream $iDS$. Table \ref{table:exp_instance_No} depicts the average number of instances per incomplete object for both real and synthetic data, where $m=1$ and $\xi=0.3$.



\begin{table}[t!]
\caption{\small The tested real/synthetic data sets and their DD rules.} 
\label{table:real_DDs}
\centering\scriptsize\vspace{-2ex}
\hspace{-3ex}
\begin{tabular}{|c|l|}
\hline
\textbf{Data Sets} & \qquad\qquad\qquad\qquad\quad \textbf{DD Rules}\\
\hline
\hline
 & $voltage \to temperature$, \{$[0,0.001], [0, 0]$\} \\
$Intel$ &   $voltage \to humidity$, \{$[0,0.001], [0, 0]$\}\\
&   $voltage \to light$, \{$[0,0.001], [0, 0]$\}\\
&   $light \to voltage$, \{$[0,0.001], [0, 9.89]$\}\\\hline
 &  $resistance4 \to resistance1$, $\{[0, 0.001], [0, 1.77]\}$\\
&   $resistance3 \to resistance2$, $\{[0, 0.001], [0, 2.615]\}$\\
& $resistance2 \to resistance3$, $\{[0, 0.001], [0, 2.79]\}$\\
& $resistance5 \to resistance4$, $\{[0, 0.001], [0, 2.39]\}$\\
$Gas$ & $resistance4 \to resistance5$, $\{[0, 0.001], [0, 2]\}$\\
& $resistance1 \to resistance6$, $\{[0, 0.001], [0, 0.38]\}$\\
& $resistance3 \to resistance7$, $\{[0, 0.001], [0, 1]\}$\\
& $temperature \to resistance8$, $\{[0, 0.001], [0, 0.06]\}$\\
& $resistance8 \to temperature$, $\{[0, 0.001], [0, 0.07]\}$\\
& $resistance8 \to humidity$, $\{[0, 0.001], [0, 0.43]\}$\\\hline
& $price\_max \to price\_sd$, $\{[0,0.001], [0,5.73]\}$ \\
& $price\_max \to price\_mean$, $\{[0,0.001], [0,4.58]\}$ \\
& $price\_mean \to price\_max$, $\{[0,0.001], [0,6.58]\}$ \\
$Bid$ & $price\_max \to price\_min$, $\{[0,0.001], [0,2.96]\}$ \\
& $sd \to mean$, $\{[0,0.001], [0,3.5]\}$ \\
& $sd \to max$, $\{[0,0.001], [0,3.21]\}$ \\
& $mean \to min$, $\{[0,0.001], [0,2.11]\}$ \\
& $max \to sd$, $\{[0,0.001], [0,2.06]\}$
\\\hline
& $sensor\_06 \to sensor\_01$, $\{[0,0.001], [0,0]\}$ \\
& $sensor\_06 \to sensor\_02$, $\{[0,0.001], [0,0]\}$ \\
& $sensor\_06 \to sensor\_03$, $\{[0,0.001], [0,0]\}$ \\
& $sensor\_06 \to sensor\_04$, $\{[0,0.001], [0,0]\}$ \\
$Pump$ & $sensor\_08 \to sensor\_05$, $\{[0,0.001], [0,0]\}$ \\
& $sensor\_07 \to sensor\_06$, $\{[0,0.001], [0,0.206]\}$ \\
& $sensor\_01 \to sensor\_07$, $\{[0,0.001], [0,0.73]\}$ \\
& $sensor\_07 \to sensor\_08$, $\{[0,0.001], [0,0.6]\}$ \\
& $sensor\_01 \to sensor\_09$, $\{[0,0.001], [0,0.65]\}$ \\
& $sensor\_08 \to sensor\_10$, $\{[0,0.001], [0,0]\}$
\\\hline
 & $B \to A$, $\{[0,0.001], [0,0.01]\}$ \\
 & $C \to B$, $\{[0,0.001], [0,0.01]\}$ \\
 & $D \to C$, $\{[0,0.001], [0,0.01]\}$ \\
$Uniform$ & $E \to D$, $\{[0,0.001], [0,0.01]\}$ \\
$Correlated$ & $F \to E$, $\{[0,0.001], [0,0.01]\}$ \\
$Anti$-$correlated$ & $G \to F$, $\{[0,0.001], [0,0.01]\}$ \\
& $H \to G$, $\{[0,0.001], [0,0.01]\}$ \\
& $I \to H$, $\{[0,0.001], [0,0.01]\}$ \\
& $J \to I$, $\{[0,0.001], [0,0.01]\}$ \\
& $A \to J$, $\{[0,0.001], [0,0.01]\}$ 
\\\hline
\end{tabular}\vspace{1ex}
\end{table}

\begin{table}[t!]\hspace{-5ex}\vspace{-4ex}
\caption{\small Average number of instances per incomplete object for real/synthetic data sets (with $\xi=0.3$ and $m=1$).} 
\label{table:exp_instance_No}
\centering\scriptsize
\begin{tabular}{|l|c|}
\hline
\qquad\textbf{data sets} & \textbf{average No. of object instances}\\
\hline
\hline
$Intel$ &  14 \\\hline
$Gas$ &  19 \\\hline
$Bid$ &  23 \\\hline
$Pump$ &  11 \\\hline
$Uniform$ & 27 \\\hline
$Correlated$ &  25 \\\hline
$Anti\textit{-}correlated$ &  27 \\\hline
\end{tabular}\vspace{1ex}
\end{table}

\begin{table}[t!]\hspace{-5ex}\vspace{-5ex}
\caption{\small The parameter settings.} 
\label{table:exp_parameter_setting}
\scriptsize
\hspace{-4ex}
\begin{tabular}{|l|c|}
\hline
\qquad\qquad\qquad\qquad\textbf{Parameters} & \textbf{Values}\\
\hline
\hline
probabilistic threshold $\alpha$ &  0.1, 0.2, \textbf{0.5}, 0.8, 0.9 \\\hline
dimensionality $d$ & 2, 3, \textbf{4}, 5, 6, 10 \\\hline
the number, $|W_t|$, of valid objects in $iDS$ & 5K, 10K, \textbf{20K}, 40K, 50K, 80K \\\hline
the size, $|R|$, of data repository $R$ & 40K, 80K, \textbf{120K}, 160K, 200K \\\hline
the number, $\theta$, of new objects per timestamp in $iDS$ & 10, 20, {\bf 30}, 40, 50, 100 \\\hline
the number, $m$, of missing attributes & \textbf{1}, 2, 3 \\\hline
the missing rate, $\xi$, of incomplete objects in $iDS$ & 0.1, 0.2, \textbf{0.3}, 0.4, 0.5 \\\hline
\end{tabular}\vspace{1ex}
\end{table}

\noindent {\bf Competitor.} We compare our Sky-iDS approach with six competitors, namely $DD+skyline$, $mul+skyline$, $con+skyline$, $DD+skyline\_tree$, $mul+skyline\_tree$, and $con+skyline\_tree$. Note that, many existing works (e.g., \cite{pei2007probabilistic,lian2008monochromatic}) for skyline on uncertain data are for static uncertain databases, and require offline building an index and online traversing the index, which is not efficient for the stream scenario. Therefore, we compare with the existing work \cite{Ding12} on skyline over uncertain data streams. The details of the six baseline methods are as follows (please refer to \cite{zhang2016sequential,van2007multiple,Ding12} for more implementation details).

\begin{itemize}
  \item[$\bullet$] $mul+skyline$: this baseline first imputes the missing attribute values via \textit{multiple imputation} \cite{royston2004multiple}, and then performs skyline query processing over imputed data streams via the algorithm in \cite{Ding12}. We implement the \textit{multiple imputation}, by first obtaining 20 possible imputed values for each missing attribute $A_j$ via Markov chain and prior distribution of attribute $A_j$ in complete objects of $R$, and then computing the final imputed value by averaging the 20 imputed values \cite{van2007multiple};
  \item[$\bullet$] $mul+skyline\_tree$: this baseline first imputes the missing attribute values via \textit{multiple imputation} \cite{royston2004multiple} (with the same implementation as the $mul+skyline$), and then performs skyline query processing via the skyline tree over imputed data streams in our work;
  \item[$\bullet$] $con+skyline$: this baseline first imputes the missing attribute values via a \textit{constraint-based imputation method} \cite{zhang2016sequential}, and then uses the skyline query processing method in \cite{Ding12};
  \item[$\bullet$] $con+skyline\_tree$: this baseline first imputes the missing attribute values via a \textit{constraint-based imputation method} \cite{zhang2016sequential}, and then performs skyline query processing via the skyline tree over imputed data streams in our work;
  \item[$\bullet$] $DD+skyline$: this baseline first imputes the missing attribute values via DD rules and data repository $R$, and then conducts the skyline query over imputed data streams via the algorithm in \cite{Ding12};
  \item[$\bullet$] $DD+skyline\_tree$: this baseline first imputes the missing attribute values via DD rules and data repository $R$, and then performs skyline query via the skyline tree over imputed data streams in our work.
\end{itemize}

\noindent {\bf Measures.} In our experiments, we will report maintenance and query times of our proposed Sky-iDS approach, which are the CPU times to incrementally maintain the skyline tree $ST$ (as discussed in Section \ref{subsubsec:dynamic_maintenance_of_skyline_tree}; including the missing data imputation via $\mathcal{I}_j$) and to retrieve actual Sky-iDS query answers (by refining candidates on the first layer of $ST$, as mentioned in Section \ref{subsubsec:refinement}), respectively.


\noindent {\bf Parameter settings.} Table \ref{table:exp_parameter_setting} depicts the parameter settings of our experiments, where default parameter values are in bold. In each set of experiments, we will vary one parameter, while setting other parameters to their default values. We ran our experiments on a machine with Intel(R) Core(TM) i7-6600U CPU 2.70 GHz and 32 GB memory. All algorithms were implemented by C++.
\vspace{-2ex}

\begin{figure}[t!]
\centering\vspace{-2ex}
\includegraphics[scale=0.25]{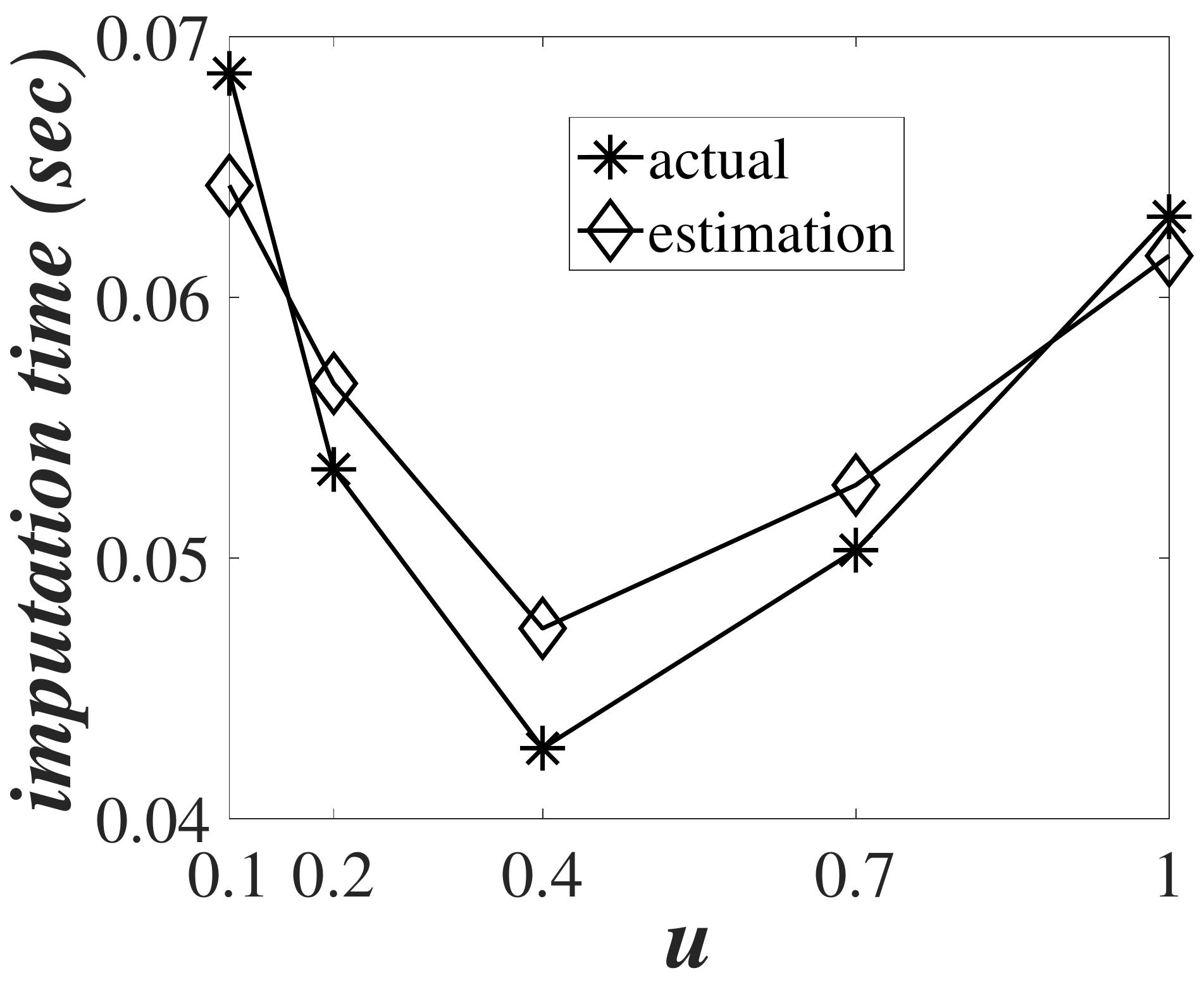}\vspace{-3ex}
\caption{\small Cost model verification for the imputation cost ($Uniform$).}
\label{fig:cost_model_verification}\vspace{1ex}
\end{figure}

\begin{figure}[t!]
\centering\vspace{-5ex}
\subfigure[][{\small real data}]{
\scalebox{0.2}[0.21]{\includegraphics{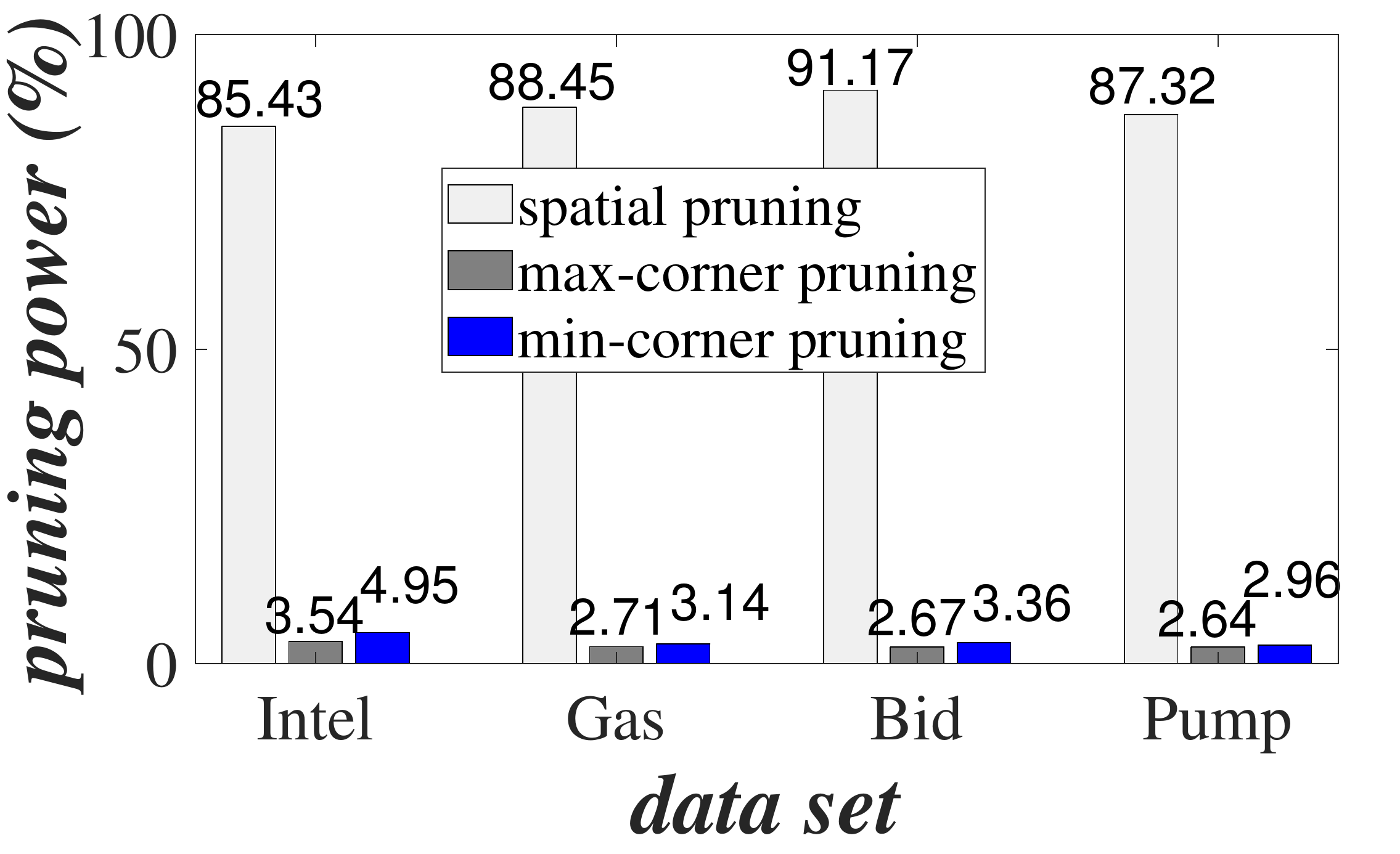}}
\label{subfig:real_pruning_power}
}%
\subfigure[][{\small synthetic data}]{\hspace{-2ex} 
\scalebox{0.2}[0.21]{\includegraphics{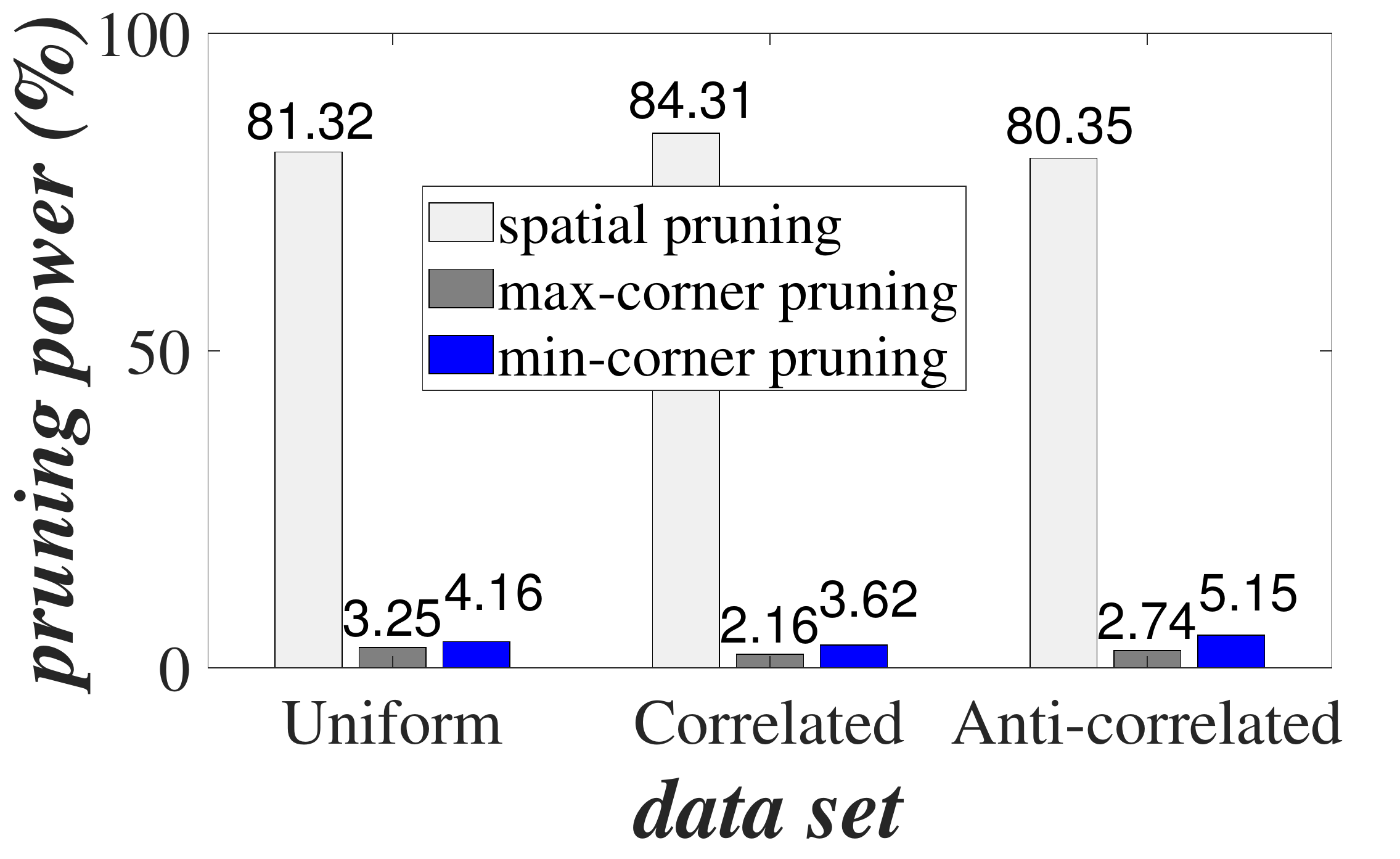}}\vspace{-1ex}
\label{subfig:syn_pruning_power}
}\vspace{-3ex}
\caption{\small Pruning power evaluation over real/synthetic data sets.} 
\label{exper:Sky-iDS_pru_pow_datasets} \vspace{1ex}
\end{figure}

\subsection{Verification of the Cost Model}
We first verify our cost model in Section \ref{sec:cost_model_for_parameter_tuning}, by comparing the estimated and actual data imputation time over $Uniform$ data set, w.r.t. different side lengths, $u$, of cells in index $\mathcal{I}_j$, where $u = 0.1, 0.2, 0.4, 0.7,$ and $1$, and $|R|=120K$. From the experimental results in Figure~\ref{fig:cost_model_verification}, we can see that our estimated imputation cost (given by Eq.~(\ref{eq:cost_model})) can closely approximate the trend of actual imputation cost, which confirms the correctness of our proposed cost model for estimating the imputation cost. As a result, we can use our cost model to select the best value of side length $u$ of cells, that minimizes the imputation cost. In Figure~\ref{fig:cost_model_verification}, the optimal $u$ value is about 0.4, which matches with the $u$ selection based on our cost model, and thus indicates the effectiveness of our cost model.

The verification results of the cost model for other data distributions (e.g., $Correlated$ and $Anti$-$Correlated$) are similar, and therefore omitted here.


\subsection{Effectiveness of Sky-iDS Pruning Methods}
\label{subsec:effectiveness_pruning_methods}
Figure \ref{exper:Sky-iDS_pru_pow_datasets} demonstrates the percentages of objects that are pruned by our three pruning rules, spatial pruning, max-corner pruning, and min-corner pruning, over real/synthetic data sets, where parameters of synthetic data sets are set to their default values. As mentioned in Section 4, we will first apply the spatial pruning, followed by max-corner and min-corner pruning rules (if the spatial pruning fails). From figures, we can see that the spatial pruning can significantly prune most of data objects for both real and synthetic data sets (i.e., 85.43\%-91.17\% for real data sets and 80.35\%-84.31\% for synthetic data sets). Then, the max-corner and min-corner pruning rules can further reduce the Sky-iDS search space. To be specific, the max-corner pruning rule can further prune 2.64\%-3.54\% and 2.16\%-3.25\% of objects from real and synthetic data, respectively, whereas the min-corner pruning rule can further filter out 2.96\%-4.95\% and 3.62\%-5.15\% of objects in real and synthetic data, respectively. Overall, our proposed three pruning methods can together prune 92.92\%-97.2\% and 88.24\%-90.09\% of data objects in real and synthetic data sets, respectively, which indicates the effectiveness of our proposed Sky-iDS approach. Note that, from our experimental results, the first layer of our proposed \textit{skyline tree} $ST$ (as mentioned in Section \ref{subsec:skyline_tree}) contains only 2.8\%-11.76\% of objects in the sliding window $W_t$, which confirms the effectiveness of our \textit{skyline tree} and shows the efficiency of our Sky-iDS refinement algorithm (Section \ref{subsubsec:refinement}).


\begin{figure}[t!]
\centering \vspace{-3ex}
\subfigure[][{\small $F$-score} ($Intel$)]{\hspace{-2ex}                  
\scalebox{0.23}[0.23]{\includegraphics{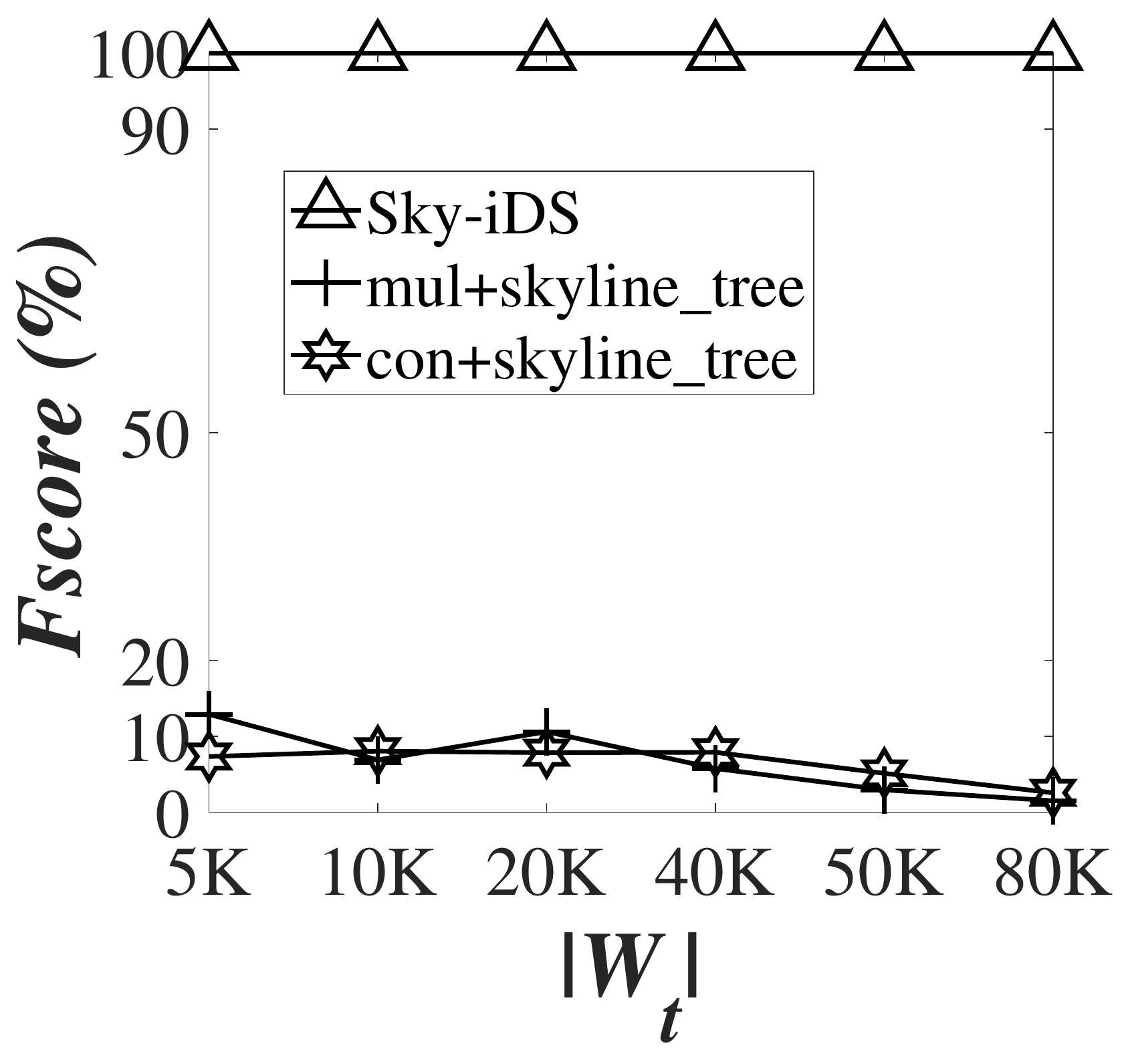}}
\label{subfig:Intel_Fscore_vs_Wt}
}\qquad
\subfigure[][{\small $F$-score} ($Gas$)]{\hspace{-5ex}                  
\scalebox{0.23}[0.23]{\includegraphics{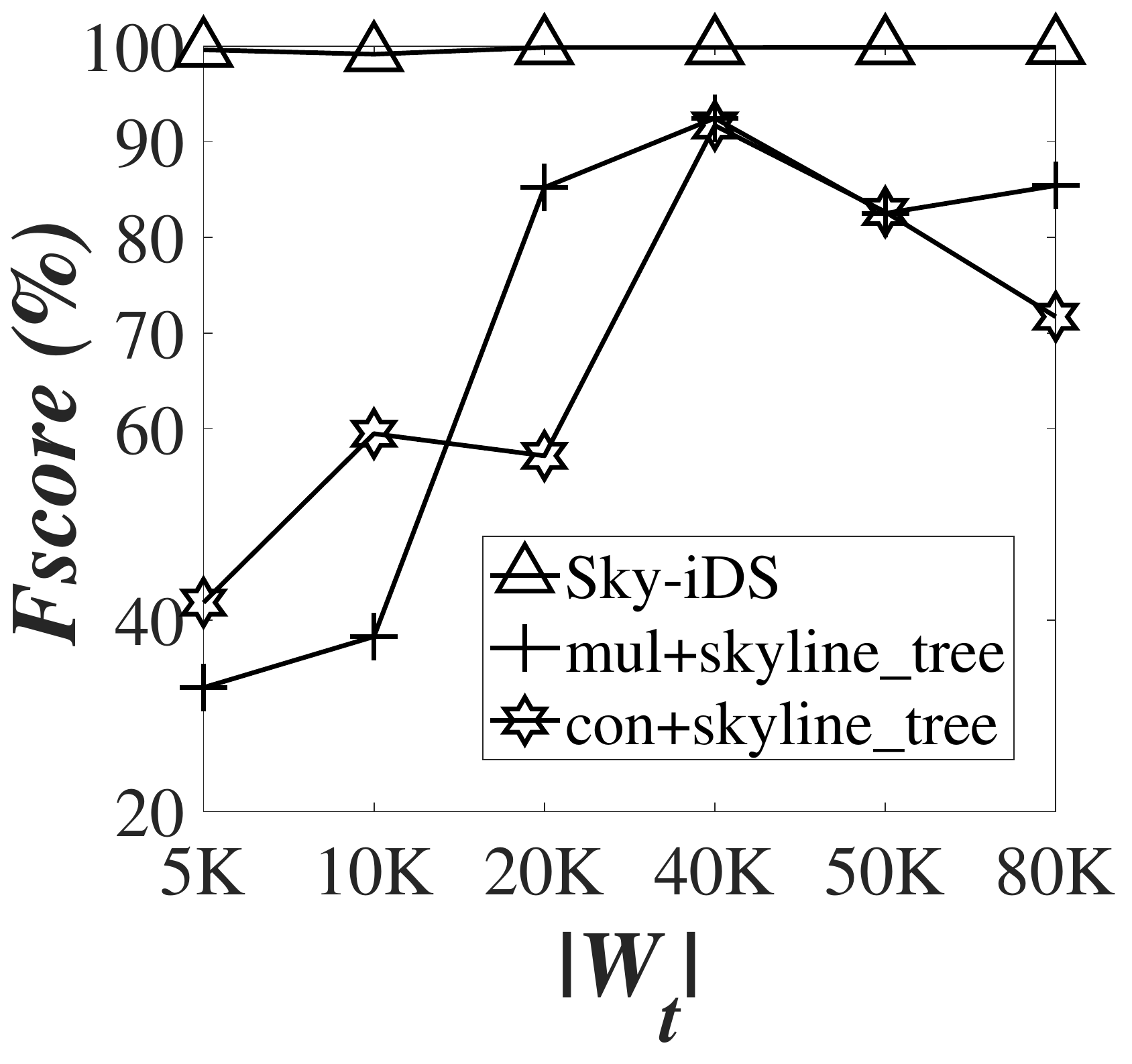}}
\label{subfig:Gas_Fscore_vs_Wt}
}\qquad \\\vspace{-2ex}
\subfigure[][{\small $F$-score} ($Bid$)]{\hspace{-1ex}
\scalebox{0.23}[0.23]{\includegraphics{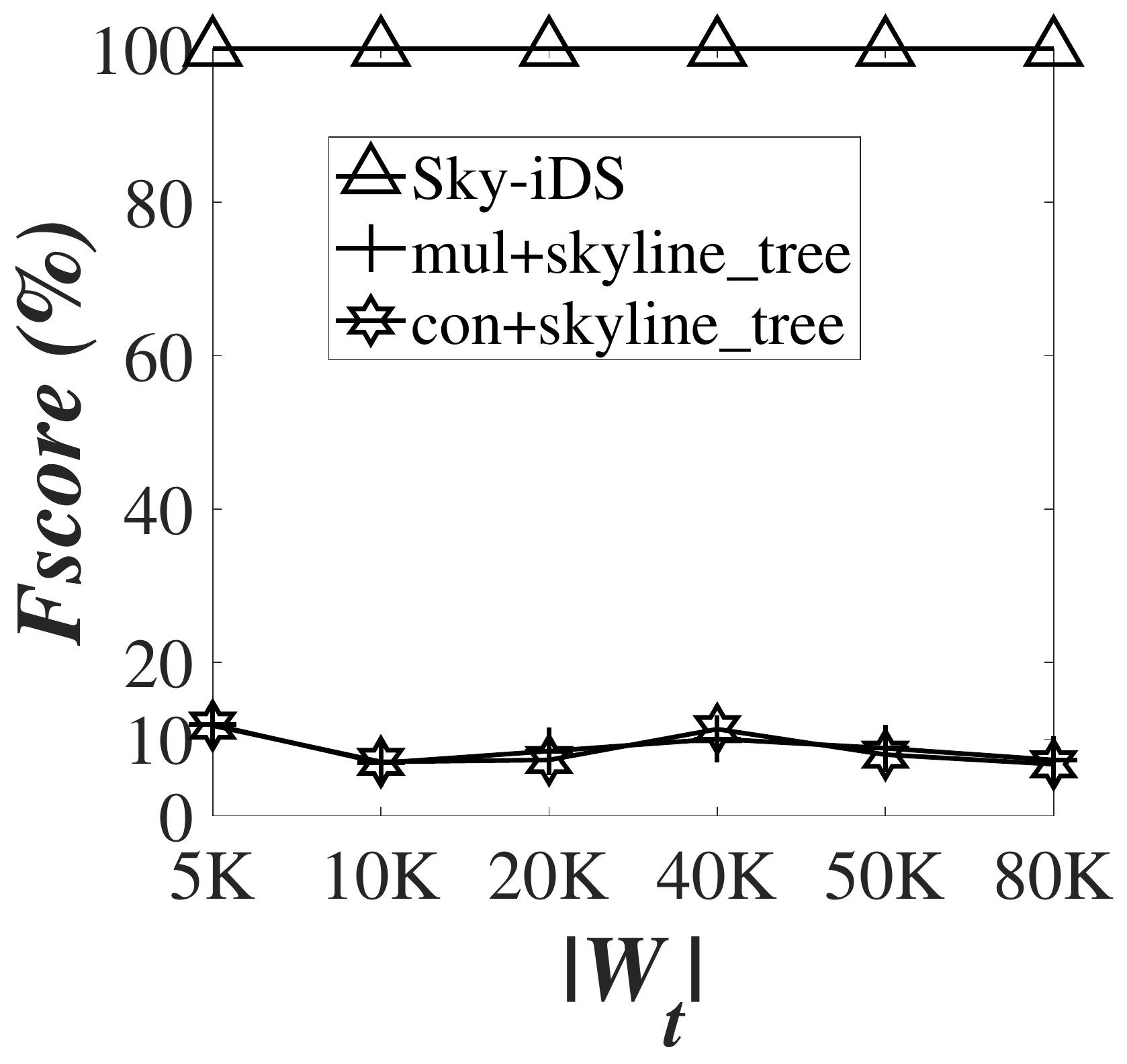}}
\label{subfig:Bid_Fscore_vs_Wt}
}\qquad
\subfigure[][{\small $F$-score} ($Pump$)]{\hspace{-4ex}
\scalebox{0.23}[0.23]{\includegraphics{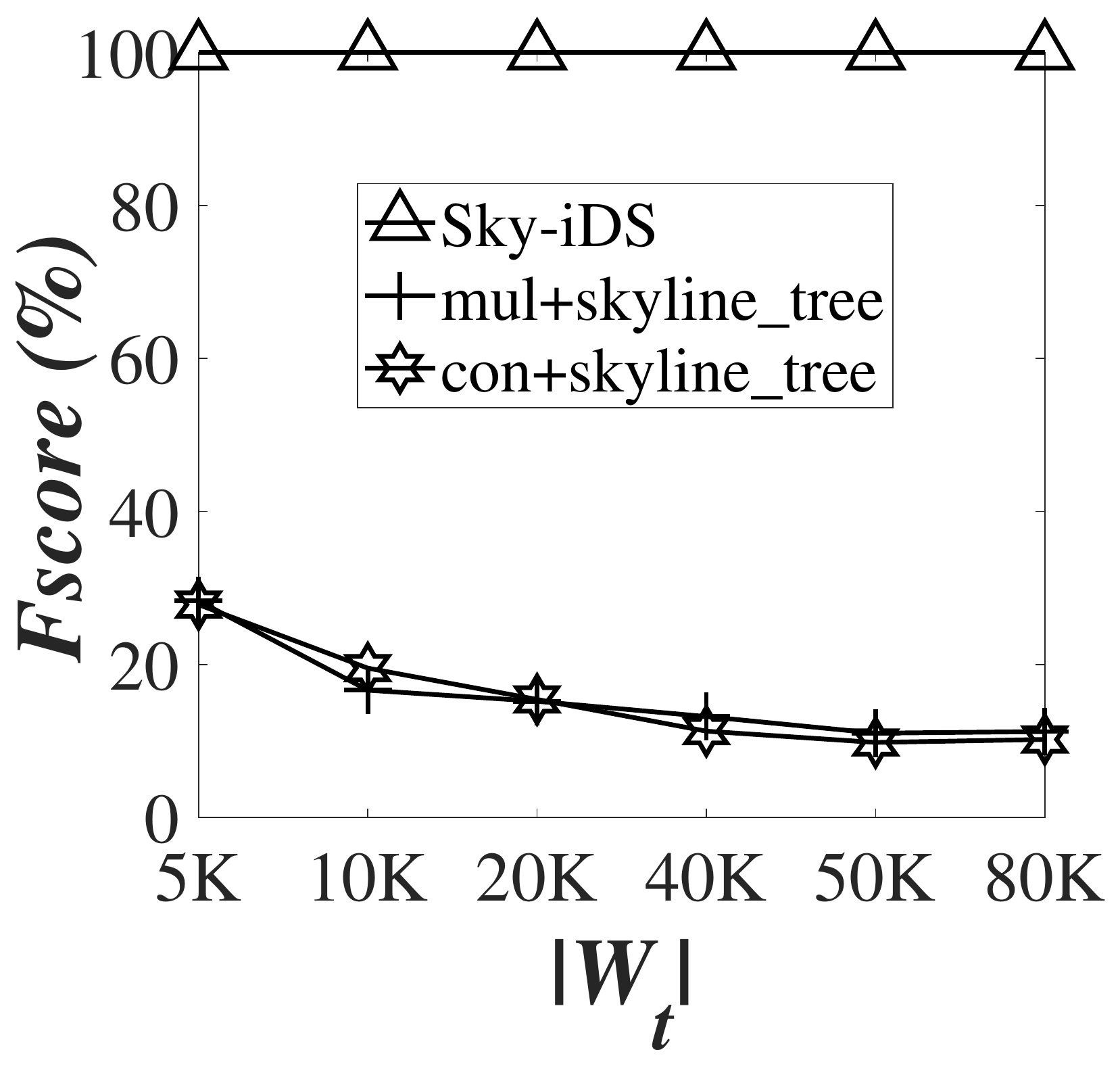}}
\label{subfig:Pump_Fscore_vs_Wt}
}\vspace{-1ex}
\caption{\small The Sky-iDS effectiveness vs. the number, $|W_t|$, of valid objects in $iDS$.} 
\label{exper:Sky-iDS_effectiveness_vs_Wt} \vspace{1ex}
\end{figure} 

\begin{figure}[t!]
\centering \vspace{-3ex}
\subfigure[][{\small $F$-score} ($Intel$)]{\hspace{-2ex}                  
\scalebox{0.23}[0.23]{\includegraphics{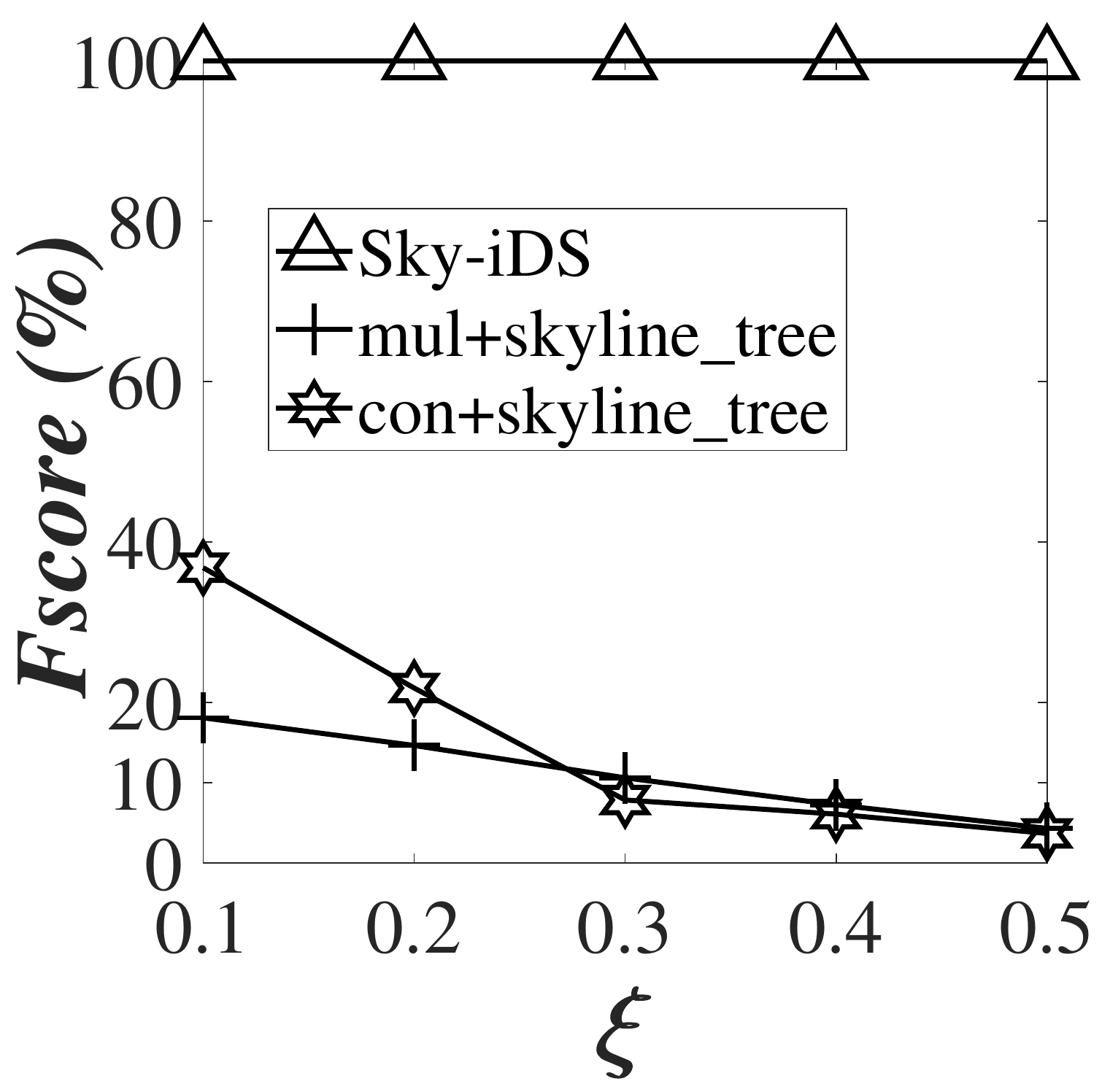}}
\label{subfig:Intel_Fscore_vs_xi}
}\qquad
\subfigure[][{\small $F$-score} ($Gas$)]{\hspace{-5ex}                  
\scalebox{0.23}[0.23]{\includegraphics{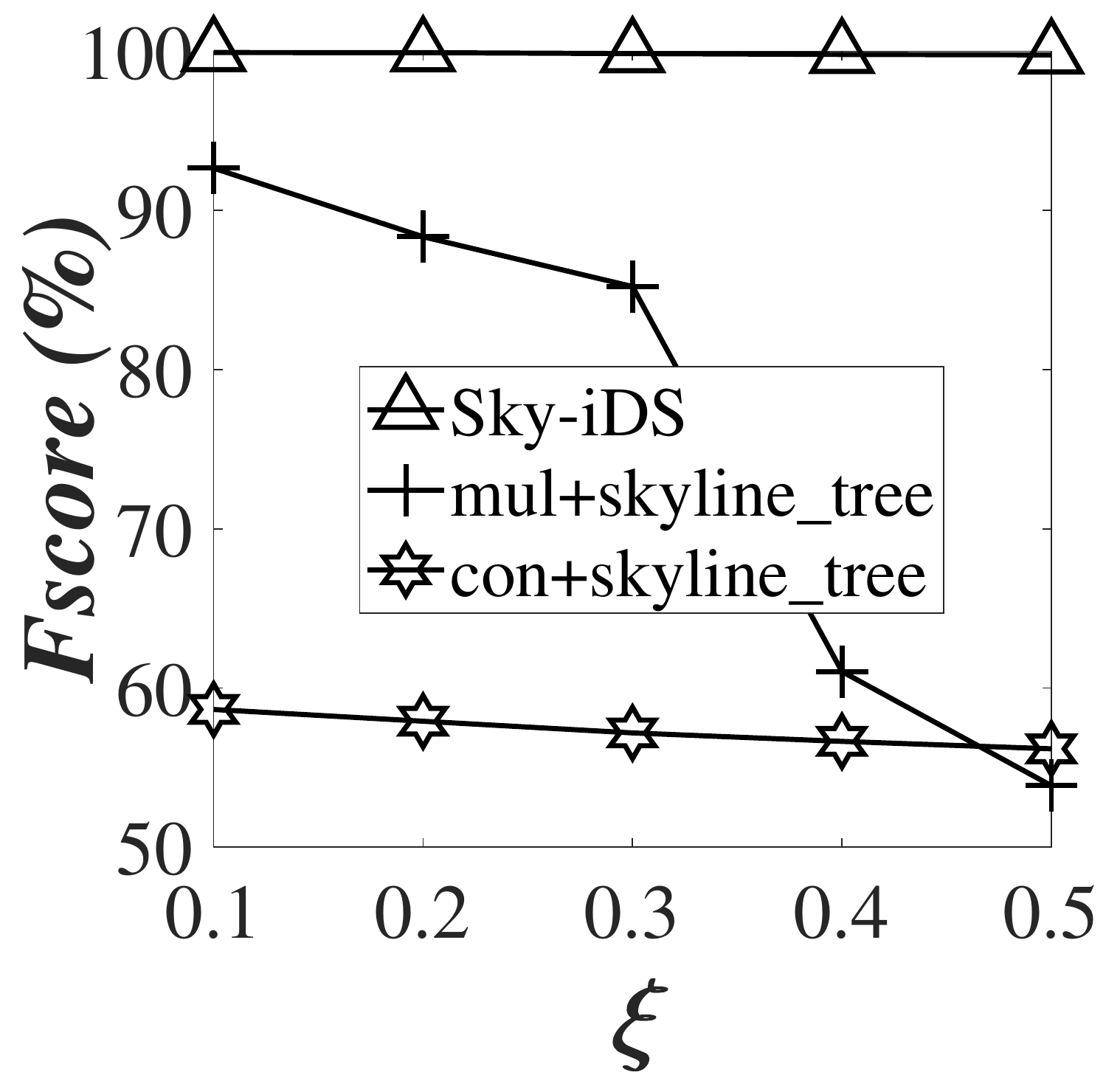}}
\label{subfig:Gas_Fscore_vs_xi}
}\qquad \\\vspace{-2ex}
\subfigure[][{\small $F$-score} ($Bid$)]{\hspace{-2ex}
\scalebox{0.23}[0.23]{\includegraphics{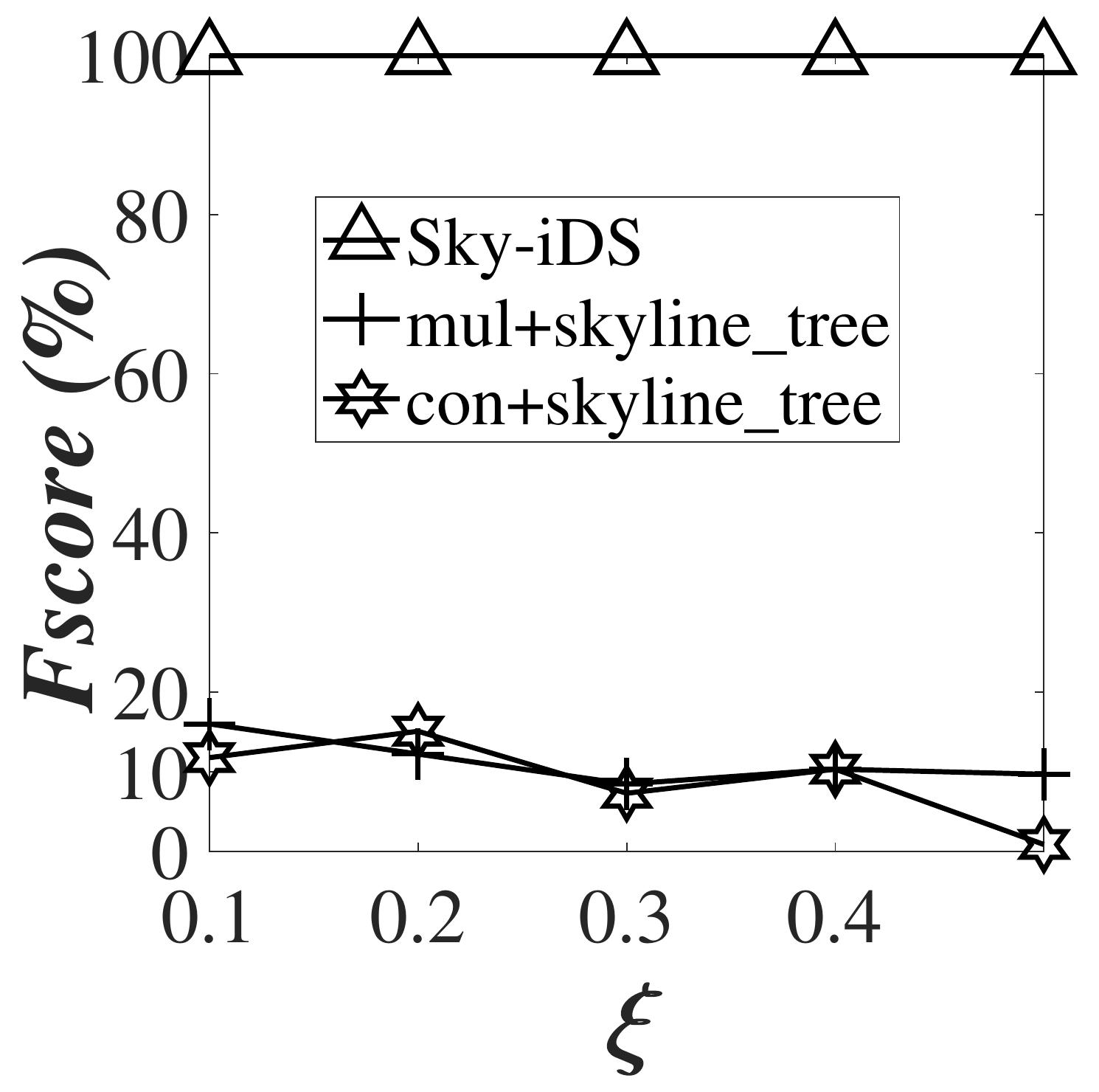}}
\label{subfig:Bid_Fscore_vs_xi}
}\qquad
\subfigure[][{\small $F$-score} ($Pump$)]{\hspace{-5ex}
\scalebox{0.23}[0.23]{\includegraphics{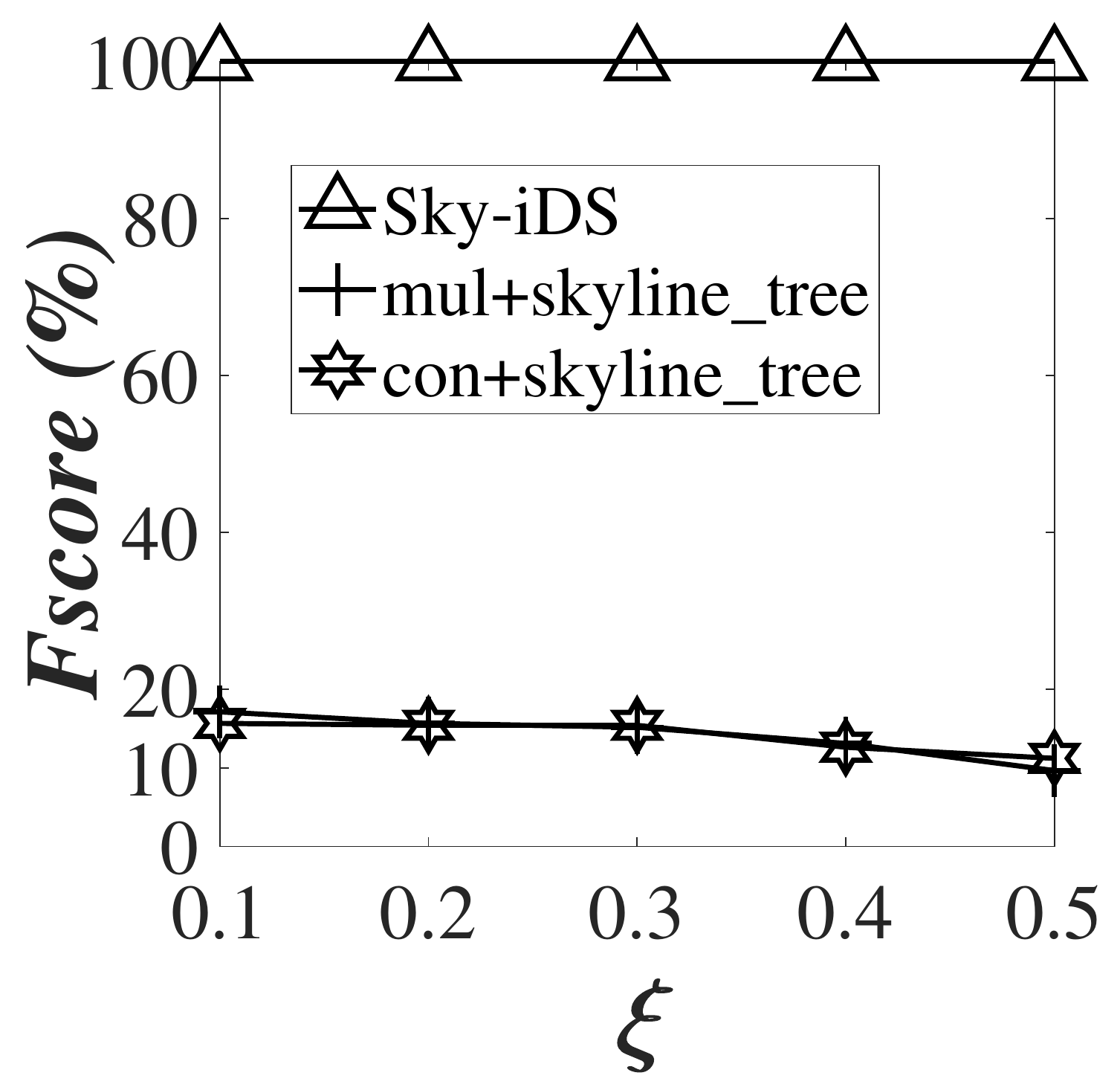}}
\label{subfig:Pump_Fscore_vs_xi}
}\vspace{-1ex}
\caption{\small The Sky-iDS effectiveness vs. the missing rate, $\xi$, of objects in $iDS$.} 
\label{exper:Sky-iDS_effectiveness_vs_xi} \vspace{1ex}
\end{figure} 

\vspace{-4ex}
\subsection{The Effectiveness of Sky-iDS Queries}
\label{subsec:accuracy}

\vspace{-2ex}In this subsection, we compare the effectiveness of our proposed Sky-iDS approach with that of $mul+skyline\_tree$ and $con+skyline\_tree$ over four real data sets (i.e., $Intel$, $Gas$, $Bid$, and $Pump$), in terms of the \textit{$F$-score}. Note that, since $DD+skyline$ and $DD+skyline\_tree$ use the same DD-based imputation method as our Sky-iDS approach, they have the same $F$-score as our Sky-iDS approach. Thus, we will not report the effectiveness of $DD+skyline$ and $DD+skyline\_tree$ here. Similarly, since they have the same $F$-score as $mul+skyline\_tree$ and $con+skyline\_tree$, we will not report the effectiveness of $mul+skyline$ and $con+skyline$, respectively. Specifically, for each (complete) real data set, we first randomly select some objects as incomplete based on the missing rate $\xi$, and then mark $m$ out of $d$ random attribute(s) as missing in the selected objects. This way, we can know the groundtruth of actual skyline query answers from complete real data, and test the accuracy of the three approaches over (masked) incomplete data sets, in terms of the \textit{$F$-score} defined as follows.
\begin{eqnarray}
F\text{-}score &=& 2 \times \frac{recall \times precision}{recall + precision},
\label{eq:F1_score}
\end{eqnarray}

\noindent where \textit{recall} is given by the number of actual skyline answers in our Sky-iDS query results divided by the total number of actual skyline answers in complete data sets, and the \textit{precision} can be calculated by the total number of actual skyline answers in our Sky-iDS query results divided by the total number of objects returned by our Sky-iDS approach.

\noindent {\bf The Sky-iDS effectiveness vs. the number, $|W_t|$, of valid objects in $iDS$.} Figure \ref{exper:Sky-iDS_effectiveness_vs_Wt} shows the query accuracy of our Sky-iDS approach and other two competitors (i.e., $mul+skyline\_tree$ and $con+skyline\_tree$) over the four real data sets, where $|W_t| = 5K$, $10K$, $20K$, $40K$, $50K$ and $80K$, and other parameters follow their default values in Table \ref{table:exp_parameter_setting}. From figures, we can see that our Sky-iDS approach can achieve high \textit{$F$-score} over real data sets with different $|W_t|$ values (i.e., close to 100\%), which significantly outperforms $mul+skyline\_tree$ and $con+skyline\_tree$. 




\noindent {\bf The Sky-iDS effectiveness vs. the missing rate, $\xi$, of objects in $iDS$.} Figure \ref{exper:Sky-iDS_effectiveness_vs_xi} demonstrates the query accuracy evaluation between our Sky-iDS approach and its competitors (i.e., $mul+skyline\_tree$ and $con+skyline\_tree$) over four real data sets, where missing rate $\xi$ varies from 0.1 to 0.5, and other parameters are set to their default values in Table \ref{table:exp_parameter_setting}. As shown in figures, as the increase of the $\xi$, the \textit{$F$-scores} of $mul+skyline\_tree$ and $con+skyline\_tree$ decrease smoothly. This is reasonable, since multiple imputation \cite{royston2004multiple} and constrained-based imputation methods \cite{zhang2016sequential} may lead to higher imputation errors with higher missing rate $\xi$. Nevertheless, Figure \ref{exper:Sky-iDS_effectiveness_vs_xi} shows that our Sky-iDS approach can still achieve high \textit{$F$-score} (close to 100\% even when $\xi=0.5$) for all real data sets, which confirms the effectiveness of our Sky-iDS approach. 

The experimental results with respect to \textit{recall} and \textit{precision} are similar, and thus will not be reported here.

\subsection{The Efficiency of Sky-iDS Queries}

\begin{figure}[t!]
\centering\vspace{-2ex}
\subfigure[][{\small real data}]{\hspace{-2ex}
\scalebox{0.21}[0.21]{\includegraphics{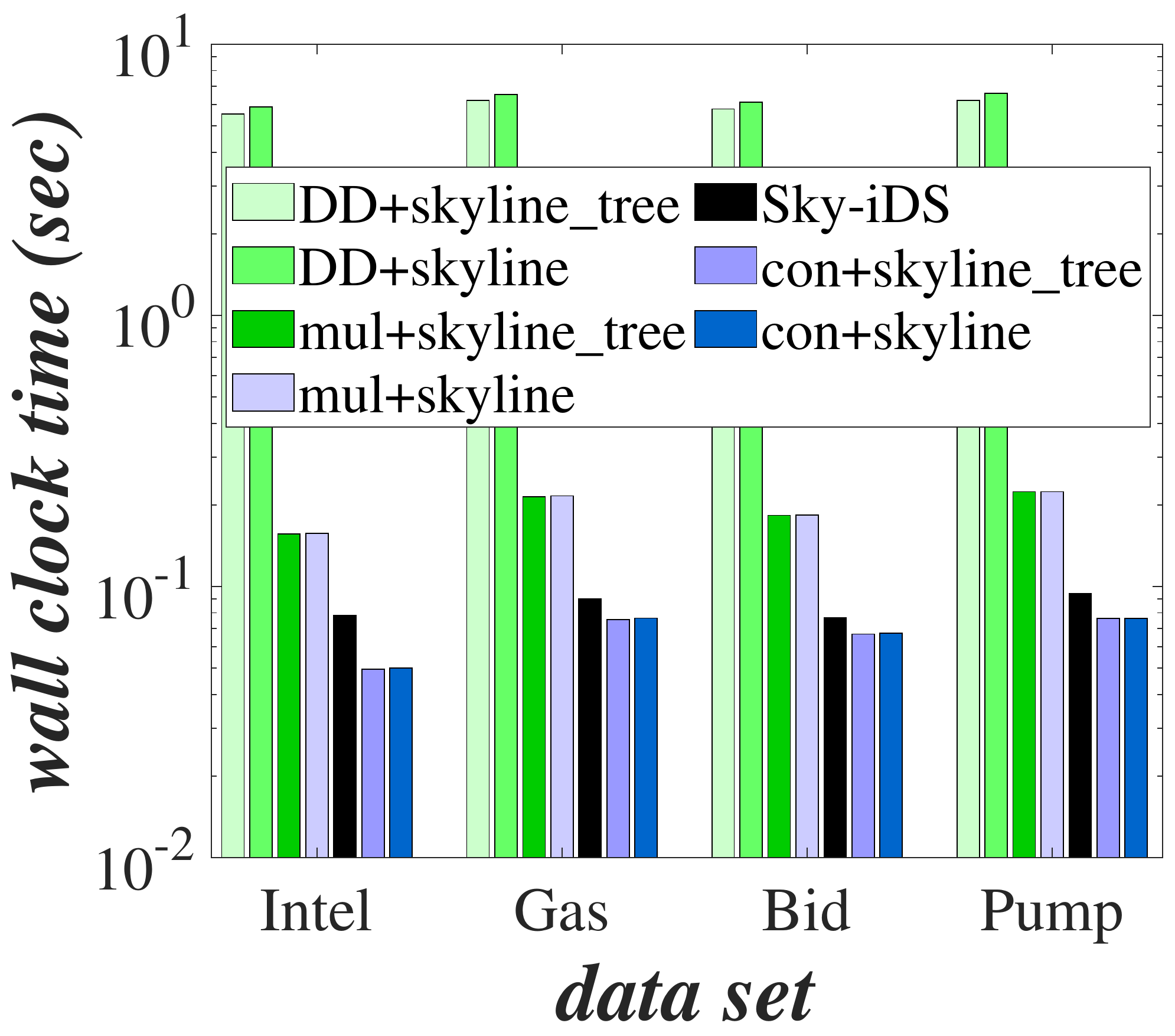}}
\label{subfig:m_cost_vs_datasets}
}%
\subfigure[][{\small synthetic data}]{\hspace{-1ex} 
\scalebox{0.21}[0.225]{\includegraphics{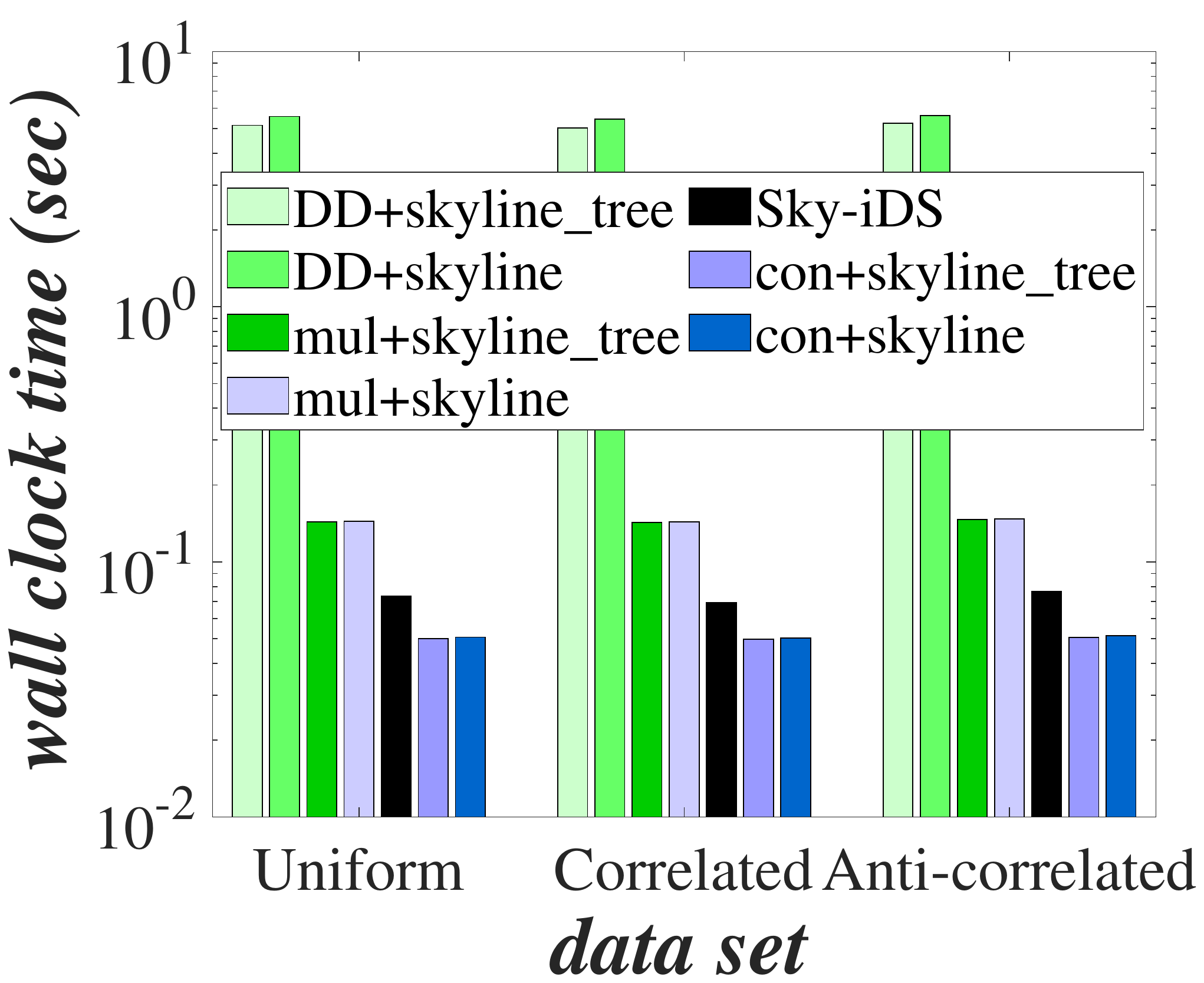}}\vspace{-1ex}
\label{subfig:q_cost_vs_datasets}
}\vspace{-3ex}
\caption{\small The efficiency vs. real/synthetic data sets.} 
\label{exper:Sky-iDS_vs_datasets} \vspace{6ex}
\end{figure}

\noindent {\bf The Sky-iDS efficiency vs. real/synthetic data sets.} Figure \ref{exper:Sky-iDS_vs_datasets} illustrates the performance of our Sky-iDS algorithm, $DD+skyline$, $mul+skyline$, $con+skyline$, $DD+skyline\_tree$, $mul+skyline\_tree$, and $con+skyline\_tree$ over both real and synthetic data sets, where parameters of synthetic data sets are set to default values. We report the overall \textit{wall clock time} of each approach, which includes both maintenance and query times. From experimental results, our Sky-iDS approach outperforms $DD+skyline$ and $DD+skyline\_tree$ algorithms by 2 orders of magnitude, has lower cost than the $mul+skyline$ and $mul+skyline\_tree$ approach, and slighter higher cost than the $con+skyline$ and $con+skyline\_tree$ approach, in terms of the wall clock time. The reason that our Sky-iDS approach is better than $DD+skyline\_tree$ and $DD+skyline$ is as follows. When Sky-iDS performs the imputation (via indexes over data repository $R$) and skyline processing (via skyline tree) at the same time, Sky-iDS can early prune incomplete objects on the level of index nodes. In contrast, $con+skyline$ and $DD+skyline\_tree$ need to impute incomplete objects to their instance level, by obtaining all samples from data repository $R$. Thus, our Sky-iDS approach outperforms $DD+skyline\_tree$ and $DD+skyline$ by two orders of magnitude, which verifies the efficiency of the ``imputation and query processing at the same time'' style of our Sky-iDS approach. Moreover, the experimental results show that our proposed Sky-iDS approach is comparable to $mul+skyline$, $mul+skyline\_tree$, $con+skyline$, and $con+skyline\_tree$, in terms of the efficiency, however, our Sky-iDS approach incurs much higher accuracy, as confirmed by Figs. \ref{exper:Sky-iDS_effectiveness_vs_Wt} and \ref{exper:Sky-iDS_effectiveness_vs_xi}.





Below, we will test the robustness of our Sky-iDS approach by varying different parameters over synthetic data sets.

\begin{figure}[t!]
\centering\vspace{-10ex}
\subfigure[][{\small maintenance time}]{\hspace{-2ex}
\scalebox{0.23}[0.23]{\includegraphics{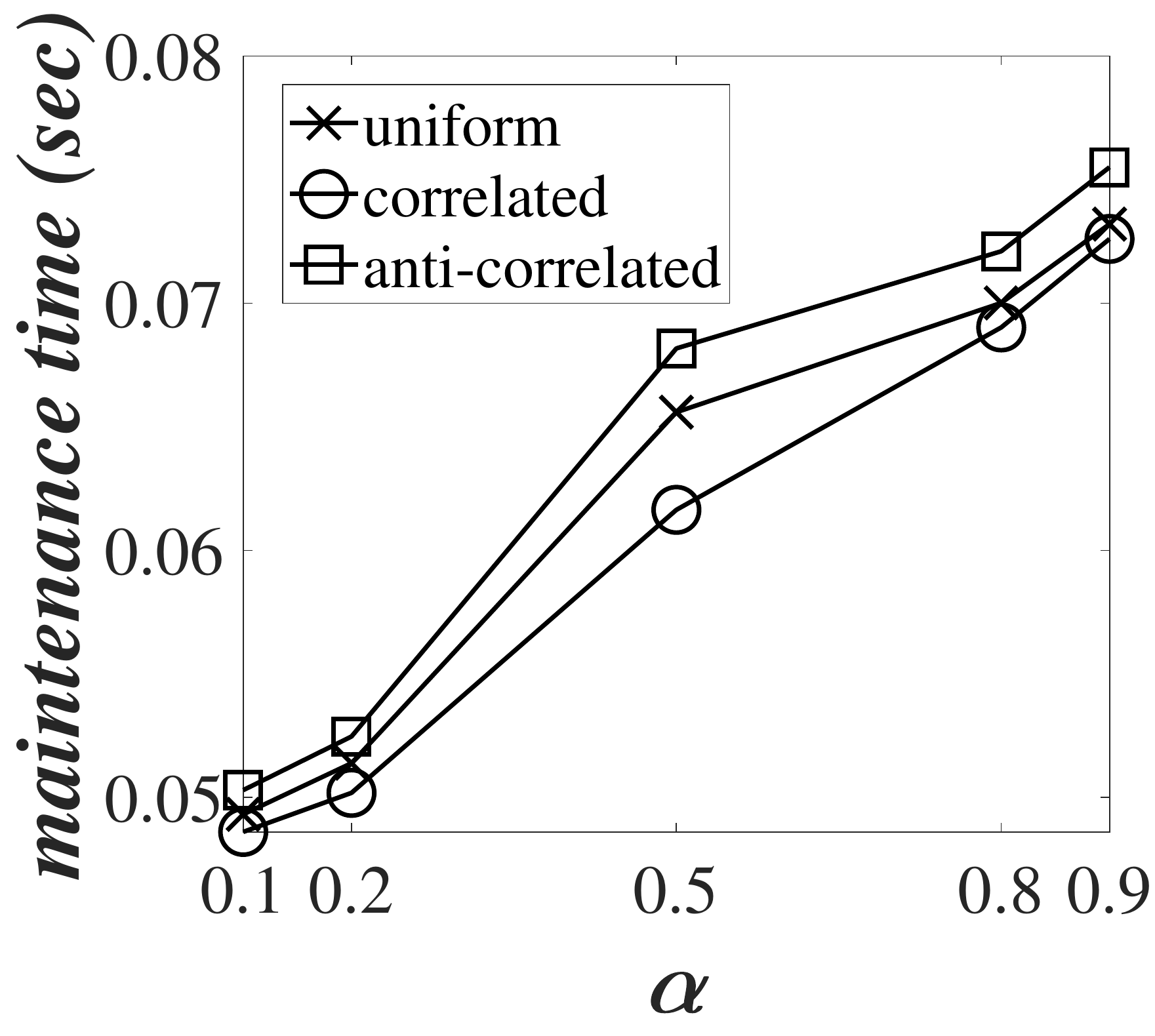}}
\label{subfig:m_cost_vs_alpha}
}%
\subfigure[][{\small query time}]{
\scalebox{0.23}[0.23]{\includegraphics{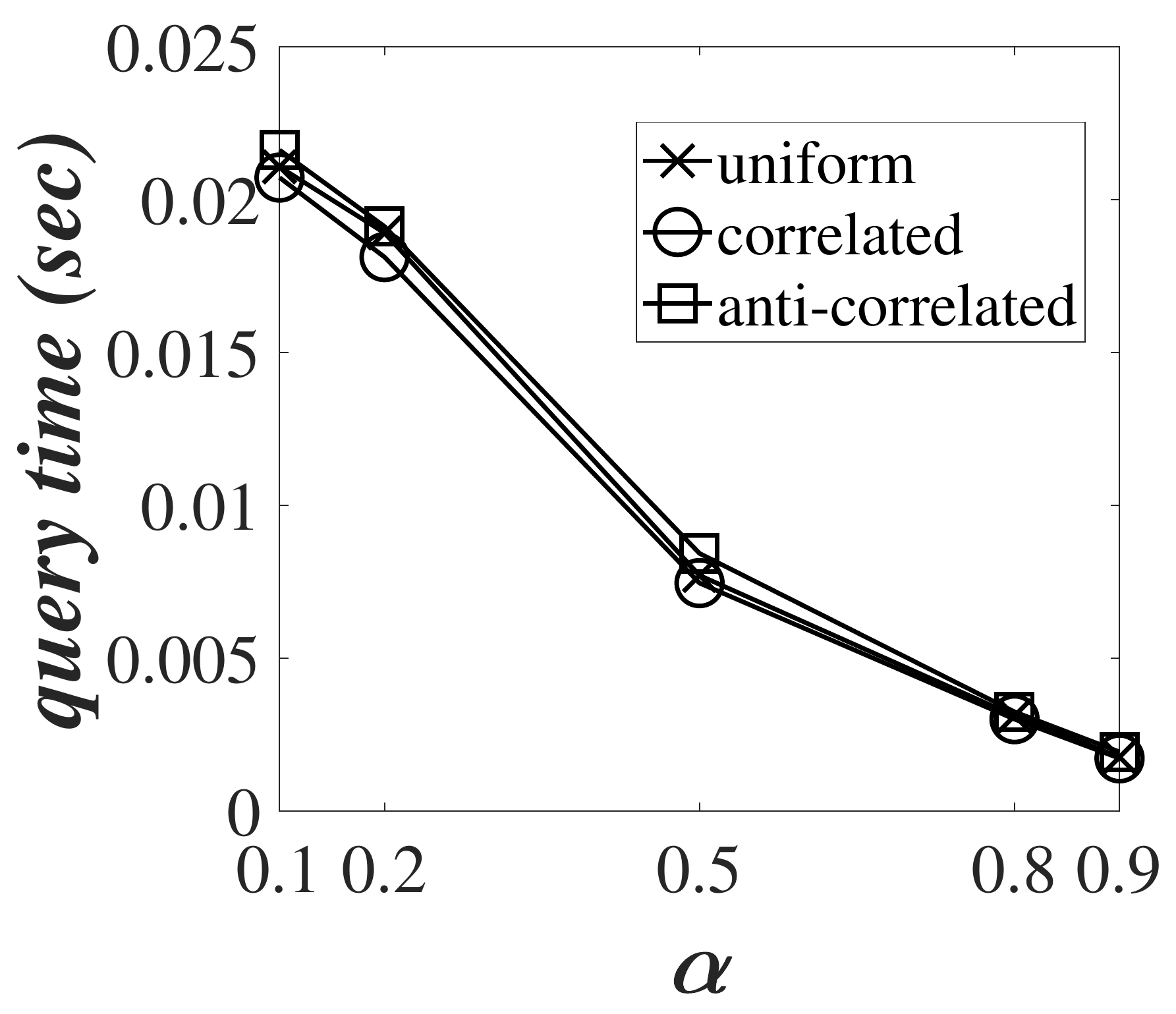}}
\label{subfig:q_cost_vs_alpha}
}\vspace{-3ex}
\caption{\small The efficiency vs. probabilistic threshold $\alpha$.} 
\label{exper:Sky-iDS_vs_alpha} \vspace{1ex}
\end{figure}

\begin{figure}[t!]
\centering\vspace{-2ex}
\subfigure[][{\small maintenance time}]{\hspace{-2ex}
\scalebox{0.21}[0.21]{\includegraphics{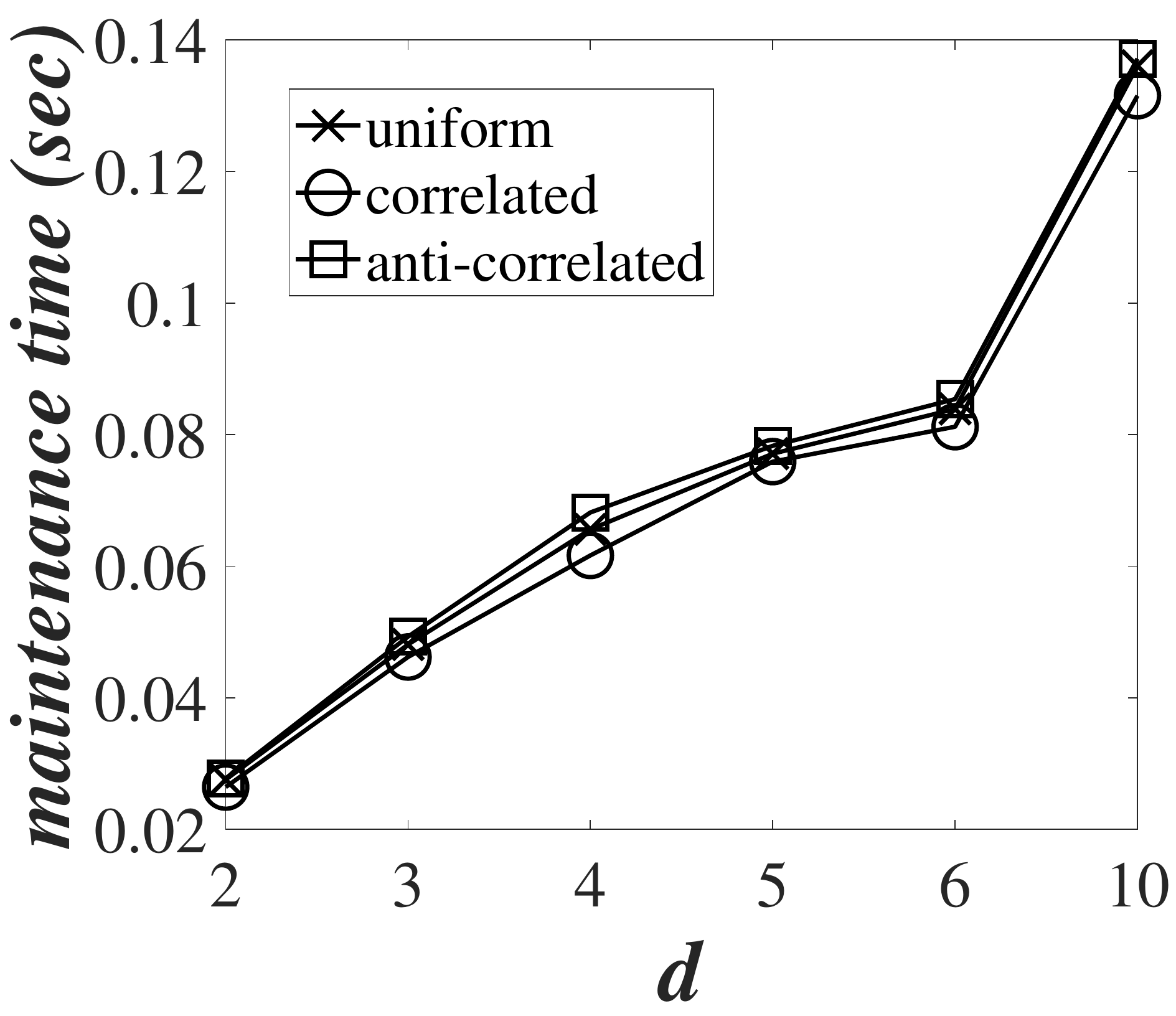}}
\label{subfig:m_cost_vs_d}
}%
\subfigure[][{\small query time}]{
\scalebox{0.21}[0.21]{\includegraphics{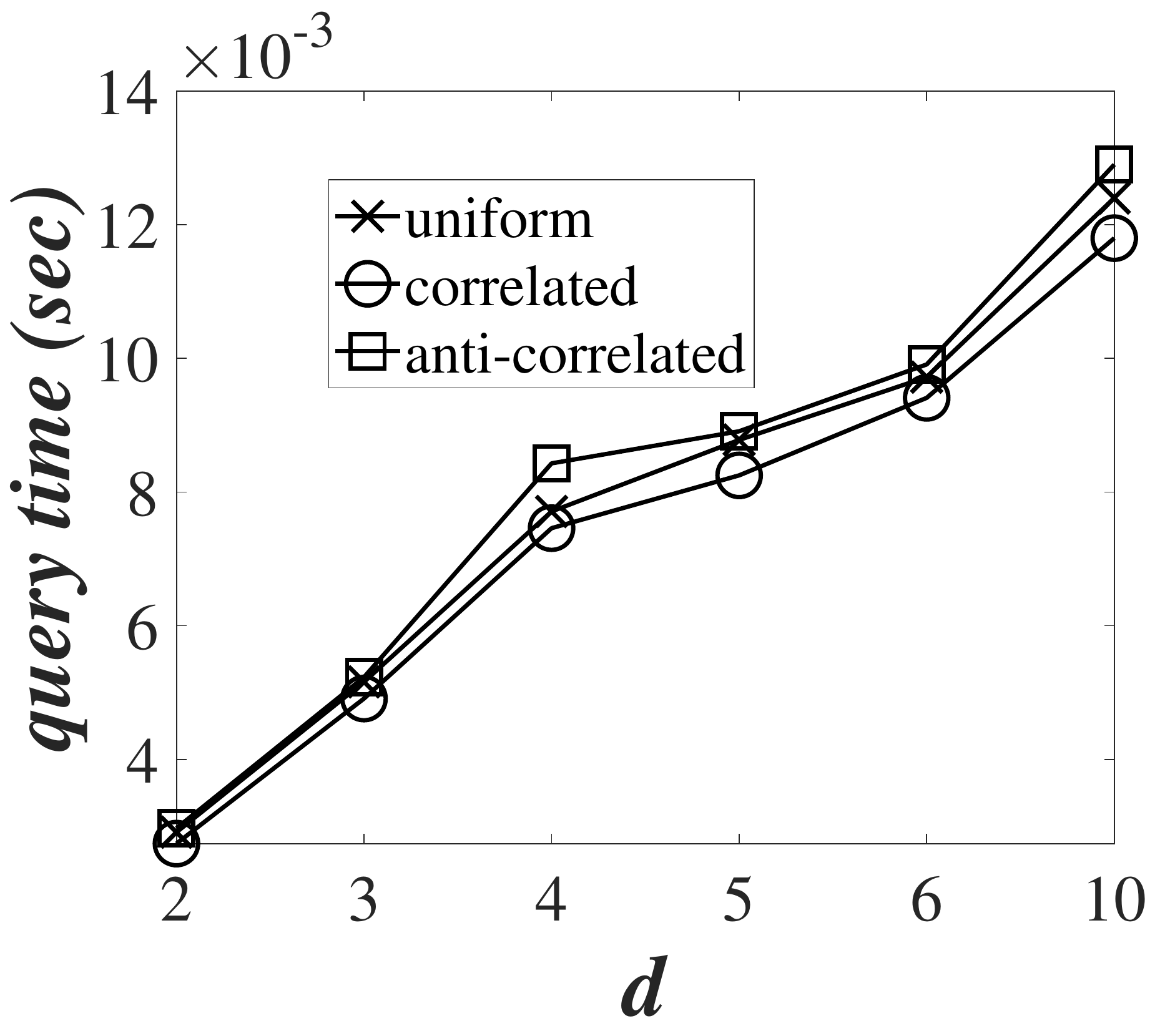}}
\label{subfig:q_cost_vs_d}
}\vspace{-3ex}
\caption{\small The efficiency vs. dimensionality $d$.} 
\label{exper:Sky-iDS_vs_d} 
\end{figure} 

\begin{figure}[t!]
\centering\vspace{-2ex}
\subfigure[][{\small maintenance time}]{\hspace{-2ex}
\scalebox{0.22}[0.22]{\includegraphics{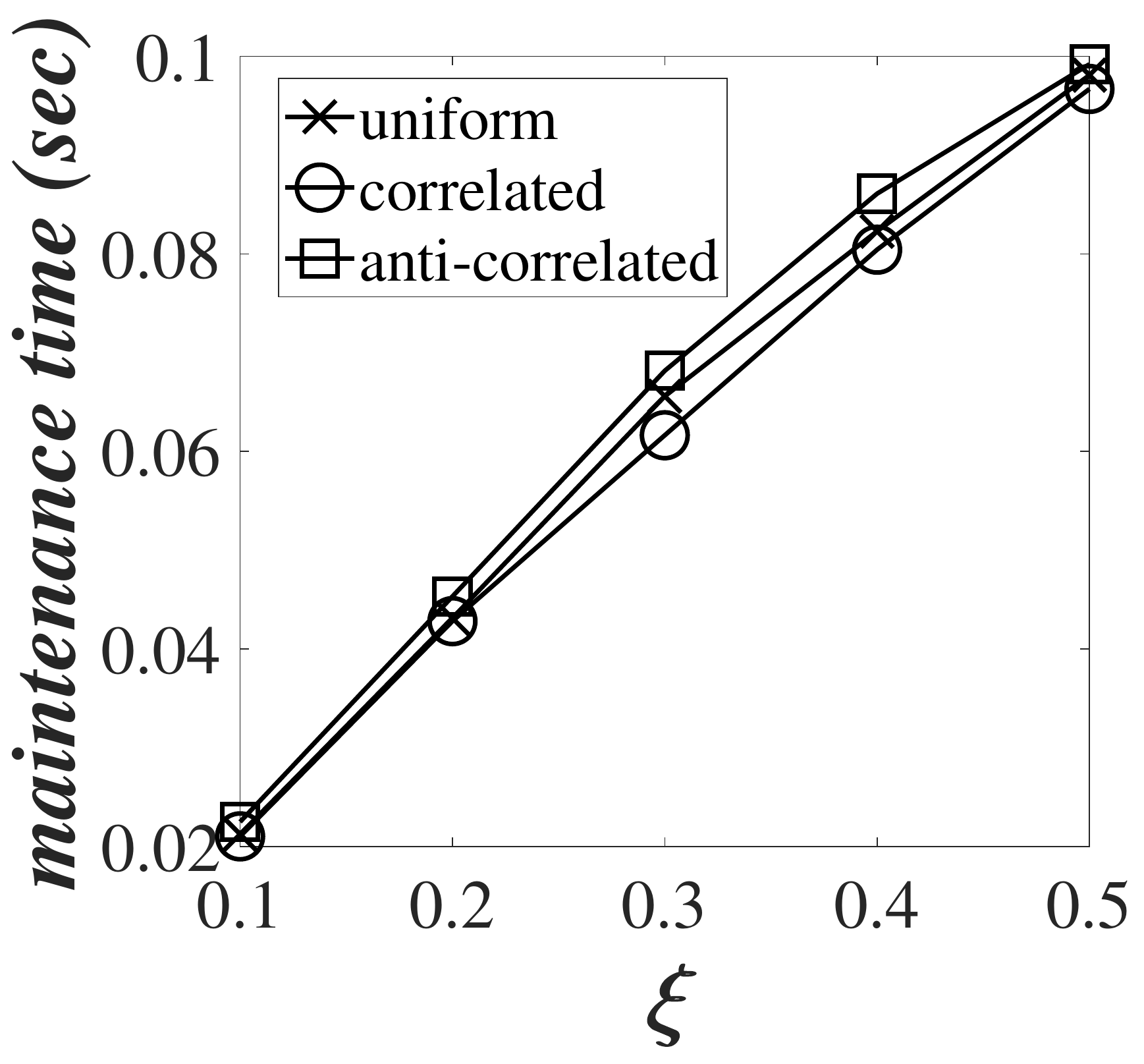}}
\label{subfig:m_cost_vs_xi}
}%
\subfigure[][{\small query time}]{
\scalebox{0.23}[0.23]{\includegraphics{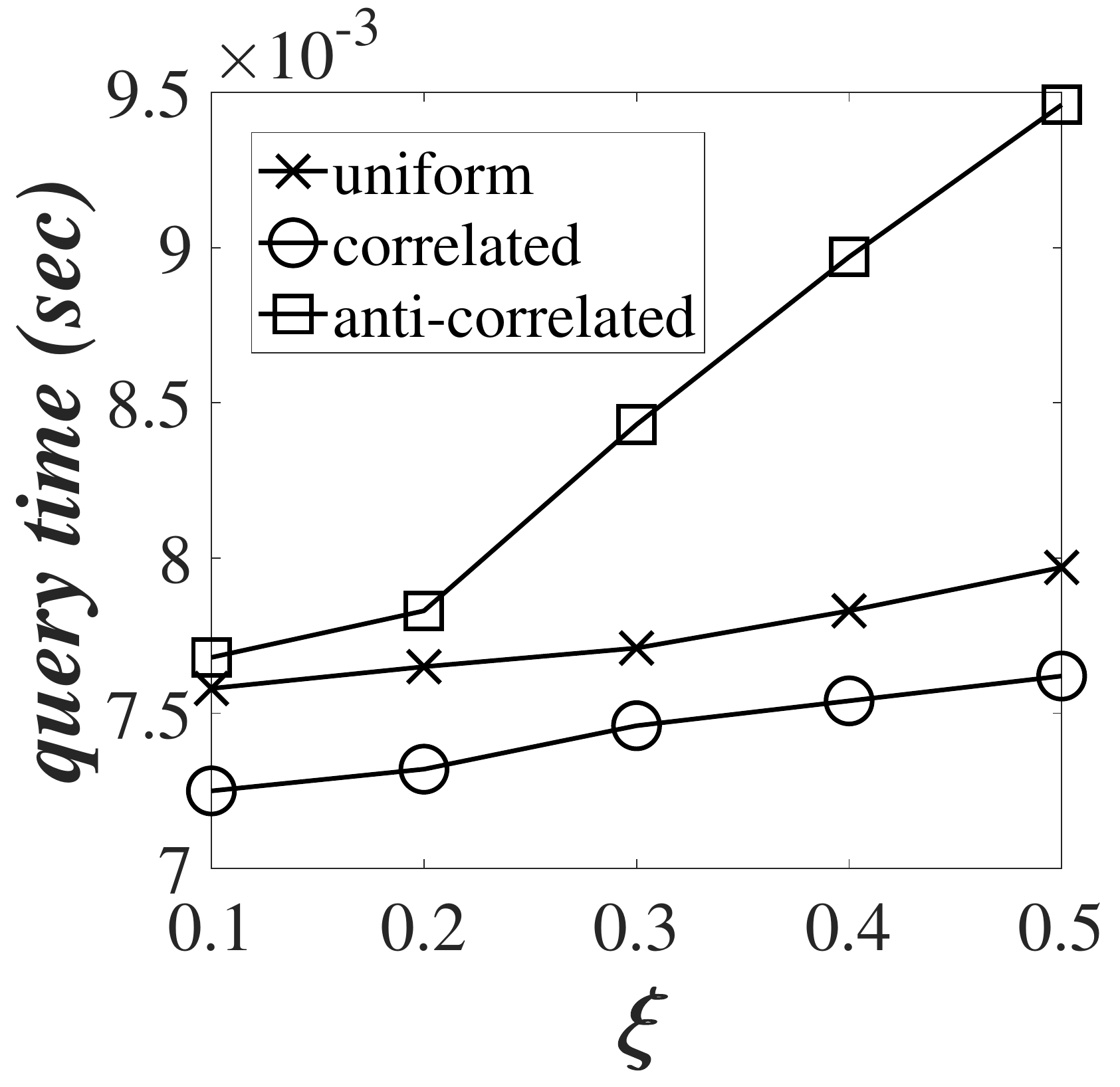}}
\label{subfig:q_cost_vs_xi}
}\vspace{-3ex}
\caption{\small The efficiency vs. missing rate $\xi$.} 
\label{exper:Sky-iDS_vs_xi} 
\end{figure} 

\noindent {\bf The Sky-iDS efficiency vs. probabilistic threshold $\alpha$.} Figure \ref{exper:Sky-iDS_vs_alpha} shows the effect of the skyline probability threshold $\alpha$ on the Sky-iDS performance over three synthetic data, where $\alpha = 0.1, 0.2, 0.5, 0.8,$ and $0.9$ and other parameters are set to default values. 
From figures, the maintenance time is low (less than 0.222 $sec$) and increases linearly for larger $\alpha$ over the three data sets, which shows good performance of our Sky-iDS approach to impute incomplete objects via indexes and incrementally maintain the skyline tree $ST$. Moreover, in Figure \ref{subfig:q_cost_vs_alpha}, when $\alpha$ increases, the query time decreases (due to the lower cost to incrementally refine skyline candidates on the first layer of $ST$) and remains small (i.e., 0.0251$\sim$0.0275 $sec$). Thus, the experimental results confirm the efficiency of our Sky-iDS approach against different $\alpha$ values.



\noindent {\bf The Sky-iDS efficiency vs. dimensionality $d$.} Figure \ref{exper:Sky-iDS_vs_d} reports the performance of our Sky-iDS approach over synthetic data sets, by varying the number, $d$, of attributes in objects from 2 to 10, where other parameters are by default. As shown in Figure \ref{subfig:m_cost_vs_d}, with the increase of dimensionality $d$, the maintenance time increases. This is because, the maintenance time includes the data imputation cost via R$^*$-tree and update time of the $ST$ index. With higher dimensionality $d$, the imputation cost via R$^*$-tree becomes higher, due to the ``dimensionality curse'' problem \cite{Berchtold96}; similarly, the updates of $ST$ need to check the dominance relationships by considering more attributes, which incurs more time cost. Thus, the maintenance cost increases for larger $d$, nevertheless, remains low (i.e., less than 0.137 $sec$).

\nop{
When $d=6$, $Uniform$ data incurs lower maintenance time compared with that when $d=5$, due to the sparseness of data that needs less effort to update the $ST$ index (i.e., new objects are more likely to be on the first layer of $ST$). 
}

Since higher dimensionality $d$ may lead to more skylines, the query cost to refine more candidates on layer 1 of $ST$ is also increasing (as shown in Figure \ref{subfig:q_cost_vs_d}). Nonetheless, for different dimensionality $d$, the query time is small (i.e., less than 0.013 $sec$).



\begin{figure}[t!]
\centering \vspace{-2ex}
\subfigure[][{\small maintenance time}]{\hspace{-2ex}
\scalebox{0.22}[0.22]{\includegraphics{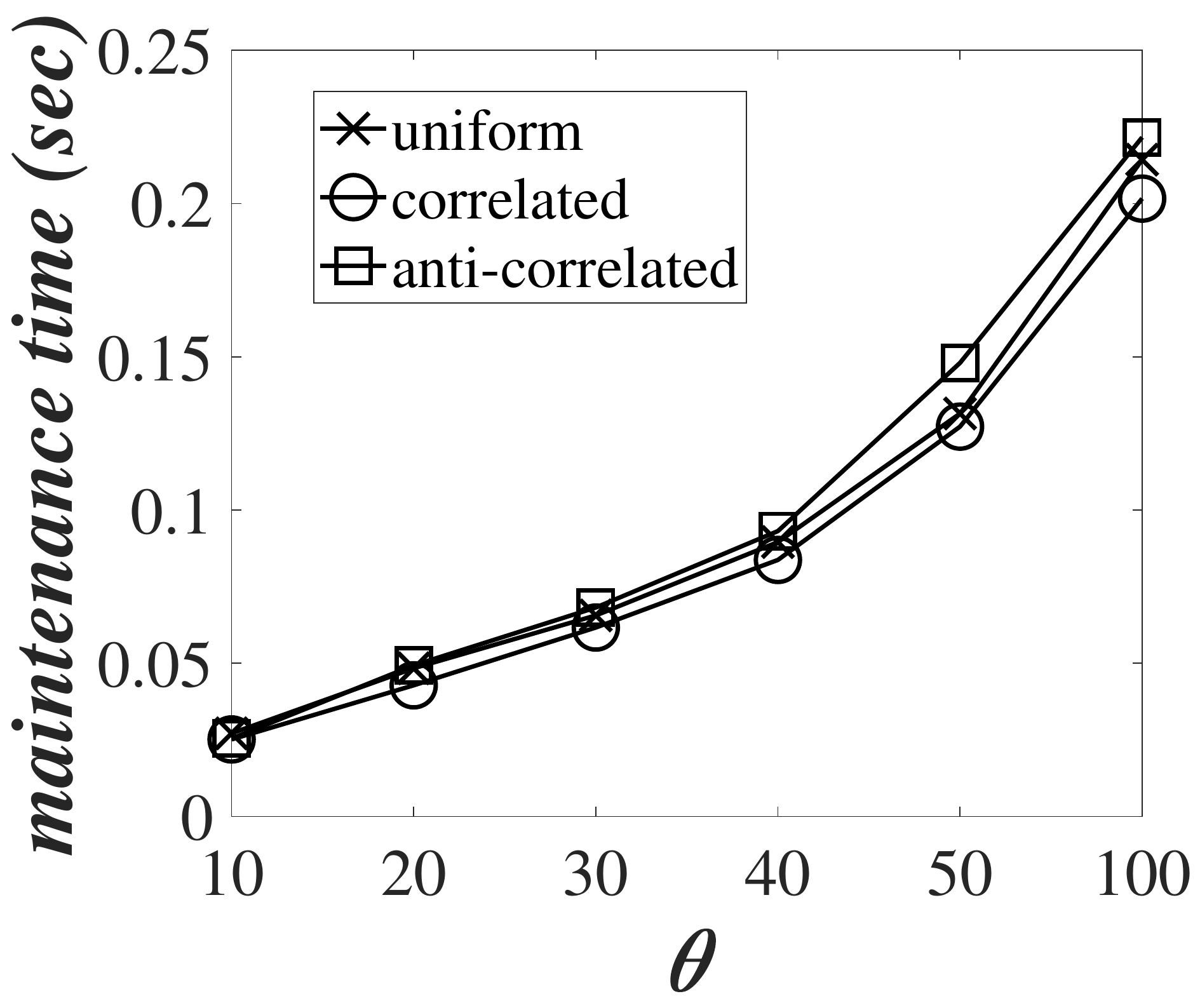}}
\label{subfig:m_cost_vs_theta}
}%
\subfigure[][{\small query time}]{\hspace{-1ex}                  
\scalebox{0.22}[0.22]{\includegraphics{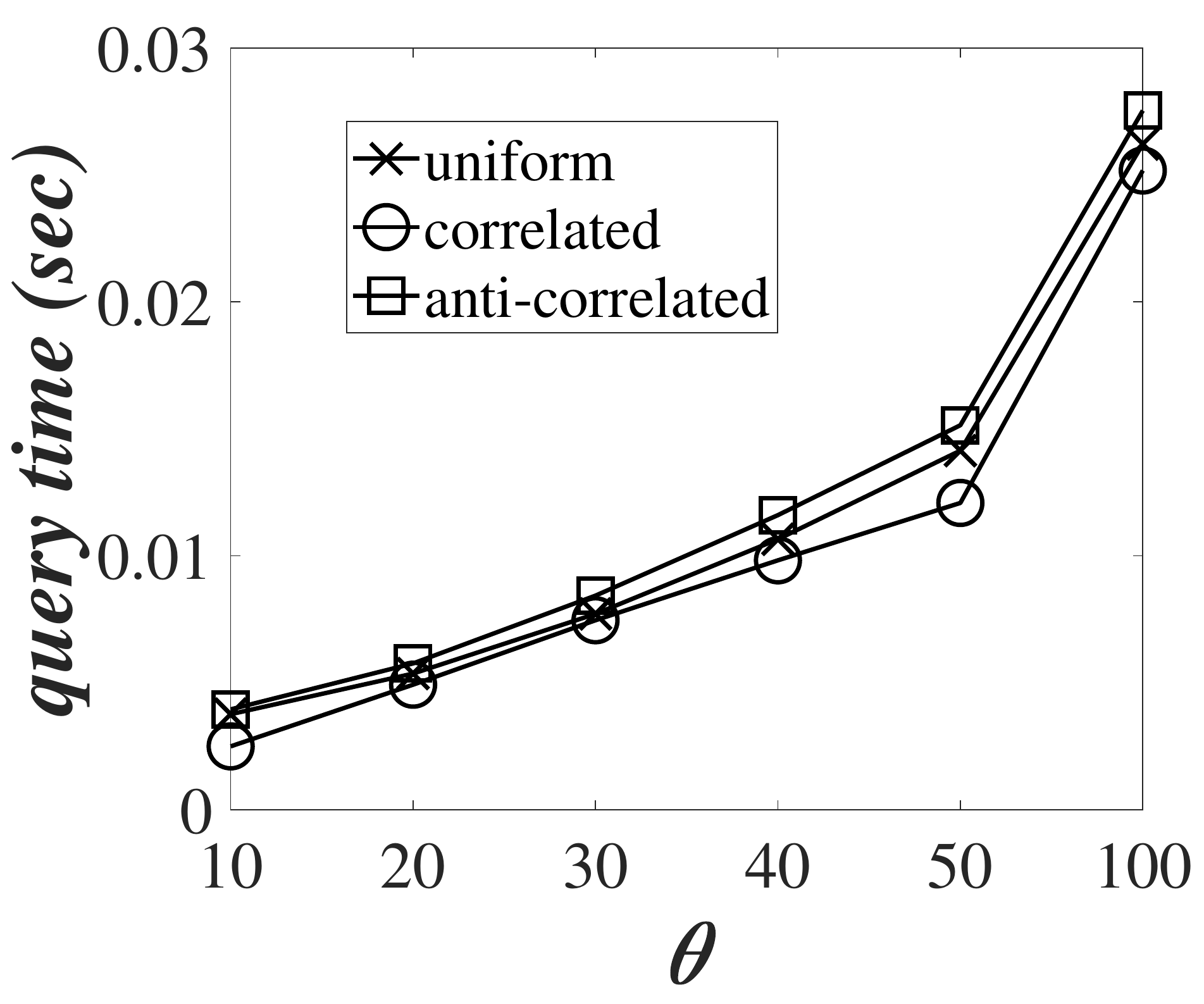}}
\label{subfig:q_cost_vs_theta}
}\vspace{-3ex}
\caption{\small The efficiency vs. the number, $\theta$, of new objects per timestamp in $iDS$.} 
\label{exper:Sky-iDS_vs_theta} \vspace{1ex}
\end{figure} 

\begin{figure}[t!]
\centering\vspace{-5ex}
\subfigure[][{\small maintenance time}]{\hspace{-2ex}
\scalebox{0.22}[0.22]{\includegraphics{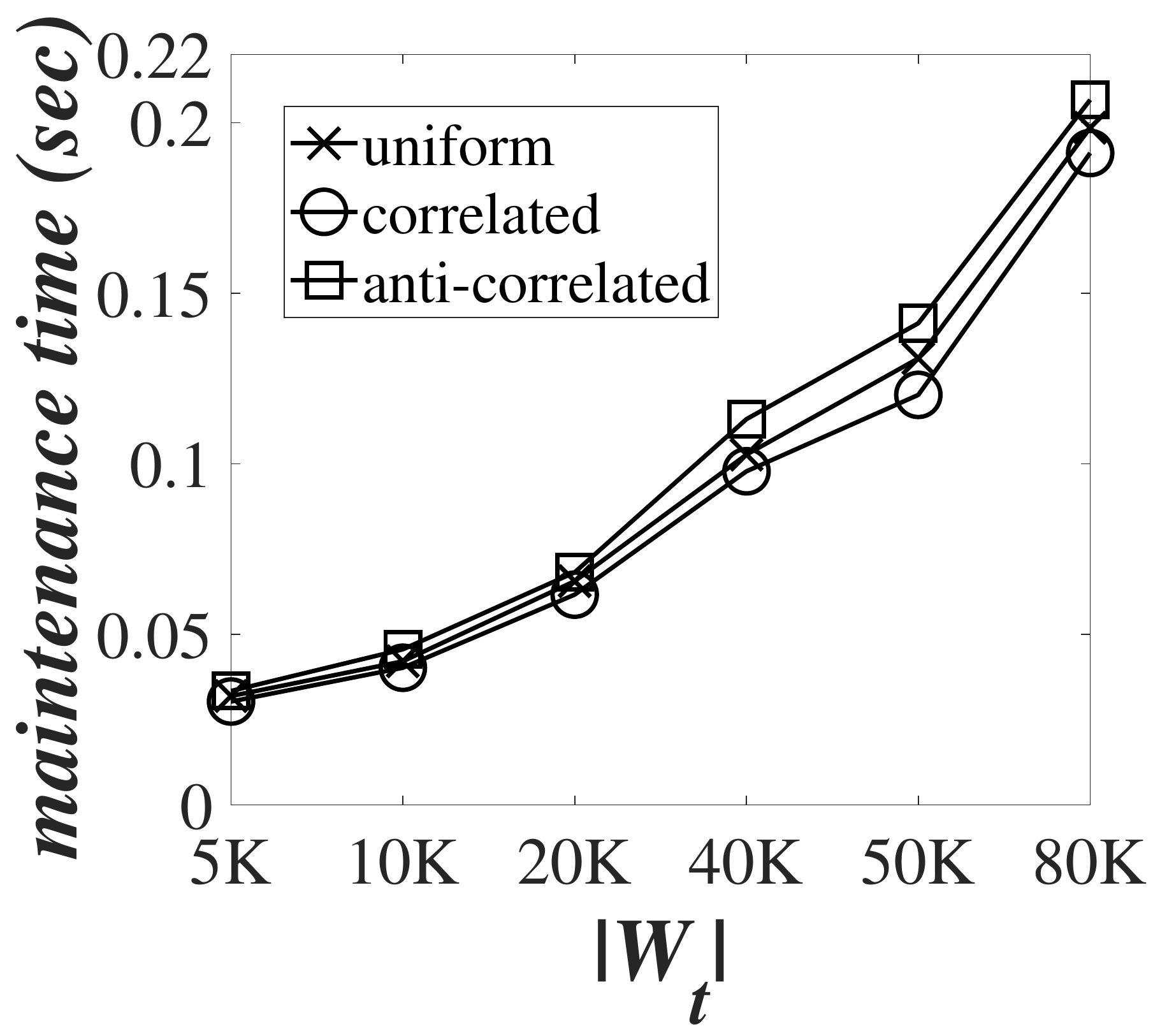}}
\label{subfig:m_cost_vs_Wt}
}%
\subfigure[][{\small query time}]{\hspace{0ex}                  
\scalebox{0.22}[0.22]{\includegraphics{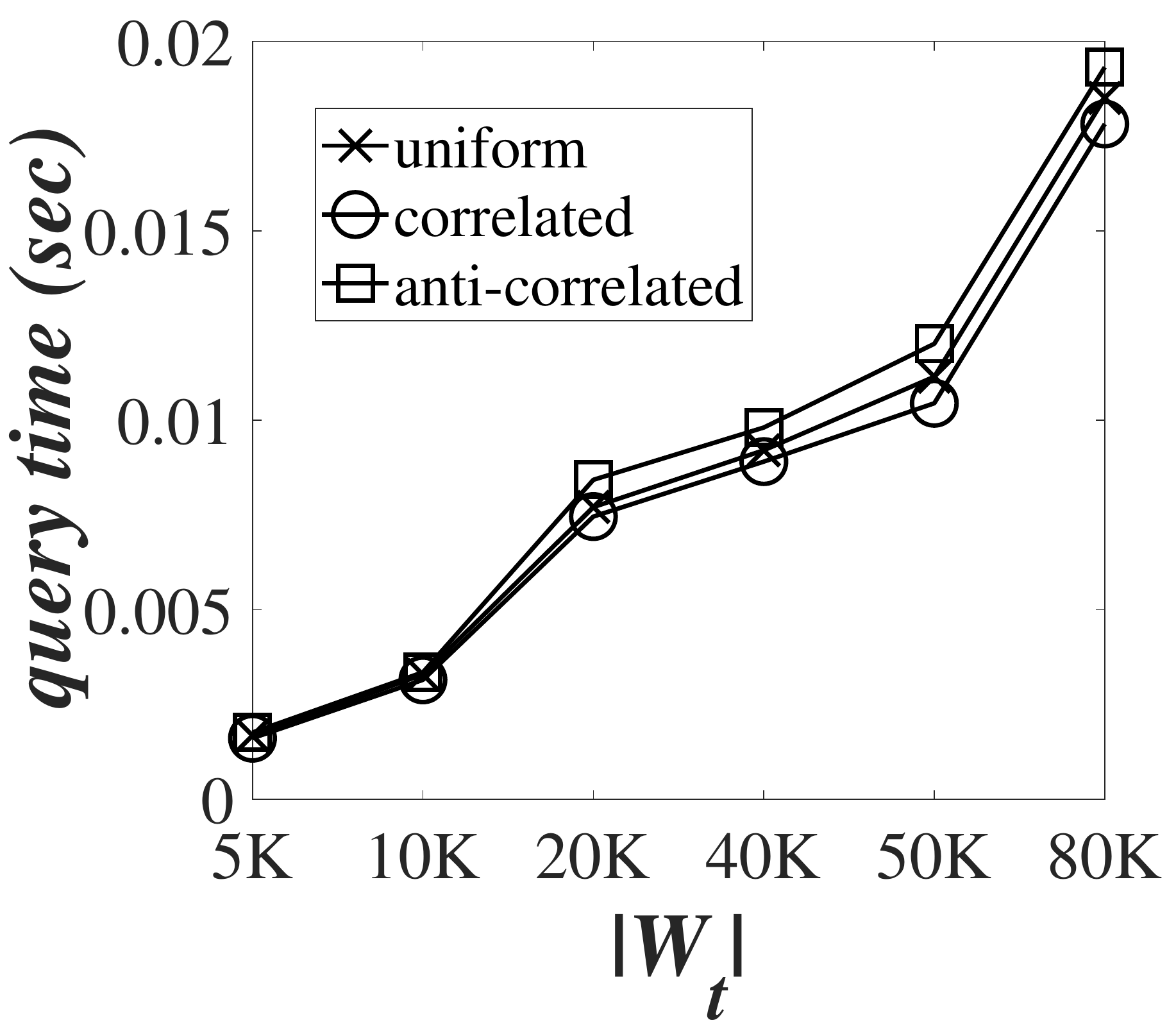}}
\label{subfig:q_cost_vs_Wt}
}\vspace{-3ex}
\caption{\small The efficiency vs. the number, $|W_t|$, of valid objects in $iDS$.} 
\label{exper:Sky-iDS_vs_Wt} 
\end{figure} 

\noindent {\bf The Sky-iDS efficiency vs. the missing rate, $\xi$, of incomplete objects in $iDS$.} Figure \ref{exper:Sky-iDS_vs_xi} evaluates the Sky-iDS performance with different missing rates, $\xi$, of incomplete objects in $iDS$, where $\xi=0.1$, $0.2$, $0.3$, $0.4$, and $0.5$, and default values are used for other parameters. As shown in Figure \ref{subfig:m_cost_vs_xi}, as the increase of $\xi$, the maintenance time increases linearly for all three data sets. This is reasonable, since more incomplete objects will need more imputation cost. Similarly, in Figure \ref{subfig:q_cost_vs_xi}, when $\xi$ increases, the query time also becomes larger for all three data sets. In particular, in Figure \ref{subfig:q_cost_vs_xi}, the $Correlated$ and $Anti\textit{-}correlated$ data sets always need the minimum and maximum query time. This is because, under ``the larger, the better'' semantics, the $Anti\textit{-}correlated$ data usually have more skylines than the $Correlated$ data \cite{Borzsonyi01}. Nevertheless, the time costs for both maintenance and query processing are still low (i.e., less than 0.1 $sec$ and 0.0095 $sec$, respectively).

\noindent {\bf The Sky-iDS efficiency vs. the number, $\theta$, of new objects per timestamp in $iDS$.} Figure \ref{exper:Sky-iDS_vs_theta} varies the number, $\theta$, of newly arriving objects per timestamp from 10 to 100, where default values are used for other parameters. In Figure \ref{subfig:m_cost_vs_theta}, when $\theta$ becomes larger, the maintenance time increases smoothly for all the three data sets. This is because, the skyline tree $ST$ is updated with more new objects per timestamp, which requires more time to impute missing attributes and maintain skyline answers (as discussed in Algorithm \ref{alg:refinement_algorithm} of Section \ref{subsubsec:refinement}). Similarly, in Figure \ref{subfig:q_cost_vs_theta}, the query time also increases with more new objects per timestamp (due to higher refinement cost). Nevertheless, both maintenance and query costs remain low (i.e., 0.201$\sim$0.221 $sec$ for dynamic maintenance and 0.0251$\sim$0.0275 $sec$ for retrieving skyline answers).


\noindent {\bf The Sky-iDS efficiency vs. the number, $|W_t|$, of valid objects in $iDS$.} Figure~\ref{exper:Sky-iDS_vs_Wt} shows the Sky-iDS performance with different numbers, $|W_t|$, of valid objects in stream $iDS$, where $|W_t|$= $5K$, $10K$, $20K$, $40K$, $50K$, and $80K$, and other parameters are set to their default values. For larger $|W_t|$ value, both maintenance and query times increase, but remain low (less than 0.2067 $sec$ and 0.01932 $sec$, respectively, even when $|W_t|=80K$). This is reasonable, with more valid objects in $iDS$, we need more efforts to maintain $ST$ index with the imputed objects and conduct the refinement over more Sky-iDS candidates. 

\begin{figure}[t!]
\centering \vspace{-1ex}
\subfigure[][{\small maintenance time}]{\hspace{-2ex}
\scalebox{0.23}[0.23]{\includegraphics{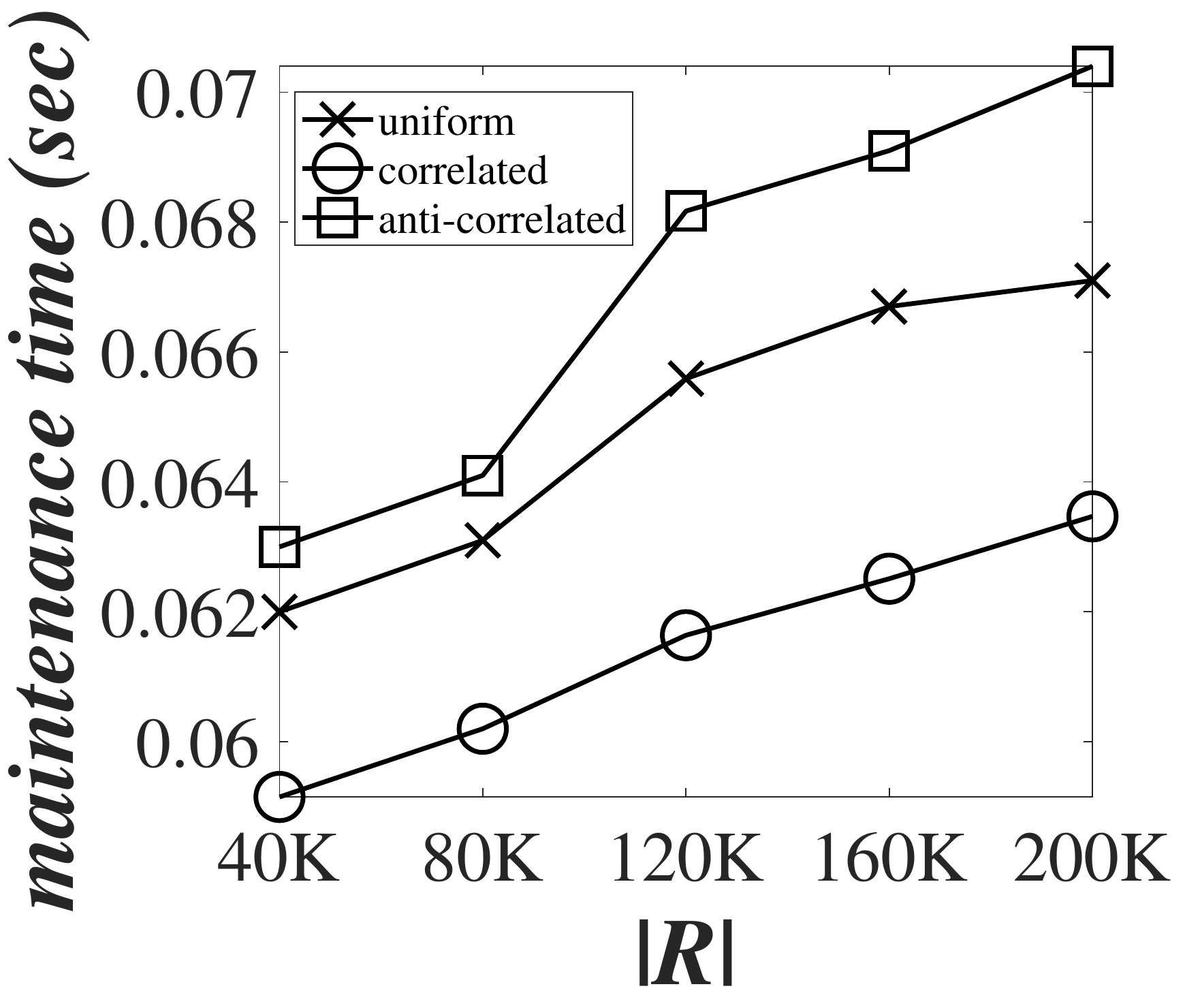}}
\label{subfig:m_cost_vs_R}
}%
\subfigure[][{\small query time}]{\hspace{-1ex}                  
\scalebox{0.24}[0.24]{\includegraphics{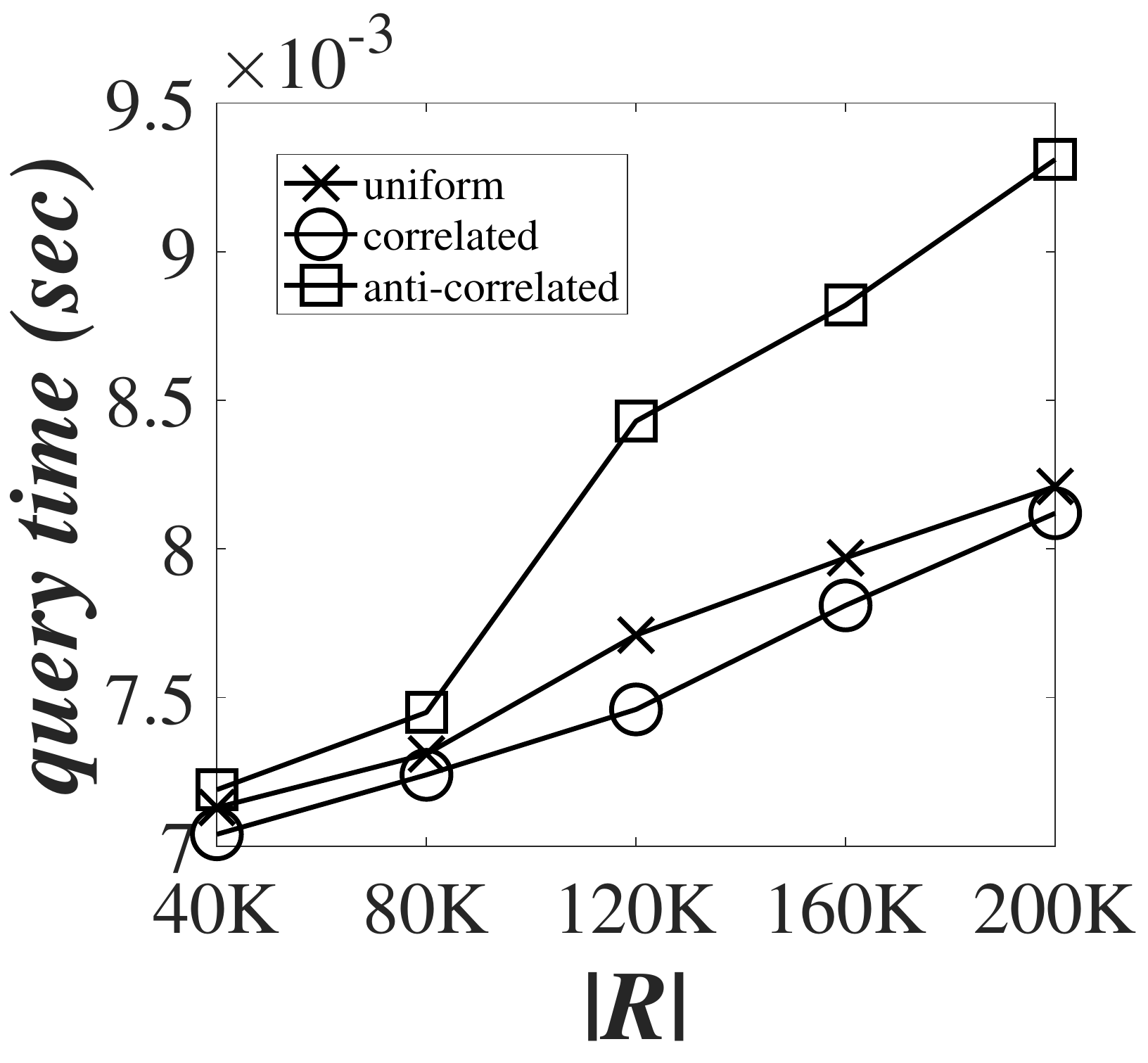}}
\label{subfig:q_cost_vs_R}
}\vspace{-3ex}
\caption{\small The efficiency vs. the size, $|R|$, of data repository.} 
\label{exper:Sky-iDS_vs_R} \vspace{1ex}
\end{figure}

\noindent {\bf The Sky-iDS efficiency vs. the size, $|R|$, of data repository $R$.} Figure~\ref{exper:Sky-iDS_vs_R} illustrates the influence of the size, $|R|$, of data repository on the performance of our Sky-iDS approach. From figures, with larger $|R|$, the maintenance time increases smoothly, since more objects in $R$ are included for data imputation. On the other hand, due to more possible imputed attribute values (resulting from larger $|R|$), the query cost to refine Sky-iDS candidates in $ST$ requires more time cost. Nonetheless, both time costs are low (i.e., around 0.0704 $sec$ for the maintenance, and 0.00931 $sec$ for the query cost, even when $|R| = 200K$). The experimental results indicate the scalability of our Sky-iDS approach against large $|R|$. 


We also did experiments on other parameters (e.g., the number, $m$, of missing attributes, coefficient $\beta$ in Eq.~(\ref{eq:cost_model}), etc.). We do not report similar experimental results here. For interested readers, please refer to Appendix \ref{sec:more_exp}. In summary, our Sky-iDS approach can achieve robust and efficient performance under various parameter settings. 

\nop{
{\color{Weilong}
\subsection{Quality Evaluation of Sky-iDS queries}

}
}

\vspace{-3ex}
\section{Related Work}
\label{sec:related_work}


\nop{
Zhang W, Lin X, Zhang Y, Wang W, Yu JX (2009) Probabilistic skyline operator over sliding windows. In: Proceedings of ICDE

On Efficient k-Skyband Query Processing over Incomplete Data
}

\noindent {\bf Stream processing.} Existing works on data streams studied many query types, including the keyword search \cite{qin2011scalable}, top-$k$ query \cite{choudhury2017monitoring,das2007ad}, join \cite{das2003approximate,hammad2008query}, aggregate queries \cite{dobra2002processing,tatbul2006window}, nearest neighbor queries \cite{koudas2004approximate,bohm2007efficiently}, skyline queries \cite{tao2006maintaining,Ding12,li2014parallelizing}, event detection \cite{zhou2014event}, and so on. These works usually assume that the underlying data (e.g., either certain or uncertain) are complete. Thus, the proposed techniques for complete data streams cannot be directly applied to our Sky-iDS problem over incomplete data stream.



\noindent {\bf Differential dependency} Differential dependency (DD) \cite{song2011differential} is a useful tool for data imputation \cite{song2015enriching}, data cleaning \cite{prokoshyna2015combining,song2016cleaning}, data repairing \cite{hao2017novel,song2014repairing,song2017graph,wang2013efficient,wang2016efficient}, and so on. Song et al. \cite{song2015enriching,song2018enriching} used DD to impute the missing attributes via extensive similarity neighbors with the same determinant attributes. Prokoshyna et al. \cite{prokoshyna2015combining} detected records violating $DD$ rules and cleaned those inconsistent records. Song et al. \cite{song2016cleaning} cleaned the dirty timestamps in data stream based on temporal constraints. Moreover, DD can be also used for constraint-based data repairs over texts \cite{hao2017novel}, events \cite{wang2013efficient,wang2016efficient}, and graphs \cite{song2014repairing}.

Many existing works on imputation methods, such as editing rule \cite{fan2010towards}, multiple imputation \cite{royston2004multiple}, smoothing-based imputation method \cite{keogh2001online}, constraint-based imputation method \cite{zhang2017time}, or regression-based imputation approach \cite{8731351}, usually impute data based on incomplete data themselves only. However, for sparse (incomplete) data sets (i.e., with many missing attributes), it is rather difficult to accurately and unbiasedly impute data attributes. For example, the supervised imputation approaches (e.g., \cite{zhang2017time}) usually require labelled data, which is not trivial how to online obtain the labelled stream data in the streaming environment. Moreover, the rule-based imputation approaches (e.g., editing rule \cite{fan2010towards}) usually requires exact matching, and we may not obtain possible candidates for missing values, especially in sparse data set. In contrast, our DD-based imputation approach utilizes an external source, a complete data repository $R$, for imputing missing attributes from incomplete data stream, which can avoid lacking of (unbiased) samples, tolerate differential differences between attribute values, and does not require any labelled data. Thus, our DD-based imputation approach can achieve unbiased and more accurate data imputation, compared with existing works. 



\noindent {\bf Skyline queries.} The skyline query was proposed by Borzsonyi et al. \cite{Borzsonyi01}. Afterwards, there are many relavant works on skyline and its variants, for example, skyline queries over certain data \cite{Papadias03,chan2006finding,tao2006maintaining,dellis2007efficient,sarkas2008categorical,zhang2009scalable,lee2014toward,awasthi2017k,bousnina2017skyline} and that on uncertain data \cite{pei2007probabilistic,lian2008monochromatic,zhang2009probabilistic,Ding12,liu2015effective}.


In the literature, Mohamed et al. \cite{khalefa2008skyline} re-defined the skyline operator over static incomplete database. In particular, they ignore the missing attributes during the dominance checking between two incomplete objects.
Based on this new skyline definition, Gao et al. \cite{gao2014processing} and Miao et al. \cite{miao2013efficient} further explored a variant of the skyline query, $k$-skyband query, which obtains those objects that are dominated by at most $k$ objects in incomplete data set. However, by neglecting incomplete dimensions, the resulting skylines may be biased (compared with skylines on all attributes). For example, given two objects, $o_1=(2, 4)$ and $o_2=(1, 9)$, with two dimensions, according to \cite{Borzsonyi01}, $o_1$ and $o_2$ cannot dominate each other. In this scenario, if the first dimension of $o_1$ is missing, that is, $o_1=(-, 4)$, based on \cite{khalefa2008skyline}, $o_2$ dominates $o_1$ (by neglecting the first missing attribute for dominance checking; the larger, the better), which may lead to biased skyline result (i.e., $o_1$ is not included).


The previous work \cite{Ding12} directly assumed that objects from data streams are uncertain, thus, skyline queries are directly conducted over uncertain objects. In contrast, we consider skyline queries over incomplete data streams, and turn incomplete objects into complete ones via \textit{differential dependencies} (DDs) \cite{song2011differential} (rather than ignoring missing attribute for dominance checking), which will result in unbiased skylines with high confidences. Most importantly, our work follows the style of ``imputation and query processing at the same time'', which is more challenging than conducting skyline queries directly over uncertain objects, and cannot borrow previous techniques for skyline computations to solve our Sky-iDS problem.

\nop{
the new skyline definition \cite{khalefa2008skyline,gao2014processing,miao2013efficient} may lead to cyclic dominance behavior (e.g., $o_1 \prec o_2$, $o_2 \prec o_3$ and $o_3 \prec o_1$), due to the non-transitive dominance relationship \cite{khalefa2008skyline}.

Moreover, Zhang et al. \cite{zhang2009scalable} builds a skyline partition tree and each object inside divides the skyline candidates into two half parts, which is a little similar to the skyline tree proposed in this paper. But instead of storing the explicit current skyline answers on the first hierarchy \cite{gao2014processing,miao2013efficient} or building partition tree as a search tree \cite{zhang2009scalable}, in our work, we build skyline tree especially for stream environment and put the candidates (superset of current skyline answers) on the first layer via light calculation, 
and then compute current skyline answers by refinement algorithm. 

}

\noindent{\bf Stream Outlier Detection and Repair.} Existing works on stream outlier detection and repair can be classified into two categories, smoothing-based \cite{keogh2001online} and constraint-based \cite{song2015screen,zhang2016sequential,zhang2017time} approaches. Without distinguishing normal data and outlier, \cite{keogh2001online} modified almost all data values, which may not be the best way to clean (repair) the outlier. To overcome this drawback, Song et al. \cite{song2015screen} proposed an approach to detect the outlier values within a sliding window, and then updated the outlier values based on a speed constraint $s$ with minimum and maximum speed changes $s_{min}$ and $s_{max})$, respectively. Zhang et al. \cite{zhang2016sequential} refined this speed-constraint approach by detecting and modifying smaller errors by narrowing the speed intervals $s$ via probability distributions of speeds and speed changes. However, \cite{song2015screen,zhang2016sequential} cannot repair outliers for data sets with consecutive errors between any two sequential data records. To solve this problem, Zhang et al. \cite{zhang2017time} proposed a supervised approach based on some labelled data on data stream. Note that, the constraint-based approaches \cite{song2015screen,zhang2016sequential,zhang2017time} detected outliers with speed change beyond the acceptable speed constraint $s$, which have different semantics from the skyline operator in this paper (i.e., skylines are records with maximum values on at least one attributes among all data within a sliding window). Nevertheless, in our experiments, we implemented a baseline method based on \cite{zhang2016sequential} and compared our imputation method with \cite{zhang2016sequential}. Specifically, \cite{zhang2017time} can not be used as the imputation method for our Sky-iDS problem, since it is not trivial how to online obtain the labelled stream data in the streaming environment.

Since these works \cite{keogh2001online,song2015screen,zhang2016sequential,zhang2017time} focus on detecting the outlier values with the high (abnormal) change rates (speeds) w.r.t. the near normal values, they cannot be applied to solve our Sky-iDS problem, which retrieves data objects not dominated by other objects in a sliding window.

\noindent {\bf Incomplete data management.} There are some previous works on incomplete data management, for example, how to model incomplete data \cite{antova2007complete,libkin2011incomplete}, how to index incomplete data \cite{ooi1998fast}, and so on. Miao et al. \cite{miao2017incomplete} did a comprehensive survey about incomplete data management.
In order to obtain complete data, some studies imputed the missing attributes by applying rule-based (exact matching over all dimensions) \cite{fan2010towards}, statistical-based (exact matching over partial dimensions)  \cite{mayfield2010eracer}, filter-based \cite{vijayakumar2007prediction}, pattern-based \cite{bohlen2017continuous}, or analysis-based \cite{royston2004multiple} imputation methods. For example, \cite{bohlen2017continuous} imputed the missing attributes in streams by finding the $k$ most similar patterns from $l$ time series. However, if the same attributes from $l$ time series are all missing, then this method cannot accomplish the imputation.  \cite{royston2004multiple} is to create multiple complete (imputed) versions of data sets and combine all these versions to impute the missing attributes. However, these generated data versions may introduce many erroneous imputed values, which may not be able to provide a stable imputation result. For \cite{fan2010towards,mayfield2010eracer,vijayakumar2007prediction}, although they can achieve explicit imputation results, they may not successfully impute the missing data, due to the sparseness of data sets \cite{song2011differential}. In contrast, in this paper, we use DDs \cite{song2011differential} and a complete data repository $R$ to impute the missing attributes.


To our best knowledge, no prior works studied the problem of conducting data imputation (via DDs) and skyline query answering, at the same time, on incomplete data in the streaming environment.

\section{Conclusions}
\label{sec:conclusions}
\vspace{-2ex}
In this paper, we study an important problem, Sky-iDS, of monitoring the skylines over incomplete data stream, which is useful in many real-world applications such as sensory data monitoring. In order to efficiently impute the missing attributes and conduct Sky-iDS queries, we propose effective data synopses and \textit{skyline tree} ($ST$) indexes to facilitate the data imputation via \textit{differential dependency} (DD) rules and skyline computations, respectively, at the same time. We also design effective pruning strategies to greatly reduce the Sky-iDS search space over the stream, and propose efficient Sky-iDS algorithms to perform ``imputation and query processing at the same time'' over incomplete data stream. Extensive experiments have demonstrated the efficiency and effectiveness of our proposed Sky-iDS processing approaches on both real and synthetic data sets under different parameter settings.

\nop{
\ifCLASSOPTIONcompsoc
  \section*{Acknowledgments}
\else
\fi
\vspace{-2ex}Xiang Lian is supported by NSF OAC No. 1739491 and Lian Startup No. 220981, Kent State University. We thank the anonymous reviewers for the useful suggestions.
}

\section*{Acknowledgments}
\vspace{-2ex}Xiang Lian is supported by NSF OAC No. 1739491 and Lian Startup No. 220981, Kent State University. We thank the anonymous reviewers for the useful suggestions.




\bibliographystyle{plain}
\let\xxx=\bibitem\def\bibitem{\par\vspace{-0.0mm}\xxx}
\bibliography{vldb_sample}  

\begin{thebibliography}{10}

\bibitem{aberer2007infrastructure}
K.~Aberer, M.~Hauswirth, and A.~Salehi.
\newblock Infrastructure for data processing in large-scale interconnected
  sensor networks.
\newblock In {\em MDM}, 2007.

\bibitem{antova2007complete}
L.~Antova, C.~Koch, and D.~Olteanu.
\newblock From complete to incomplete information and back.
\newblock In {\em SIGMOD}, 2007.

\bibitem{awasthi2017k}
A.~Awasthi, A.~Bhattacharya, S.~Gupta, and U.~Singh.
\newblock K-dominant skyline join queries: Extending the join paradigm to
  k-dominant skylines.
\newblock In {\em ICDE}, 2017.

\bibitem{beckmann1990r}
N.~Beckmann, H.~Kriegel, R.~Schneider, and B.~Seeger.
\newblock The {R}*-tree: an efficient and robust access method for points and
  rectangles.
\newblock In {\em SIGMOD}, 1990.

\bibitem{belussi1998self}
A.~Belussi and C.~Faloutsos.
\newblock Self-spacial join selectivity estimation using fractal concepts.
\newblock {\em TOIS}, 1998.

\bibitem{Berchtold96}
S.~Berchtold, D.~Keim, and H.~Kriegel.
\newblock The x-tree: An index structure for high-dimensional data.
\newblock In {\em VLDB}, 1996.

\bibitem{bohm2007efficiently}
C.~Bohm, B.C. Ooi, C.~Plant, and Y.~Yan.
\newblock Efficiently processing continuous k-nn queries on data streams.
\newblock In {\em ICDE}, 2007.

\bibitem{Borzsonyi01}
S.~Borzsony, D.~Kossmann, and K.~Stocker.
\newblock The skyline operator.
\newblock In {\em ICDE}, 2001.

\bibitem{bousnina2017skyline}
F.~Bousnina, S.~Elmi, M.~Chebbah, M.~Tobji, A.~HadjAli, and B.~Yaghlane.
\newblock Skyline operator over tripadvisor reviews within the belief functions
  framework.
\newblock In {\em ICDE}, 2017.

\bibitem{chan2006finding}
C.~Chan, HV~Jagadish, K.~Tan, A.~Tung, and Z.~Zhang.
\newblock Finding k-dominant skylines in high dimensional space.
\newblock In {\em SIGMOD}, 2006.

\bibitem{choudhury2017monitoring}
FM~Choudhury, Z.~Bao, JS~Culpepper, and T.~Sellis.
\newblock Monitoring the top-m rank aggregation of spatial objects in streaming
  queries.
\newblock In {\em ICDE}, 2017.

\bibitem{cranor2003gigascope}
C.~Cranor, T.~Johnson, O.~Spataschek, and V.~Shkapenyuk.
\newblock Gigascope: a stream database for network applications.
\newblock In {\em SIGMOD}, 2003.

\bibitem{Dalvi07}
N.~Dalvi and D.~Suciu.
\newblock Efficient query evaluation on probabilistic databases.
\newblock {\em VLDB}, 2007.

\bibitem{das2003approximate}
A.~Das, J.~Gehrke, and M.~Riedewald.
\newblock Approximate join processing over data streams.
\newblock In {\em SIGMOD}, 2003.

\bibitem{das2007ad}
G.~Das, D.~Gunopulos, N.~Koudas, and N.~Sarkas.
\newblock Ad-hoc top-k query answering for data streams.
\newblock In {\em VLDB}, 2007.

\bibitem{das2009randomized}
A.~Das~Sarma, A.~Lall, D.~Nanongkai, and J.~Xu.
\newblock Randomized multi-pass streaming skyline algorithms.
\newblock {\em VLDB}, 2009.

\bibitem{dellis2007efficient}
E.~Dellis and B.~Seeger.
\newblock Efficient computation of reverse skyline queries.
\newblock In {\em VLDB}, 2007.

\bibitem{dhanabal2015study}
L~Dhanabal and SP~Shantharajah.
\newblock A study on nsl-kdd dataset for intrusion detection system based on
  classification algorithms.
\newblock {\em IJARCCE}, 2015.

\bibitem{Ding12}
X.~Ding, X.~Lian, L.~Chen, and H.~Jin.
\newblock Continuous monitoring of skylines over uncertain data streams.
\newblock {\em Inf. Sci.}, 2012.

\bibitem{dobra2002processing}
A.~Dobra, M.~Garofalakis, J.~Gehrke, and R.~Rastogi.
\newblock Processing complex aggregate queries over data streams.
\newblock In {\em SIGMOD}, 2002.

\bibitem{fan2010towards}
W.~Fan, J.~Li, S.~Ma, N.~Tang, and W.~Yu.
\newblock Towards certain fixes with editing rules and master data.
\newblock {\em VLDB}, 2010.

\bibitem{gao2014processing}
Y.~Gao, X.~Miao, H.~Cui, G.~Chen, and Q.~Li.
\newblock Processing k-skyband, constrained skyline, and group-by skyline
  queries on incomplete data.
\newblock {\em EXPERT SYST APPL}, 2014.

\bibitem{golab2003issues}
L.~Golab and T.~{\"O}zsu.
\newblock Issues in data stream management.
\newblock {\em ACM SIGMOD Record}, 2003.

\bibitem{hammad2008query}
MA~Hammad, WG~Aref, and AK~Elmagarmid.
\newblock Query processing of multi-way stream window joins.
\newblock {\em VLDB}, 2008.

\bibitem{hao2017novel}
S.~Hao, N.~Tang, G.~Li, J.~He, N.~Ta, and J.~Feng.
\newblock A novel cost-based model for data repairing.
\newblock In {\em ICDE}. IEEE, 2017.

\bibitem{igbe2016distributed}
O.~Igbe, I.~Darwish, and T.~Saadawi.
\newblock Distributed network intrusion detection systems: An artificial immune
  system approach.
\newblock In {\em CHASE}. IEEE, 2016.

\bibitem{keogh2001online}
E.~Keogh, S.~Chu, D.~Hart, and M.~Pazzani.
\newblock An online algorithm for segmenting time series.
\newblock In {\em ICDE}, 2001.

\bibitem{khalefa2008skyline}
M.~Khalefa, M.~Mokbel, and J.~Levandoski.
\newblock Skyline query processing for incomplete data.
\newblock In {\em ICDE}, 2008.

\bibitem{koudas2004approximate}
N.~Koudas, B.C. Ooi, K.~Tan, and R.~Zhang.
\newblock Approximate nn queries on streams with guaranteed error/performance
  bounds.
\newblock In {\em VLDB}, 2004.

\bibitem{lee2014toward}
J.~Lee and S.~Hwang.
\newblock Toward efficient multidimensional subspace skyline computation.
\newblock {\em VLDB}, 2014.

\bibitem{li2014parallelizing}
X.~Li, Y.~Wang, X.~Li, and Y.~Wang.
\newblock Parallelizing skyline qeries over ncertain data streams with sliding
  window partitioning and grid index.
\newblock {\em KAIS}, 2014.

\bibitem{lian2008monochromatic}
X.~Lian and L.~Chen.
\newblock Monochromatic and bichromatic reverse skyline search over uncertain
  databases.
\newblock In {\em SIGMOD}, 2008.

\bibitem{lian2010generic}
X.~Lian and L.~Chen.
\newblock A generic framework for handling uncertain data with local
  correlations.
\newblock {\em VLDB}, 2010.

\bibitem{libkin2011incomplete}
L.~Libkin.
\newblock Incomplete information and certain answers in general data models.
\newblock In {\em PODS}, 2011.

\bibitem{lin2005stabbing}
X.~Lin, Y.~Yuan, W.~Wang, and H.~Lu.
\newblock Stabbing the sky: Efficient skyline computation over sliding windows.
\newblock In {\em ICDE}, 2005.

\bibitem{liu2015effective}
M.~Liu and S.~Tang.
\newblock An effective probabilistic skyline query process on uncertain data
  streams.
\newblock {\em EUSPN/ICTH}, 2015.

\bibitem{mayfield2010eracer}
C.~Mayfield, J.~Neville, and S.~Prabhakar.
\newblock Eracer: a database approach for statistical inference and data
  cleaning.
\newblock In {\em SIGMOD}, 2010.

\bibitem{miao2013efficient}
X.~Miao, Y.~Gao, L.~Chen, G.~Chen, Q.~Li, and T.~Jiang.
\newblock On efficient $k$-skyband query processing over incomplete data.
\newblock In {\em DASFAA}, 2013.

\bibitem{miao2017incomplete}
X.~Miao, Y.~Gao, S.~Guo, and W.~Liu.
\newblock Incomplete data management: a survey.
\newblock {\em Frontiers of Computer Science}, 2017.

\bibitem{ooi1998fast}
B.C. Ooi, C.H. Goh, and K.~Tan.
\newblock Fast high-dimensional data search in incomplete databases.
\newblock In {\em VLDB}, 1998.

\bibitem{Papadias03}
D.~Papadias, Y.~Tao, G.~Fu, and B.~Seeger.
\newblock An optimal and progressive algorithm for skyline queries.
\newblock In {\em SIGMOD}, 2003.

\bibitem{pei2007probabilistic}
J.~Pei, B.~Jiang, X.~Lin, and Y.~Yuan.
\newblock Probabilistic skylines on uncertain data.
\newblock In {\em VLDB}, 2007.

\bibitem{prokoshyna2015combining}
N.~Prokoshyna, J.~Szlichta, F.~Chiang, RJ~Miller, and D.~Srivastava.
\newblock Combining quantitative and logical data cleaning.
\newblock {\em PVLDB}, 2015.

\bibitem{qin2011scalable}
L.~Qin, J.X. Yu, and L.~Chang.
\newblock Scalable keyword search on large data streams.
\newblock {\em VLDB}, 2011.

\bibitem{royston2004multiple}
P.~Royston.
\newblock Multiple imputation of missing values.
\newblock {\em The Stata Journal}, 2004.

\bibitem{sarkas2008categorical}
N.~Sarkas, G.~Das, N.~Koudas, and A.~Tung.
\newblock Categorical skylines for streaming data.
\newblock In {\em SIGMOD}, 2008.

\bibitem{song2016cleaning}
S.~Song, Y.~Cao, and J.~Wang.
\newblock Cleaning timestamps with temporal constraints.
\newblock {\em PVLDB}, 2016.

\bibitem{song2011differential}
S.~Song. and L.~Chen.
\newblock Differential dependencies: Reasoning and discovery.
\newblock {\em TODS}, 2011.

\bibitem{song2014repairing}
S.~Song, H.~Cheng, J.X. Yu, and L.~Chen.
\newblock Repairing vertex labels under neighborhood constraints.
\newblock {\em PVLDB}, 2014.

\bibitem{song2017graph}
S.~Song, B.~Liu, H.~Cheng, J.X. Yu, and L.~Chen.
\newblock Graph repairing under neighborhood constraints.
\newblock {\em VLDBJ}, 2017.

\bibitem{song2018enriching}
S.~Song, Y.~Sun, A.~Zhang, L.~Chen, and J.~Wang.
\newblock Enriching data imputation under similarity rule constraints.
\newblock {\em TKDE}, 2018.

\bibitem{song2015enriching}
S.~Song, A.~Zhang, L.~Chen, and J.~Wang.
\newblock Enriching data imputation with extensive similarity neighbors.
\newblock {\em VLDB}, 2015.

\bibitem{song2015screen}
S.~Song, A.~Zhang, J.~Wang, and PS~Yu.
\newblock Screen: Stream data cleaning under speed constraints.
\newblock In {\em SIGMOD}, 2015.

\bibitem{srivastava2000web}
J.~Srivastava, R.~Cooley, M.~Deshpande, and P.~Tan.
\newblock Web usage mining: Discovery and applications of usage patterns from
  web data.
\newblock {\em SIGKDD}, 2000.

\bibitem{tao2006maintaining}
Y.~Tao and D.~Papadias.
\newblock Maintaining sliding window skylines on data streams.
\newblock {\em TKDE}, 2006.

\bibitem{tatbul2006window}
N.~Tatbul and S.~Zdonik.
\newblock Window-aware load shedding for aggregation queries over data streams.
\newblock In {\em VLDB}, 2006.

\bibitem{van2007multiple}
S.~Van~Buuren.
\newblock Multiple imputation of discrete and continuous data by fully
  conditional specification.
\newblock {\em Statistical methods in medical research}, 2007.

\bibitem{vijayakumar2007prediction}
N.~Vijayakumar and B.~Plale.
\newblock Prediction of missing events in sensor data streams using kalman
  filters.
\newblock In {\em sensorKDD}, 2007.

\bibitem{wang2013efficient}
J.~Wang, S.~Song, X.~Zhu, and X.~Lin.
\newblock Efficient recovery of missing events.
\newblock {\em PVLDB}, 2013.

\bibitem{wang2016efficient}
J.~Wang, S.~Song, X.~Zhu, X.~Lin, and J.~Sun.
\newblock Efficient recovery of missing events.
\newblock {\em TKDE}, 2016.

\bibitem{bohlen2017continuous}
K.~Wellenzohn, MH~B{\"o}hlen, A.~Dign{\"o}s, J.~Gamper, and H.~Mitterer.
\newblock Continuous imputation of missing values in streams of
  pattern-determining time series.
\newblock 2017.

\bibitem{Xue06}
W.~Xue, Q.~Luo, L.~Chen, and Y.~Liu.
\newblock Contour map matching for event detection in sensor networks.
\newblock In {\em SIGMOD}, 2006.

\bibitem{8731351}
A.~{Zhang}, S.~{Song}, Y.~{Sun}, and J.~{Wang}.
\newblock Learning individual models for imputation.
\newblock In {\em ICDE}, 2019.

\bibitem{zhang2016sequential}
A.~Zhang, S.~Song, and J.~Wang.
\newblock Sequential data cleaning: a statistical approach.
\newblock In {\em SIGMOD}, 2016.

\bibitem{zhang2017time}
A.~Zhang, S.~Song, J.~Wang, and PS~Yu.
\newblock Time series data cleaning: From anomaly detection to anomaly
  repairing.
\newblock {\em VLDB}, 2017.

\bibitem{zhang2009scalable}
S.~Zhang, N.~Mamoulis, and D.~Cheung.
\newblock Scalable skyline computation using object-based space partitioning.
\newblock In {\em SIGMOD}, 2009.

\bibitem{zhang2009probabilistic}
W.~Zhang, X.~Lin, Y.~Zhang, W.~Wang, and J.X. Yu.
\newblock Probabilistic skyline operator over sliding windows.
\newblock In {\em ICDE}, 2009.

\bibitem{zhou2014event}
X.~Zhou and L.~Chen.
\newblock Event detection over twitter social media streams.
\newblock {\em VLDB}, 2014.

\end{thebibliography}




\nop{
\section{Derivation of Eq.~(\ref{eq:eq8})}
\label{section:A}
\begin{qnarray}
&\overline{nb}(\epsilon_{DD},shape) \\
&=((Vol(\epsilon_{DD}),shape)/(Vol(\epsilon_{DD}),hypercube))^{\frac{D_2}{d}} \\
&\times (N-1) \times 2^{D_2} \times (\epsilon_{DD})^{D_2} \\
&=(N-1) \times 2^{D_2} \times (\frac{l_{DD}.z}{2})^{D_2} \times \left[ \prod_{x=1}^d \frac{l_x.z}{l_{DD}.z} \right]^{\frac{D_2}{d}} \\
&=(N-1) \times (\prod_{x=1}^d l_x.z)^{\frac{D_2}{d}}
\end{qnarray}
}

\noindent {\bf \large Appendix}
\vspace{-2ex}
\section{Proofs of Lemmas for Pruning Strategies}
\label{sec:proof_lemmas_pruning_strategies}
\vspace{-2ex}
\subsection{Proof of Lemma~\ref{lemma:lem1}}
\label{subsec:proof_lemmas_pruning_strategies1}

\vspace{-2ex}
\textit{Proof: }
As shown in Figure \ref{subfig:pruningRule_1}, since $o'^p.min$ is the minimum corner of the imputed object $o'^p$, it holds that imputed samples of $o'^p$ is dominating $o'^p.min$, that is, $o'^p \preccurlyeq o'^p.min$. Similarly, we also have $o_i^p.max \preccurlyeq o_i^p$. Due to the lemma assumption that $o'^p.min$ $\prec o_i^p.max$, by dominance transition, we can derive $o'^p \preccurlyeq o'^p.min \prec o_i^p.max \preccurlyeq o_i^p$. Thus, we have $Pr\{o'^p \prec o_i^p\} = 1$ (or $Pr\{o'^p \prec o_{il}\} = 1$ for any instance $o_{il} \in o_i^p$). According to Eq.~(\ref{eq:eq14}), it holds that $P_{Sky\text{-}iDS}(o_i^p) = 0$. Moreover, since $o'.exp \ge o_i.exp$ holds (i.e., object $o'$ expires after $o_i^p$ from the lemma assumption), it indicates that $o_i^p$ can never be the skyline due to the existence of object $o'^p$. Hence, object $o_i^p\in iDS$ can be safely pruned, which completes the proof.
$\hfill\square$

\vspace{-4ex}
\subsection{Proof of Lemma~\ref{lemma:lem2}}
\label{subsec:proof_lemmas_pruning_strategies2}

\vspace{-1ex}
\textit{Proof: }
From Eq.~(\ref{eq:eq14}), we can derive a probability upper bound as follows.

\begin{scriptsize}
\begin{eqnarray}
P_{Sky\text{-}iDS}(o_i^p) &\leq& \sum_{\forall o_{il} \in o_i^p} o_{il}.p \cdot (1- Pr\{o'^p \prec o_{il}\})\notag\\
&=& 1- \sum_{\forall o_{il} \in o_i^p} o_{il}.p \cdot Pr\{o'^p \prec o_{il}\}.\label{eq:max_corner_prob}
\end{eqnarray}
\end{scriptsize}
Since $o_i^p.max \preccurlyeq o_{il}$ ($o_{il}\in o_i^p$) and $Pr\{o'^p \prec o_i^p.max\} \geq 1-\alpha$ hold, we have $Pr\{o'^p \prec o_{il}\} \geq Pr\{o'^p \prec o_i^p.max\} \geq 1-\alpha$. By substituting this probability into Eq.~(\ref{eq:max_corner_prob}), we can obtain: $P_{Sky\text{-}iDS}(o_i^p) \leq 1- \sum_{\forall o_{il} \in o_i^p} o_{il}.p \cdot (1-\alpha) = \alpha$. Moreover, since $o'.exp \ge o_i.exp$ holds,  $o_i^p$ always has the skyline probability less than $\alpha$ during its life time, due to the existence of object $o'$. Thus, object $o_i$ can be safely pruned. 
$\hfill\square$

\vspace{-4ex}
\subsection{Proof of Lemma~\ref{lemma:lem3}}
\label{subsec:proof_lemmas_pruning_strategies3}

\vspace{-2ex}
\textit{Proof: }
Similar to the proof of Lemma \ref{lemma:lem2}, since $o'^p \preccurlyeq o'^p.min$ and $Pr\{o'^p.min \prec o_i^p\} \ge 1-\alpha$ hold, we have $Pr\{o'^p \prec o_i^p\} \ge Pr\{o'^p.min \prec o_i^p\} \ge 1-\alpha$.  By substituting this probability into Eq.~(\ref{eq:max_corner_prob}), we can obtain: $P_{Sky\text{-}iDS}(o_i^p) \leq 1- Pr\{o'^p \prec o_i^p\} = \alpha$. Thus, since object $o_i$ expires before object $o'$ (i.e., $o'.exp \ge o_i.exp$), object $o_i$ always has the skyline probability lower than $\alpha$ during its life time. Hence, object $o_i$ can be safely pruned. 
$\hfill\square$


\section{Proofs of Properties for Skyline Tree ST}
\label{subsec:proof_propeties_1_3}
\vspace{-2ex}
\subsection{Proof of Property 1 of ST}
\vspace{-2ex}
\textit{Proof: }
We can prove this property by showing that no such an imputed object $o_i^p$ exists, where $o_i^p$ is a valid object not within skyline tree $ST$ but is actually a skyline or may become a skyline later.

First, assume that the object $o_i^p$ is a current skyline. According to Definition \ref{def:sky-iDS}, we can obtain $P_{Sky\text{-}iDS}(o_i^p)>\alpha$. By substituting this probability into Eq.~(\ref{eq:max_corner_prob}), we have $\sum_{\forall o_{il} \in o_i^p}$ $o_{il}.p \cdot Pr\{n^p \prec o_{il}\} < 1-\alpha$, that is, $Pr\{n^p \prec o_i^p\}<1-\alpha$. Thus, no object $t^p$ in $ST$ dominates $o_i^p$ with probability not smaller than $(1-\alpha)$, and then object $o_i^p$ should be on the first layer of $ST$.

Second, assume that the object $o_i^p$ is dominated by some objects $n^p \in ST$, and may become the skyline after these objects $n^p$ expire (i.e., $n^p.exp < o_i^p.exp$). In this case, object $n^p$ should be the child of one of these objects $n^p$, since $Pr\{n^p \prec o_i^p\} \ge 1-\alpha$ and $n^p.exp < o_i^p.exp$. Therefore, the $ST$ index contains all the objects $o_i^p \in pDS$ that have the chance to be skylines before they expire.
$\hfill\square$

\vspace{-4ex}
\subsection{Proof of Property 2 of ST}
\vspace{-2ex}
\textit{Proof: }
Given an imputed object $o_i^p \in ST$, if it is not on the first layer of $ST$, $o_i^p$ will be dominated by its non-empty parent node (object) $n^p \in ST$ with probability $Pr\{n^p \prec o_i^p\} \ge 1-\alpha$. By substituting this probability into Eq.~(\ref{eq:max_corner_prob}), we can obtain
$P_{Sky\text{-}iDS}(o_i^p)\le 1-\sum_{\forall o_{il} \in o_i^p} o_{il}.p \cdot Pr\{n^p \prec o_{il}\}=1-Pr\{n^p \prec o_{il}\} \le \alpha$, that is, $P_{Sky\text{-}iDS}(o_i^p)\le \alpha$, which violates the Sky-iDS definition in Definition \ref{def:sky-iDS}.
Hence, object $o_i^p$ cannot become a skyline before its parent node expires from stream $iDS$.
$\hfill\square$

\vspace{-4ex}
\subsection{Proof of Property 3 of ST}
\vspace{-2ex}
\textit{Proof: }
According to \textit{Property 2}, we can get objects $n^p$ not on the first layer all have the skyline probabilities not bigger than $\alpha$ ($P_{Sky\text{-}iDS}(n^p) \le \alpha$). So current skyline objects must be all on the first layer of $ST$, in other words, the set of objects on the first layer of $ST$ is a superset of Sky-iDS answers.
$\hfill\square$


\section{The Correctness of Insertion/ Deletion and Refinement Algorithms}
\label{sec:proof_of_lemma_for_dm_of_st_and_ra}
\vspace{-2ex}
For insertion algorithm in Algorithm \ref{alg:alg_insertion}, (1) in line 4, $o_i^p$ can be inserted into layer 1, since no object $n^p$ on layer 1 of $ST$ can dominate $o_i^p$ with high probability (i.e., $\ge \alpha$), if $o_i^p$ can dominate any object $n^p$ on layer 1 with high probability (Lemma~\ref{lemma:lem4} as given later in Section \ref{subsec:lemma4}). (2) in line 24, we only check the descendant nodes of $n^p$ (with dominance probability $Pr\{n^p \prec o_i^p\} \ge 1-\alpha$), since the descendant of other objects (except $n^p$) cannot dominate $o_i^p$ with high confidence (Lemma~\ref{lemma:lem6}, as given later in Section \ref{subsec:lemma6}). (3) in line 30, based on Lemma~\ref{lemma:lem5} (as given later in Section \ref{subsec:lemma5}), if the object $o_i^p$ cannot be inserted into $ST$, $o_i^p$ cannot be used for maintaining $ST$. This is because $o_i^p$ cannot prune any object in $ST$. (4) in line 31, we start from layer $o_i^p.layer$ to maintain the $ST$, since $o_i^p$ cannot dominate objects on layers where its ancestors stay (Property 4 in Section~\ref{sub:sec:property4_of_ST}). (5) in lines 33-40, $o_i^p$ can inherit the descendant of $n^p$, due to the dominance transitivity (Lemma~\ref{lemma:lem6} in Section \ref{subsec:lemma6}).


\vspace{-4ex}
\subsection{Property 4 of ST}
\label{sub:sec:property4_of_ST}
\vspace{-2ex}
\textit{Property 4.} {\bf (Layer Dominance Rule)}
For any object $o_i^p$ on layer $o_i^p.layer$ of $ST$, $o_i^p$ cannot dominate any other objects $n^p$ with probability not smaller than $1-\alpha$ not only on layers $o_i^p.layer$ (sibling objects), but also on layers where its ancestor stays ($n^p.layer$ $< o_i^p.layer$).

\textit{Proof: }
Similar to the proof of \textit{Property 1} of $ST$, we can prove \textit{Property 4} by showing that no such an object $o_i^p$ exists, where $o_i^p$  dominates some of its sibling objects (on layer $o_i^p.layer$) or some objects on its ancestors' layers (on layers $\le o_i^p.layer$) with probability not smaller than $1-\alpha$.

First, based on the definition of the skyline tree in Section~\ref{subsec:skyline_tree}, any object cannot dominate its sibling objects with probability not smaller than $1-\alpha$.

Second, assume that object $o_i^p$ actually dominates some objects, $n^p$, of its ancestors' sibling objects (i.e., $n^p.layer < o_i^p.layer$) with probability not smaller than $1-\alpha$. There are two cases: 
\vspace{-1ex}
\begin{itemize}
\item Case (1): objects $o_i^p$ may expire earlier than $n^p$ (denoted as ), or; 
\item Case (2): objects $o_i^p$ may expire later than $n^p$. 
\end{itemize}
\vspace{-1ex}
Based on the definition of $ST$ in Section~\ref{subsec:skyline_tree}, for Case (1), $n^p$ should be one of the descendant objects of $o_i^p$; for Case (2), $n^p$ can be pruned by $o_i^p$, and $n^p$ should not stay in $ST$. Therefore, object $o_i^p$ cannot dominate any objects $n^p$ on layers $n^p.layer$ with probability not smaller than $1-\alpha$, where $n^p.layer \le o_i^p.layer$.
$\hfill\square$


\vspace{-4ex}
\subsection{Lemma~\ref{lemma:lem4}}
\label{subsec:lemma4}
\vspace{-2ex}
\begin{lemma} {\bf (Skyline Candidate)}
Given a new object $o_i^p$ from the data stream $iDS$, if no object $n^p$ on layer 1 of $ST$ can dominate $o_i^p$ with probability not smaller than $(1-\alpha)$, then $o_i^p$ is a skyline candidate and can be inserted into layer 1 of $ST$.
\label{lemma:lem4}
\end{lemma}

\vspace{-2ex}
\textit{Proof: }
If no object on layer 1 of $ST$ can dominate $o_i^p$ with probability greater than or equal to $(1-\alpha)$, based on the parent-child constraint condition in Section~\ref{subsec:skyline_tree}, object $o_i^p$ do not have parent node. In other words, object $o_i^p$ is a skyline candidate, and should be inserted into the first layer of $ST$.
$\hfill\square$

\vspace{-4ex}
\subsection{Lemma~\ref{lemma:lem5}}
\label{subsec:lemma5}
\vspace{-2ex}
\begin{lemma} {\bf (Pruned Object)}
Given a new object $o_i^p$ from the data stream $iDS$, if $o_i^p$ cannot be inserted into $ST$, then $o_i^p$ cannot be used for pruning any object in $ST$.
\label{lemma:lem5}
\end{lemma}

\vspace{-2ex}
\textit{Proof: }
If a new object $o_i^p$ cannot be inserted into $ST$, it means that $o_i^p$ is pruned by some object, $n^p$, in $ST$, where $Pr\{n^p \prec o_i^p\} \ge 1-\alpha$ and $n^p.exp \ge o_i^p.exp$ hold. However, $o_i^p$ may dominate some objects, $q^p$, in $ST$ with probability not smaller than $1-\alpha$. Even in this case, $o_i^p$ cannot prune any object in $ST$. We will prove this in two cases: (1) $o_i^p$ expires earlier than $q^p$ (i.e., $o_i^p.exp < q^p.exp$), and (2) $o_i^p$ expires later than or at the same time as $q^p$  (i.e., $o_i^p.exp \ge q^p.exp$).

First, for Case (1), if $o_i^p.exp < q^p.exp$ holds, object $q^p$ is also dominated by object $n^p$, but $q^p$ will expire after $n^p$, so $q^p$ is still a potential skyline candidate after $n^p$ expires and still stays in $ST$.

Second, for Case (2), if $o_i^p.exp \ge q^p.exp$ holds, $q^p$ should have already been pruned by object $n^p$ before $o_i^p$ arrives at the data stream. Therefore, object $o_i^p$ cannot prune any object in $ST$.
$\hfill\square$

\vspace{-4ex}
\subsection{Lemma~\ref{lemma:lem6}}
\label{subsec:lemma6}
\vspace{-2ex}
\begin{lemma} {\bf (Dominance Transitivity)}
When a new object $o_i^p$ is inserted into $ST$, $o_i^p$ may dominate a set of objects $n^p$ in $ST$ with probabilities not smaller than $1-\alpha$. If $o_i^p.exp \ge n^p.exp$ holds, objects $n^p$ cannot be skylines till they expire, and can be removed from $ST$; if $o_i^p < n^p.exp$ holds, then $o_i^p$ may be the new parent of $n^p$.
\label{lemma:lem6}
\end{lemma}

\vspace{-2ex}
\textit{Proof: }
If a new object $o_i^p$ is inserted into $ST$, based on the definition of $ST$ in Section~\ref{subsec:skyline_tree}, no object in $ST$ can both satisfy the two conditions: (1) dominates $o_i^p$ with probability not smaller than $1-\alpha$; (2) expires from data stream after $o_i^p$. 

In this case, assume that $o_i^p$ can dominate some objects, $n^p$, in $ST$ with probability not smaller than $1-\alpha$. Based on the expiration time of $o_i^p$ and $n^p$, we also have two cases: Case (3) $o_i^p.exp \ge n^p.exp$; and Case (4) $o_i^p.exp < n^p.exp$. 

For Case (3), objects $n^p$ cannot have the chance to be skylines in their lifetimes due to the existence of $o_i^p$, and thus can be removed from $ST$.

For Case (4), we need to further check the expiration time between $o_i^p$ and the parent node, $a^p$, of object $n^p$. If $a^p.exp \ge o_i^p$ holds, $a^p$ will still be the parent of $n^p$ (since $a^p$ is the last object to expire from $ST$ among objects that dominate $o_i^p$ with probability not smaller than $1-\alpha$). Similarly, if $o_i^p.exp > a^p$ holds, $o_i^p$ will replace $a^p$ to be the new parent of $o_i^p$.
$\hfill\square$


\vspace{-4ex}
\subsection{Lemma~\ref{lemma:lem7}}
\label{subsec:lemma7}
\vspace{-2ex}
\begin{lemma} {\bf (Full Update)}
If a new object $o_i$ arrives from the data stream $W_t$ at timestamp $t$, some skyline objects in $A_{t-1}$ at timestamp $(t-1)$ may not be skylines any more due to this new object $o_i$. Therefore, the current skyline anwer set $A_t$ should be re-calculated.
\label{lemma:lem7}
\end{lemma}

\vspace{-2ex}
\textit{Proof: }
Based on Eq.~(\ref{eq:eq14}), to check whether or not an object to be a skyline, we need to consider each valid object in data stream. At timestamp $t$, if a new object $o_i$ arrives from data streams, the Sky-iDS probabilities $P_{Sky\text{-}iDS} (\cdot)$ of skylines in $A_{t-1}$ may decrease (by multiplying the probability w.r.t. $o_i$ $\leq 1$ in Eq.~(\ref{eq:eq14})). Thus, some skylines in $A_{t-1}$ at timestamp $(t-1)$ may fail to be skyline at timestamp $t$, and all objects in $A_{t-1}$ should be re-checked. 
$\hfill\square$

\vspace{-4ex}
\subsection{Lemma~\ref{lemma:lem8}}
\label{subsec:lemma8}
\vspace{-2ex}
\begin{lemma} {\bf (Partial Update)}
When some objects $o_i^p$ expire from $W_{t-1}$ at timestamp $t$, as long as no new object is added to $W_t$, the remaining skyline objects in $A_{t-1}$ are still skyline objects (i.e., $A_{t-1} \subseteq A_t$). Objects on the first layer of $ST$, but not in $A_{t-1}$, may have chance to be skylines and need to be re-checked.
\label{lemma:lem8}
\end{lemma}
\vspace{-1ex}
\textit{Proof: }
When some objects expire in $W_{t-1}$ and no new object arrives at timestamp $t$, all the remaining valid objects in $ST$ will have the same or larger Sky-iDS probabilities (based on Eq.~(\ref{eq:eq14}), by removing some probability terms w.r.t. the expired objects). In this case, the remaining objects in the skyline answer set $A_{t-1}$ are still skylines (i.e., in $A_t$) at timestamp $t$. Other candidates on the first layer of $ST$ may also have chances to be skylines at timestamp $t$, and thus should be re-checked.
$\hfill\square$

\section{Derivation of Cost Model}
\label{sec:derivation_of_cost_model}

\subsection{Cost Model}
\label{extra_cost_model} 

\nop{
As shown in Figure \ref{fig:extracost}, the query rectangle and total cost is based on a specific \textit{DD} from conceptual lattices.
For different DDs in conceptual lattices, the responding query ranges and total cost are different.
In the sequel, we first explore how to calculate the total cost for one DD, and then deduce the cost model based on all imputed DDs in conceptual lattices.
We use $\Omega^{'}$ and $|\Omega^{'}|$ represent the set and number of DDs in d different lattices.

To obtain the cost model based on a DD $Y \to A_j$, we need to estimate the number of data points falling into the areas of query ranges and the actual accessed cells.
Inspired by \cite{lian2010generic}, we use power law \cite{belussi1998self} to obtain the approximate estimation.
And according to \cite{belussi1998self}, we can obtain this approximation estimation by using the volume ratio between query shape (query range) and a standard hypercube (or square for 2-dimension) taking as its side length the length on X Axis of query shape.
}

\vspace{-2ex}We provide a cost model to tune the parameter $u$ (i.e., the side length of each cell in the grid) for index $\mathcal{I}_j$ over $R$ (discussed in Section \ref{sec:index_over_R_for_imp}). We formally define the total cost, $Cost$, of accessing the grid, which contains two types of costs, $cost_{cell}$ and $cost_{extra}$, that access cells and false alarms, respectively. 

\begin{scriptsize}
\begin{eqnarray}
Cost=\beta \cdot cost_{cell} + (1-\beta) \cdot cost_{extra},\label{eq:cost_model}
\end{eqnarray}
\end{scriptsize}

\noindent where $\beta$ is a parameter to make a trade-off between the two costs $cost_{cell}$ and $cost_{extra}$. Note that, for $cost_{extra}$, we can use the \textit{power law} \cite{belussi1998self} to estimate the number of false alarms that should be checked with extra cost.

\nop{
Therefore, to select optimal $u$ that minimizes the total cost $Cost$, we will take the derivative of $Cost$ to $u$, and let it be 0, that is, $\frac{\partial Cost}{\partial u}=0$.
}

As shown in Figure \ref{fig:extracost}, to impute the missing attribute $A_j$, we will access all grid cells in index $\mathcal{I}_j$ that intersect with query range $Q$ (inferred from DDs), and retrieve objects in these grid cells that fall into $Q$. Note that, here we may need extra efforts to refine objects in those cells that partially overlap with $Q$ (i.e., the region with the sloped lines in Figure \ref{fig:extracost}).  

Intuitively, when the size, $u$, of grid cells is large (e.g., the entire data space is just one cell in the extreme case), the number of cells we need to access and check is small, but it takes more extra time to refine candidates for cells partially intersecting with $Q$ (i.e., regions with the sloped lines). On the other hand, when the cell size, $u$, is small, we need to check more cells (with higher cost), but refine fewer false alarms (due to smaller area of the region with extra cost). Thus, our goal is to select the best $u$ value such that the total cost is minimized (making a balance between the costs of checking cells and refining false alarms). 

\nop{
So we formally define the total cost, $Cost$, as shown in Eq. (\ref{eq:cost_model}). }
To explore how to calculate $Cost$,
we first explore how to calculate the extra cost for one \textit{DD}, $Y \to A_j$ (for $1 \le j \le d$), and then deduce the cost model based on all imputed DDs from $d$ conceptual lattices $Lat_j$.
To obtain the extra cost model based on a single \textit{DD}, we need to estimate the number of data points falling into the areas of query ranges and the actually accessed cells.
Inspired by \cite{lian2010generic}, we use power law \cite{belussi1998self} to obtain the approximate estimation.
According to \cite{belussi1998self}, we can obtain this approximate estimation by using the volume ratio between query shape (query range) and a standard hypercube (or square for 2-dimension) taking as its side length the length on x-axis of the query shape.

As mentioned in Section \ref{imputing_of_DD}, a DD can be represented as $\{Y \to A_j, \phi[Y A_j]\}$, where $\phi[Y A_j]$ is the differential function of the DD on determinant attribute set $Y$ and dependent attribute $A_j$.
As discussed in Section \ref{subsec:data_imputation_via_DDs}, for each attributes $A_x \in Y$, we use $A_x.I$ to represent the difference interval tolerated by DD on attribute $A_x$, where $A_x.I = [o,\epsilon_{A_x}]$.
Based on the tolerance intervals $A_x.I$ of attributes $A_x$ in determinant attribute set $Y$ of a DD, we can deduce the edge lengths, denoted by $l_x.query$ and $l_x.actual$, of query shape and actual accessed shape, respectively, of incomplete object on attribute $A_x$.

For the query shape of the incomplete object $o_i$ based on DD $Y \to A_j$, its length $l_x.query$ equal to: $2\epsilon_{A_x}$ when $A_x \in Y$ and $l_x$ when $A_x \notin Y$, where $l_x$ is the length of dataset space on attribute $A_x$. 
\nop{
Since we assume the addressing space is fixed, we represent the length $l_j$ as $C_j \cdot u$, which is the constant multiple ($C_j$) of the length of grid cell $u$. 
}
That is, we have:

\begin{scriptsize}
\begin{eqnarray}
l_x.query=
\begin{cases}
2\epsilon_{A_x}& A_x \in Y;\\
l_x &A_x \notin Y.
\end{cases}
\label{eq:eq5}
\end{eqnarray}
\end{scriptsize}

For actually accessed shape, in the worst case, its lengths $l_x.actual$ equal to $(\left\lfloor \frac{2\epsilon_{A_x}}{u}\right\rfloor +2)\cdot u$ when $A_x \in Y$ and equal to $l_x$ when $A_x \notin Y$. That is, we obtain:

\begin{scriptsize}
\begin{eqnarray}
l_x.actual=
\begin{cases}
\left(\left\lfloor \frac{2\epsilon_{A_x}}{u}\right\rfloor + 2\right)\cdot u& A_x \in Y;\\
l_x& A_x \notin Y.
\end{cases}
\label{eq:eq6}
\end{eqnarray}
\end{scriptsize}

For the side length, denoted as $l_{DD}$, of the contrastive d-dimensional hypercube, it equals to $2\epsilon_{A_1}$ when $A_1 \in Y$ or $l_1$ when $A_1 \notin Y$, where $A_1$ represents the $A_1$ Axis.
Furthermore, we divide $l_{DD}$ into $l_{DD}.query$ and $l_{DD}.actual$, which represent the lengths of standard contrastive hypercubes of query shape and actual accessed shape, respectively.
The reason that we use the length of query shape in $A_1$ Axis to represent all sides of contrastive hypercube is due to the self-similarity of data space \cite{belussi1998self}.
The calculation of side lengths of the contrastive hypercube of query shape and actual accessed shape is shown in Eqs.~(\ref{eq:eq7}) and (\ref{eq:eq8}), separately.

\vspace{-2ex}
\begin{scriptsize}
\begin{eqnarray}
l_{DD}.query=
\begin{cases}
2\epsilon_{A_1}& A_1 \in Y;\\
l_1& A_1 \notin Y;
\end{cases}
\label{eq:eq7}
\end{eqnarray}
\end{scriptsize}
\vspace{-2ex}
\begin{scriptsize}
\begin{eqnarray}
l_{DD}.actual=
\begin{cases}
(\left\lfloor \frac{2\epsilon_{A_1}}{u}\right\rfloor + 2)\cdot u& A_1 \in Y;\\
l_1& A_1 \notin Y.
\end{cases}
\label{eq:eq8}
\end{eqnarray}
\end{scriptsize}

With Eqs.~(\ref{eq:eq5})$\sim$(\ref{eq:eq8}), according to the power law \cite{belussi1998self}, we can obtain the approximate estimation of the number of data points falling into the query shape or accessed shape, which is shown in Eq.~(\ref{eq:eq9}).

\begin{scriptsize}
\begin{eqnarray}
&&\overline{nb}(\epsilon_{DD},shape)  \\
&=&\left(\frac{Vol(\epsilon_{DD},shape)}{Vol(\epsilon_{DD},hypercube)}\right)^{\frac{D_2}{d}} \times (N-1) \times 2^{D_2} \times (\epsilon_{DD})^{D_2} \nonumber\\
&=&(N-1) \times 2^{D_2} \times \left(\frac{l_{DD}.z}{2}\right)^{D_2} \times \left( \prod_{x=1}^d \frac{l_x.z}{l_{DD}.z} \right)^{\frac{D_2}{d}} \nonumber\\
&=&(N-1) \times \left(\prod_{x=1}^d l_x.z\right)^{\frac{D_2}{d}},\nonumber
\label{eq:eq9}
\end{eqnarray}
\end{scriptsize}

\noindent where $\epsilon_{DD}$ is the radius of the shape enclosed by intervals of DD, $D_2$ is the correlation fractal dimension of data space \cite{belussi1998self}, and $z$ can be set as $query$ when \textit{shape} is the query shape or $actual$ when \textit{shape} is the actually accessed shape.
\nop{
The explicit deducing of Eq.~(\ref{eq:eq9}) can be found in Appendix A.
}

With Eq.~(\ref{eq:eq9}), we can get the number of searching points in the extra cost area, which is given by $\overline{nb}(\epsilon_{DD},actual) - \overline{nb}(\epsilon_{DD},query)$, where $\epsilon_{DD}$ is the radius based on range shape (e.g. query or actual).
Combing all DDs, we can propose our cost model as follows.

\begin{scriptsize}
\begin{eqnarray}
Cost &=& \beta \cdot cost_{cell} + (1-\beta) \cdot cost_{extra} \label{eq:eq10}\\
&=& \sum_{DD \in \Omega'} \left( \beta \cdot t_{cell} \cdot \prod_{A_x \in Y} \left(\left\lfloor \frac{2\epsilon_{A_x}}{u}\right\rfloor + 2\right) \cdot \prod_{A_x \notin Y} \frac{l_x}{u} \right. \nonumber\\
&& \left. + (1-\beta) \cdot t_{s_r} \cdot ( \overline{nb}(\epsilon_{DD},actual) - \overline{nb}(\epsilon_{DD},query)) \right),\nonumber
\end{eqnarray}
\end{scriptsize}

\noindent where $\Omega^{'}$ is the set of DDs in $d$ different lattices, $t_{s_r}$ and $t_{cell}$ are the unit time costs to search a single complete object $s_r \in R$ and a single cell, respectively, and
$\beta$ is the coefficient of the trade-off between the time costs $t_{s_r}$ and $t_{cell}$.
Since $\left\lfloor \frac{2\epsilon_{A_x}}{u}\right\rfloor$ is within $\left(\frac{2\epsilon_{A_x}}{u}-1,\frac{2\epsilon_{A_x}}{u}\right]$, we use $\frac{2\epsilon_{A_x} - \Delta_x}{u}$ to replace $\left\lfloor \frac{2\epsilon_{A_x}}{u}\right\rfloor$, where $\Delta_x \in [0,u)$.
To get the optimal value of $u$, we take into consideration the worst case of the value of $\left\lfloor \frac{2\epsilon_{A_x}}{u}\right\rfloor$.
That is, we set $\Delta_x = 0$, and then get $\left\lfloor \frac{2\epsilon_{A_x}}{u}\right\rfloor = \frac{2\epsilon_{A_x}}{u}$.
We take Eqs.~(\ref{eq:eq5})$\sim$(\ref{eq:eq9}) into Eq.~(\ref{eq:eq10}), and we obtain Eq.~(\ref{eq:eq11}).

\vspace{-2ex}
\begin{scriptsize}
\begin{eqnarray}
Cost
&=& \sum_{DD \in \Omega^{'}} \left( t_{cell} \cdot \beta \cdot \prod_{A_x \in Y} \left(\frac{2\epsilon_{A_x}}{u}+2 \right) \cdot \prod_{A_x \notin Y} \frac{l_x}{u} \right.\nonumber\\
&& \left.+ (N-1) \cdot t_{s_r} \cdot (1-\beta) \cdot \prod_{A_x \notin Y} l_x^{\frac{D_2}{d}} \right. \nonumber \\
&&\left. \cdot \left(\prod_{A_x \in Y} (2\epsilon_{A_x}+2u)^{\frac{D_2}{d}} - \prod_{A_x \in Y} (2\epsilon_{A_x})^{\frac{D_2}{d}} \right)  \right),\label{eq:eq11}
\end{eqnarray}
\end{scriptsize}

\noindent where $\beta$, $t_{s_r}$ and $t_{cell}$ are given in Eq.~(\ref{eq:eq10}), $D_2$ is the correlation fractal dimension of data space \cite{belussi1998self}.

\nop{
where $a=\lbrace \sum_{j=0}^{r_i}  \left[\sum_{g=1}^{|C_{r_i}^j|} \prod_{h \in o_g} 2\epsilon_{ih} \right] \cdot (2u)^{r_i-j} \rbrace \cdot (l_j)^{d-r_i}$,
$r_i$ is the number of attributes in determined attribute set $o_{dom}^i$ of $DD_i$,
$|C_{r_i}^j|$ is the number of all possible combinations of choosing $|j|$ different elements from $|r_i|$ elements,
and $o_g$ is one combination set of the $|C_{r_i}^j|$ combination sets. 
Especially, $o_{C_{r_i}^0} = \O$.
}

With the cost model in Eq.~(\ref{eq:eq11}), we can find the optimal length $u$ of cells in grid index that can reduce the extra cost to the lowest level by calculating the derivative of Eq.~(\ref{eq:eq11}) to $u$ by $\frac{\partial Cost}{\partial u}=0$, that is,

\begin{scriptsize}
\begin{eqnarray}
&&\frac{\partial Cost}{\partial u} 
=  b \cdot \sum_{DD \in \Omega^{'}}  c \cdot \left( \frac{1}{u^d}  \cdot \sum_{x=0}^{|Y|} f 
  \cdot e \cdot u^{e-1} - \frac{d}{u^{d+1}} \cdot \sum_{x=0}^{|Y|} f \cdot u^{e} \right) \nonumber\\
&& + a \cdot \sum_{DD \in \Omega^{'}} \left( c^{\frac{D_2}{d}} \cdot (\sum_{x=0}^{|Y|} f \cdot u^{e})^{\frac{D_2-d}{d}}  \cdot \sum_{x=0}^{|Y|} f \cdot e \cdot u^{e-1} \right)= 0,\label{eq:eq12}
\end{eqnarray}
\end{scriptsize}

\noindent where $|Y|$ is the number of attributes on determinant attribute set $Y$ of DD, $Y \to A_j$, $a=(N-1)\cdot t_{s_r} \cdot \beta \cdot \frac{D_2}{d}$, $b=t_{cell} \cdot (1-\beta)$, $c=2^{|Y|} \cdot \prod_{A_x \notin Y} l_x$, $f = {Y \choose x} \cdot \prod_{\forall A_x} \epsilon_{A_x}$ and $e=|Y|-x$.

Especially, ${Y \choose x}$ is to choose $x$ ($x \le |Y|$) different attributes $A_x$ from the determinant attribute set $Y$ of DD $Y \to A_j$, $\prod_{\forall A_x} \epsilon_{A_x}$ is to compute the product of ranges $\epsilon_{A_x}$ of attributes $A_x$ within a choosing, 
and $f$ is to compute the product sum of all possible choosing.

Eq.~(\ref{eq:eq11}) is actually a parabola with a positive binomial coefficient, and the optimal value of $u$ refers to the minimum value of Eq.~(\ref{eq:eq11}).
So there is only one solution of Eq.~(\ref{eq:eq12}).
It is complicated to directly compute Eq.~(\ref{eq:eq12}), and instead we use Algorithm \ref{alg:approximation_u} to get the approximation of optimal value of $u$ within error $\eta$ (set by user).

 \begin{algorithm}[t!]\scriptsize
\KwIn{Eq.~(\ref{eq:eq12}) with all needed coefficients, and the precision $\eta$} 
\KwOut{the approximation of optimal $u$ within error $\eta$}
$u.max \leftarrow$ the maximum domain value for all dimensions in the data space \\
$u.min \leftarrow 0$ \\
\While{$u.max-u.min \ge 2\eta$}{
    $u = \frac{u.max+u.min}{2}$ \\
    Substitute $u$ into Eq.~(\ref{eq:eq12}), and obtain result $rlt=\frac{\partial Cost}{\partial u}$ \\
    \If{$rlt > 0$}{
        $u.max \leftarrow u$
    }\Else{
        $u.min \leftarrow u$
    }
}
$u \leftarrow \frac{u.min + u.max}{2}$
\caption{Calculation of $u$ in Eq.~(\ref{eq:eq12})}
\label{alg:approximation_u}
\end{algorithm}

\subsection{Selection of the \textsl{\large u} Value}
\label{approximation_of_u}
\vspace{-2ex}
As depicted in Algorithm \ref{alg:approximation_u}, we first set the initial value of $u.max$ to the maximum domain value for all dimensions in the data space (line 1), and then set 0 as the initial smallest possible value $u.min$ (line 2). As long as the error between $u.max$ and $u.min$ is not within $2\eta$, Algorithm \ref{alg:approximation_u} uses binary search to shrink the range between the lower and upper bounds of $u$ (lines 3-10). Especially, after substituting $u$ into Eq.~(\ref{eq:eq12}), Algorithm \ref{alg:approximation_u} checks whether or not the solution is larger than 0 (lines 5-9). If the answer is yes, Algorithm \ref{alg:approximation_u} shrinks the upper bound $u.max$ of $u$ (line 7). If the answer is no, Algorithm \ref{alg:approximation_u} shrinks the lower bound $u.min$ of $u$ (line 9). When the difference between $u.max$ and $u.min$ is within error bound $2\eta$, we set the mean of $u.max$ and $u.min$ as the approximation of the optimal $u$ value (line 10), such that $\frac{\partial Cost}{\partial u} \approx 0$.


 \begin{figure}[t!]\scriptsize
\centering\vspace{-4ex}
\subfigure[][{\small maintenance time}]{\hspace{-2ex}
\scalebox{0.21}[0.21]{\includegraphics{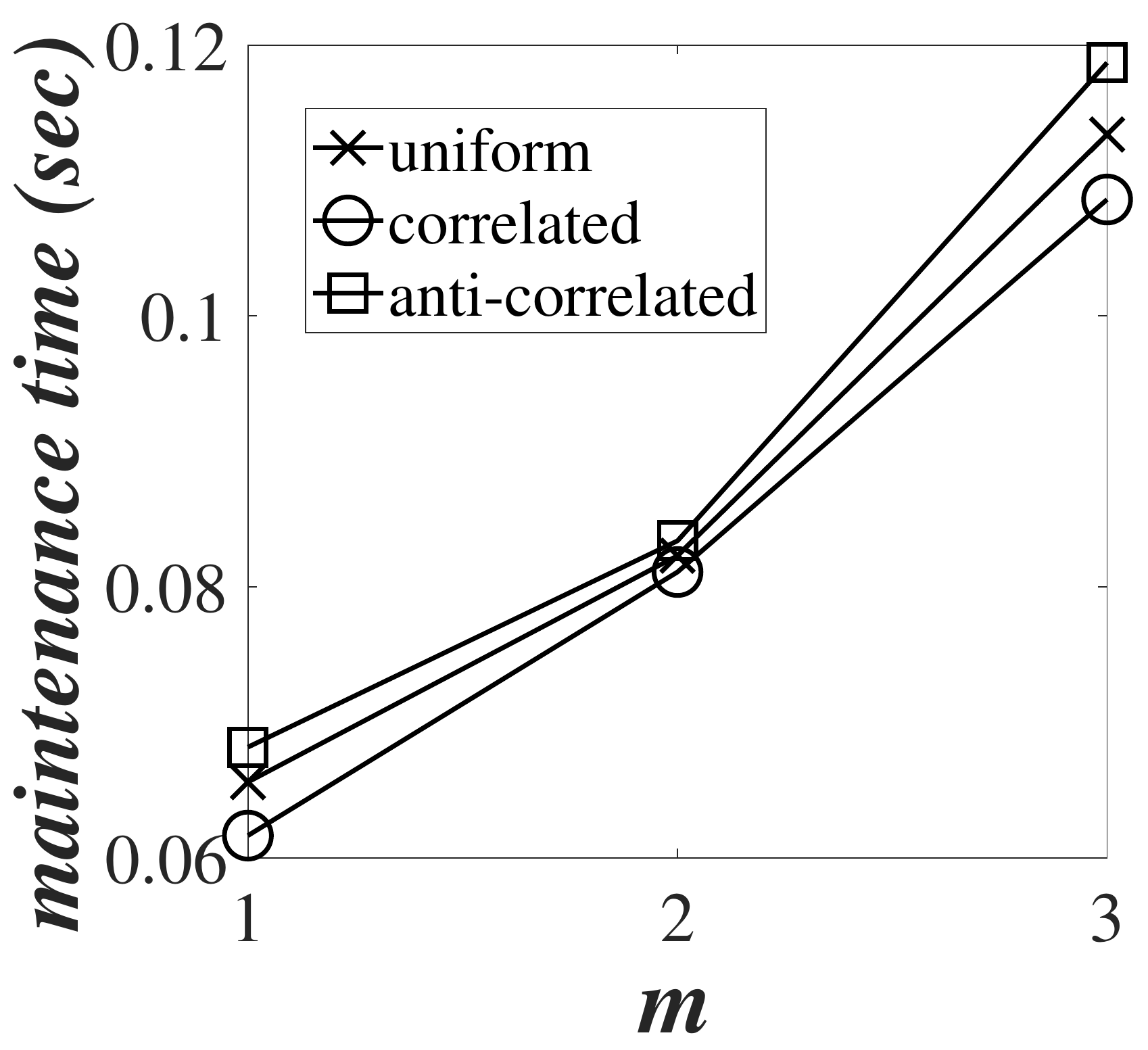}}
\label{subfig:m_cost_vs_m}
}%
\subfigure[][{\small query time}]{\hspace{-1ex}           
\scalebox{0.21}[0.21]{\includegraphics{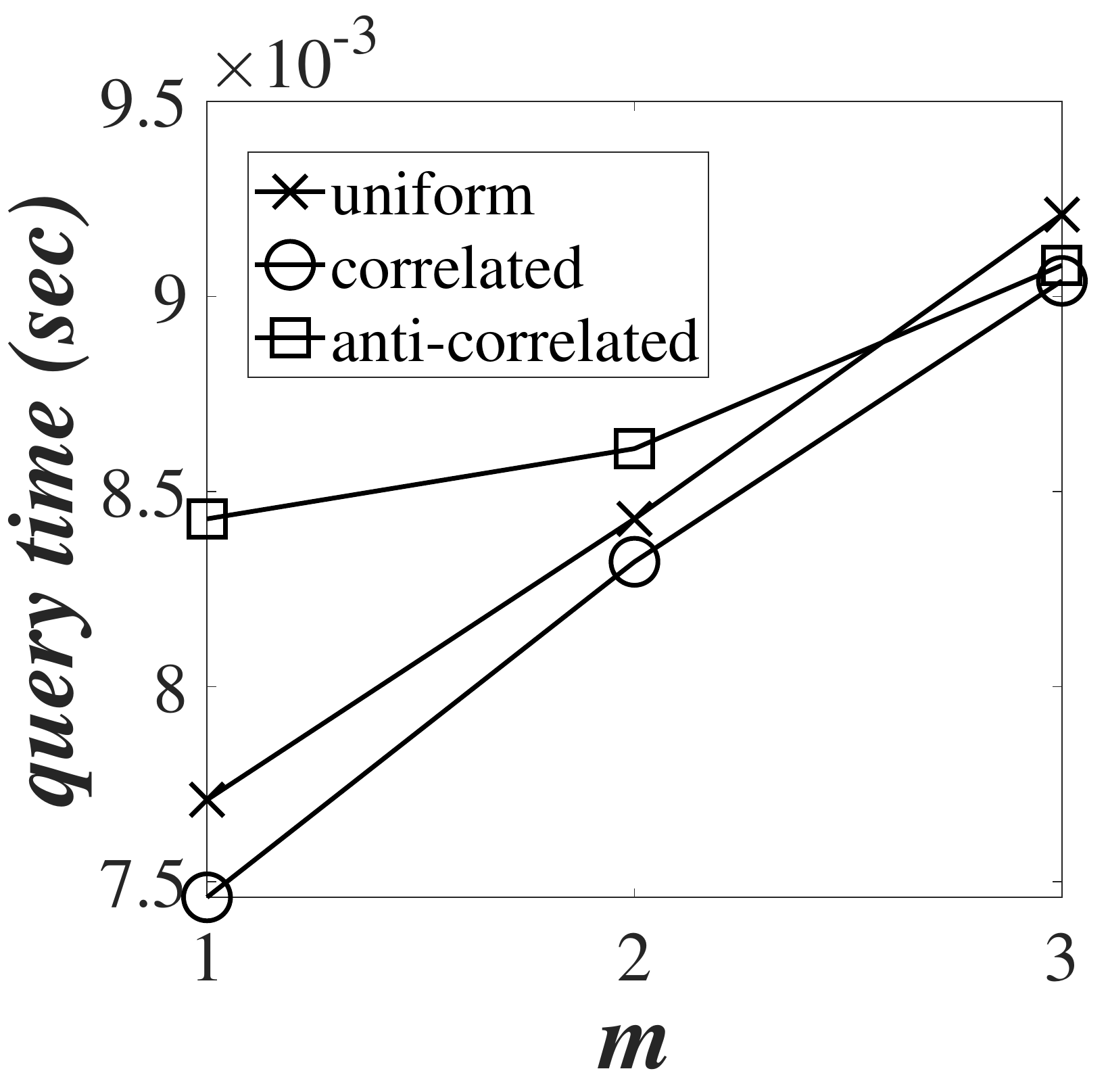}}
\label{subfig:q_cost_vs_m}
}\vspace{-4ex}
\caption{\small The efficiency vs. No., $m$, of missing attributes.} 
\label{exper:Sky-iDS_vs_m} 
\end{figure}
\begin{figure}[t!]
\centering\vspace{-4ex}
\subfigure[][{\small maintenance time}]{\hspace{-2ex}
\scalebox{0.2}[0.2]{\includegraphics{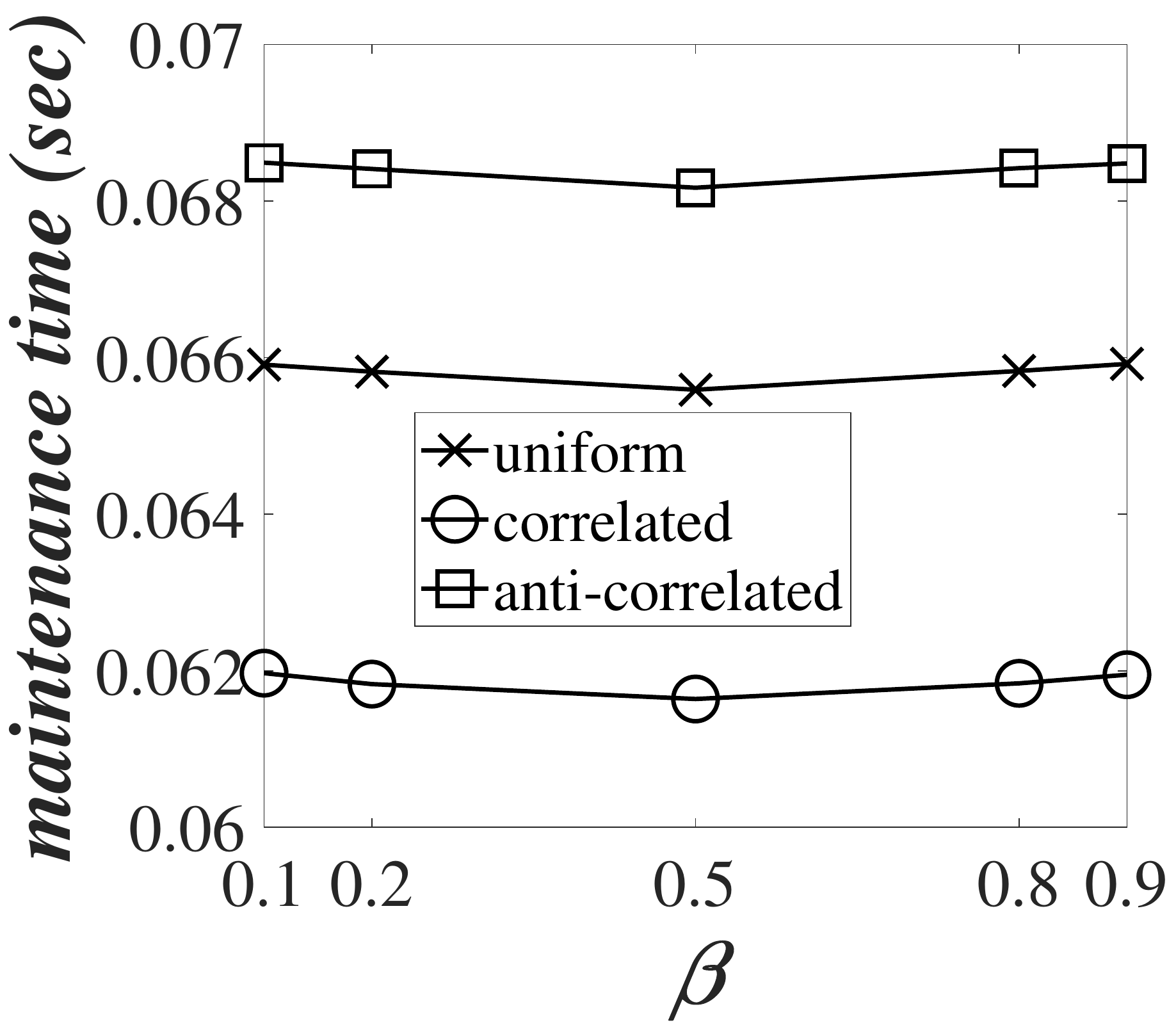}}
\label{subfig:m_cost_vs_beta}
}
\subfigure[][{\small query time}]{\hspace{-1ex}         
\scalebox{0.2}[0.2]{\includegraphics{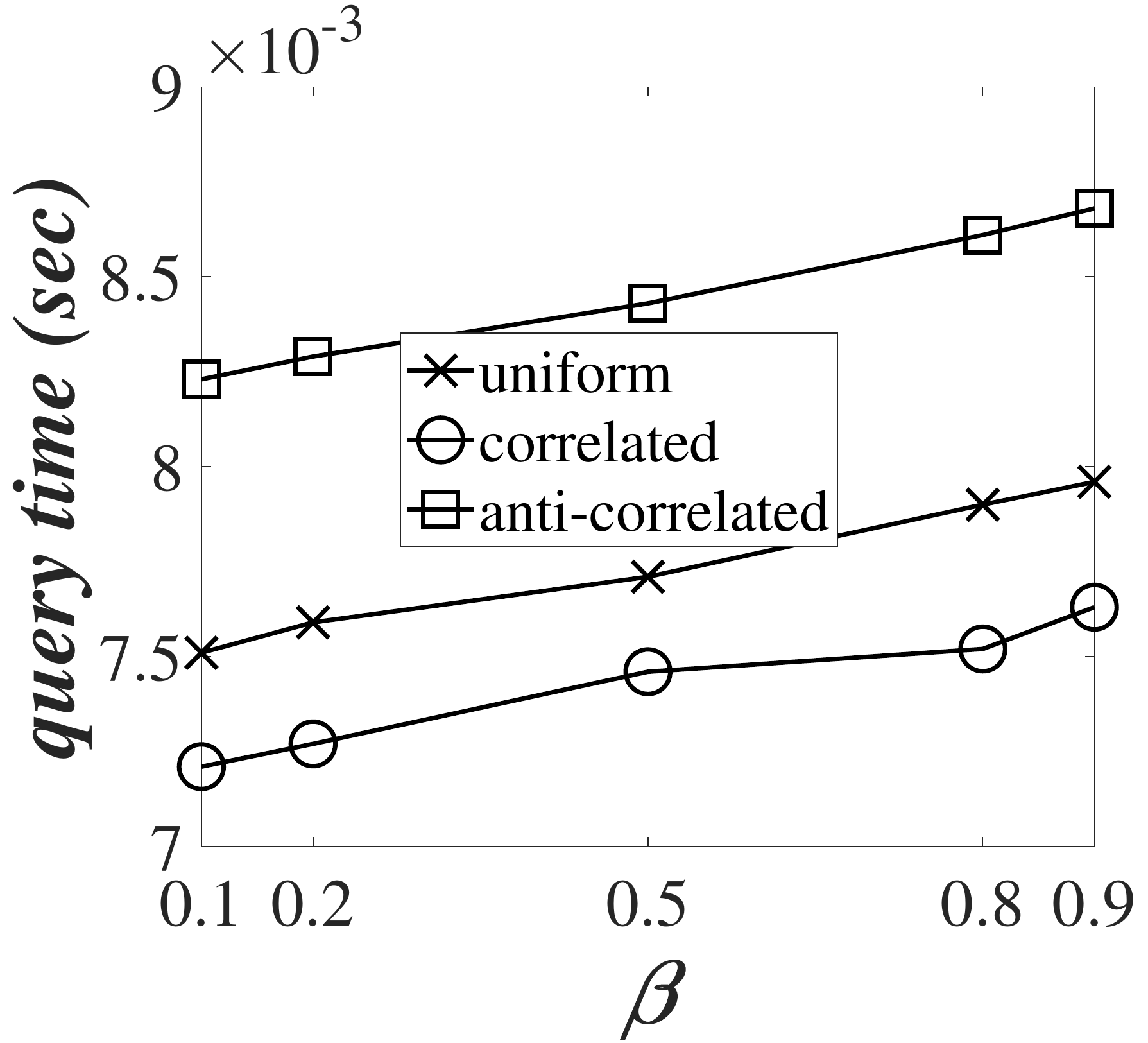}}
\label{subfig:q_cost_vs_beta}
}\vspace{-4ex}
\caption{\small The efficiency vs. the coefficient, $\beta$.} 
\label{exper:Sky-iDS_vs_beta} \vspace{-0ex}
\end{figure}

\section{More Experimental Results}
\label{sec:more_exp}

\noindent {\bf The Sky-iDS performance vs. the number, $m$, of missing attributes.} Figure~\ref{exper:Sky-iDS_vs_m} varies the number, $m$, of missing attributes of data objects from 1 to 3 (other parameters are set to default values), and shows the effect of parameter $m$ for our Sky-iDS approach. From experimental results, with more missing attributes, the maintenance and query times increase. This is because, we need to impute more attributes in objects, and refine skyline candidates with more uncertain attributes. Nevertheless, the time costs remain low (i.e., less than 0.1187 $sec$ for the maintenance, and 0.00921 $sec$ for the query cost).



\noindent {\bf The Sky-iDS performance vs. the coefficient, $\beta$, of the cost model.} Figure~\ref{exper:Sky-iDS_vs_beta} shows the effect of coefficient parameter, $\beta$, in the cost model on the Sky-iDS performance, where $\beta = 0.1, 0.2,$ $0.5, 0.8,$ and $0.9$, and other parameters are by default. From the figures, when $\beta$ is small or large (e.g., 0.1 or 0.9), the maintenance time is large; when $\beta$ is set to around 0.5, the maintenance time is the lowest. In particular, $Correlated$ has the lowest maintenance time, due to lower imputation cost over sparse data.

In Figure \ref{subfig:q_cost_vs_beta}, with larger $\beta$, the query cost increases smoothly. This is because, when $\beta$ is set to a smaller value, the cell length, $u$, will be large, which requires to access more samples $s_r$ in the data repository $R$ to impute. This can help obtain more accurate imputed objects, and fewer imputed objects will show up on the first layer of $ST$, which needs smaller query time. For all the three data sets, with different $\beta$ values, the query time remains low (i.e., 0.00763 $\sim$ 0.00868 $sec$). This is because not many new skyline candidates need to be incrementally updated in skyline answer set $A_t$.


\end{document}